\documentclass[usenatbib]{mn2e} \usepackage{graphicx}
\usepackage{ifthen} \usepackage{url}

\def\ltsima{$\; \buildrel < \over \sim \;$}
\def\lta{\lower.5ex\hbox{\ltsima}}
\def\gtsima{$\; \buildrel > \over \sim \;$}
\def\simgt{\lower.5ex\hbox{\gtsima}}
\def\kms{{\rm\,km\,s^{-1}}}

\def\msun{{\rm\,M_\odot}}
\def\lsun{{\rm\,L_\odot}}

\def\pc{{\rm\,pc}}

\newcommand{\fmmm}[1]{\mbox{$#1$}}

\newcommand{\scnp}{\mbox{\fmmm{''}}}
\newcommand{\mcnd}{\mbox{\fmmm{'}\hskip-0.3em .}}
\newcommand{\mcnp}{\mbox{\fmmm{'}}}

\def\AA{$\; \buildrel \circ \over {\rm A}$}

\def\deg{^\circ}

\def\s{\ifmmode \widetilde \else \~\fi}
\def\={\overline}

\def\spose#1{\hbox to 0pt{#1\hss}}

\def\lta{\mathrel{\spose{\lower 3pt\hbox{$\mathchar"218$}}
     \raise 2.0pt\hbox{$\mathchar"13C$}}}
\def\gta{\mathrel{\spose{\lower 3pt\hbox{$\mathchar"218$}}
     \raise 2.0pt\hbox{$\mathchar"13E$}}}
\def\Dt{\spose{\raise 1.5ex\hbox{\hskip3pt$\mathchar"201$}}}    
\def\dt{\spose{\raise 1.0ex\hbox{\hskip2pt$\mathchar"201$}}}    

\def\dotsfill{\leaders\hbox to 1em{\hss.\hss}\hfill}

\def\FeH{{\rm[Fe/H]}}

\def\andix{And\,IX}
\def\andxi{And\,XI}
\def\andxii{And\,XII}
\def\andxiii{And\,XIII}
\def\ec4{EC4}

\loadboldmathitalic 
\title[]{A Keck/DEIMOS spectroscopic survey of the
  faint M31 satellites $\andix, \andxi, \andxii,$ and $\andxiii^{1,2}$}
\author[M.\ L.\ M.\ Collins et al.]  {M. L. M. Collins $^{3}$,
  S. C. Chapman$^{3}$, M. J. Irwin$^{3}$, N. F. Martin$^{4}$,
  R. A. Ibata$^{5}$, \newauthor D.\ B.\ Zucker $^{6,7}$,
  A.\ Blain$^{8}$, A.\ M.\ N. Ferguson$^{9}$, G. F. Lewis$^{10}$,
  A.\ W.\ McConnachie$^{11}$, \newauthor
  J.\ Pe\~narrubia$^{3}$\\ $^{3}$ Institute of Astronomy, Madingley
  Road, Cambridge, CB3 0HA, U.K.\\ $^{4}$ Max-Planck-Institut for
  Astronomy, K\"onigstuhl 17, D-69117, Heidelberg, Germany\\ $^{5}$
  Observatoire de Strasbourg, 11, rue de l'Universit\'e, F-67000,
  Strasbourg, France\\ $^{6}$ Macquarie University NSW 2109, Australia
  \\$^{7}$Anglo-Australian Telescope, PO Box 296, Epping, NSW 1710, Australia\\$^{8}$ California Institute of Technology, 1200 E. California
  Blvd, MC 105-24, Pasadena, CA 91125 USA\\ $^{9}$ Institute for
  Astronomy, University of Edinburgh, Royal Observatory, Blackford
  Hill, Edinburgh, UK EH9 3HJ\\ $^{10}$ Sydney Institute for Astronomy,
  School of Physics, A29, University of Sydney, NSW 2006,
  Australia\\ $^{11}$ NRC Herzberg Institute of Astrophysics, 5071 West
  Saanich Road, Victoria, V9E 2E7, Canada\\ }

\date{\today}
\begin{document} 

\maketitle 
\begin{abstract} 
 We present the first spectroscopic analysis of the faint M31
 satellite galaxies, \andxi\ and \andxiii, as well as a reanalysis of
 existing spectroscopic data for two further faint companions,
 \andix\ (correcting for an error in earlier geometric modelling) and
 \andxii. By combining data obtained using the DEep Imaging
 Multi-Object Spectrograph (DEIMOS) mounted on the Keck~II telescope
 with deep photometry from the Suprime-Cam instrument on Subaru, we
 have identified the most probable members for each of the satellites
 based on their radial velocities (precise to several $\kms$ down to
 $i\sim22$). Using both the photometric and spectroscopic data, we
 have also calculated global properties for the dwarfs, such as
 systemic velocities, metallicities and half-light radii. We find each
 dwarf to be very metal poor ([Fe/H]$\sim-2$ both photometrically and
 spectroscopically, from their stacked spectrum), and as such, they
 continue to follow the luminosity-metallicity relationship
 established with brighter dwarfs.  We are unable to resolve
 dispersion for \andxi\ due to small sample size and low S:N, but we
 set a one sigma upper limit of $\sigma_v<$4.5 kms$^{-1}$.  For
 \andix, \andxii\ and \andxiii\ we resolve velocity dispersions of
 $\sigma_v$=4.5$^{+3.6}_{-3.4}$, 2.6$^{+5.1}_{-2.6}$ and
 9.7$^{+8.9}_{-4.5}$ kms$^{-1}$, though we note that the dispersion for
 \andxiii\ is based on just three stars. We derive masses within the
 half light radii for these galaxies of
 6.2$^{+5.3}_{-5.1}\times10^6\msun$,
 2.4$^{+6.5}_{-2.4}\times10^6\msun$ and
 1.1$^{+1.4}_{-0.7}\times10^7\msun$ respectively. We discuss each
 satellite in the context of the Mateo relations for dwarf spheroidal
 galaxies, and in reference to the Universal halo profiles established
 for Milky Way dwarfs \citep{walker09b}. Both \andix\ and XII fall
 below the universal halo profiles of \citet{walker09b}, indicating
 that they are less massive than would be expected for objects of
 their half-light radius. When combined with the findings of
 \citet{mcconnachie06b}, which reveal that the M31 satellites are
 twice as extended (in terms of both half-light and tidal radii) as
 their Milky Way counterparts, these results suggest that the
 satellite population of the Andromeda system could inhabit halos that
 with regard to their central densities -- are significantly different
 from those of the Milky Way.
\end{abstract}

\section{Introduction}
\footnotetext[1]{The data presented herein were obtained at the
  W.M. Keck Observatory, which is operated as a scientific partnership
  among the California Institute of Technology, the University of
  California and the National Aeronautics and Space
  Administration. The Observatory was made possible by the generous
  financial support of the W.M. Keck Foundation.}
\footnotetext[2]{Based in part on data collected at Subaru Telescope, which is operated by the National Astronomical Observatory of Japan.}

In recent years, the low luminosity population of dwarf spheroidal
galaxies (dSphs) has drastically increased in number, with $\sim$30
faint satellites discovered both within the halo of our Milky Way (MW)
and in that of our ``sister'' galaxy, M31
(e.g. \citealt{zucker04,willman05b,martin06,zucker06a,zucker06b,zucker07,irwin08,mcconnachie08,belokurov08,martin09}). Whilst
in principle these minuscule stellar associations should be some of
the simplest objects of the galactic family - which spans roughly
seven orders of magnitude in mass - to understand, the analysis of
faint dSphs (M$_V>-8$) within the MW has dissolved the notion that
they form in a simple, homogeneous fashion. Detailed studies (both
photometric and spectroscopic) of such objects reveal several unusual
properties for the population. For example, they highlight a
significant departure from the well-established mass-luminosity
relationship for dSphs \citep{mateo98}. A number of these newly
discovered objects are also quite controversial in their nature,
possessing properties that are common to both dSphs and globular
clusters and are therefore difficult to classify as either, such as
the unusual objects, Willman 1 and Segue 1
\citep{willman05a,belokurov07,niederste09}. In addition, comparisons of
the MW and M31 dSph systems have revealed significant differences
between the populations, where the classical M31 dwarfs
(M$_V<-8$) are observed to be at least twice as extended as
their Milky Way counterparts, in terms of both their half-light and
tidal radii \citep{mcconnachie06b}. In addition, the spatial
distribution of the M31 satellites around their host is more extended
than the Galactic satellite distribution, with approximately half of
the Galactic satellites found within 100 kpc of the Galaxy, compared
to roughly 200 kpc for M31 \citep{mcconnachie06a, koch06}. Such
differences indicate that environment plays a significant role in the
evolution of dSphs, the full extent of which is not known. 

Another significant point of interest concerning dwarf galaxies is the
well known discrepancy between the observed number of these
satellites, and the number predicted by $\Lambda$CDM models in
numerical simulations (e.g. \citealt{moore99,klypin99}), which still
amounts to one to two orders of magnitude. Survey completeness is
likely to contribute to at least part of this deficit (see
e.g. \citealt{willman04,simon07,tollerud08,koposov08,mcconnachie09}); as in the
Galactic case, observational efforts are hampered by obscuration from
the disk and the bulge. It is thought that future
all-sky surveys may find a wealth of ultra-faint satellites that would
reduce the gap between observation and theory. Other explanations
assume that many of these halos would remain dark to the present day,
owing to suppression of star formation as the result of a
photo-ionizing background \citep{somerville02}. These halos would
therefore represent a significant proportion of the satellite halos
found in simulations, reconciling these with observations.

In an attempt to better understand this unusual population of
galaxies, and to unite theory with observation, spectroscopic surveys
of faint dwarf galaxies (particularly within the MW via the Sloan
Digital Sky Survey (SDSS)) have been carried out. Using the velocity
information from the spectra, one can measure the systemic velocities
and intrinsic velocity
dispersions for these objects, which can be used to estimate the mass of these systems
(e.g. \citealt{illingworth76,richstone86,walker09b,wolf09}). However,
methods such as these make various assumptions about the anisotropy and
virial equilibrium of the systems that are not necessarily
correct. Nonetheless, such methods have been shown to be good
indicators of the instantaneous mass of the system
(e.g. \citealt{oh95,piatek95,munoz08}), and can therefore provide an
insight into the mass distribution of the satellites. Historically,
the results of such studies indicated the existence of a mass limit of
$\sim10^7\msun$ under which no dwarf galaxies are found, supporting
the theory that these satellites would indeed only inhabit massive
dark matter halos. The existence of such a mass limit would confirm
the empirical relation defined by \citet{mateo98} for satellites that
are brighter than Draco ($M_V=-8.8$) and Ursa Minor ($M_V=-8.9$) in
the Local Group: $M/L=2.5+ (L/10^7\lsun)$, where $M/L$ is the
mass-to-light ratio of the galaxies and $L$ is their luminosity. Work
by \citet{strigari08}, investigating the mass contained within the
central 300~pc of dSphs, reinforced this relation, demonstrating that,
despite the studied dSphs spanning five orders of magnitude in
luminosity, they appear to show a lower dynamical mass limit of
$\sim10^7\msun$. However, such a finding is in stark contrast with the
masses derived for some of the recently discovered ultra-faint dSphs,
e.g. Segue 1 and Leo V, whose masses of $\sim10^5\msun$
\citep{geha09,walker09a} fall well below this bound. This discrepancy
has been addressed in work by \citet{walker09b}, where they evaluate
the mass of dSphs within the half-light radius (r$_{1/2}$). Their analysis
shows that these ultra faint dSphs may simply be more deeply embedded
within dark matter halos that are similar to those inhabited by their
larger and brighter counterparts (also discussed in
\citealt{penarrubia08a}), which could suggest the existence of a
``universal'' dark matter halo. Similar results are shown in
\citet{wolf09}, where they conclude that all the Milky Way dSphs are
consistent with inhabiting dark matter halos with a mass
M$_{halo}\sim3\times10^9\msun$. The recently discovered faint dSphs
described above populate the faint end of the satellite luminosity
function, and their spectroscopic analysis would allow us to see if
these objects are indeed highly dark matter dominated, and whether they 
conform to such a mass limit.

Until recently, only a handful of the M31 dSphs have been analysed
spectroscopically (And II, \andix, \andxii, And XIV, And XV, And XVI
and And X
\citep{cote99,chapman06,chapman07,majewski07,letarte09,kalirai09},
with only one of these (And XII by Chapman et al. 2007) probing the
fainter end of the luminosity function (M$_V>$-8). A recent paper by
\citet{kalirai10} has extended the number of bright dSphs surveyed, 
presenting new data for And I, II and III, but the kinematic properties of M31's lowest
luminosity dSphs are still unknown. Spectroscopic observations of
these objects would allow us to see how they compare to both their
brighter M31 brethren, and their MW cousins of a comparable
luminosity.

As a step towards this end, we have initiated a kinematic survey of
the trio of faint dwarfs discovered by \citet{martin06}, \andxi,
\andxii\ and \andxiii\ using the DEep Imaging Multi-Object
Spectrograph (DEIMOS) on Keck~II. We combine these kinematics with
deep photometry from Suprime-Cam on the Subaru telescope. The analysis
of our observations is performed here for these objects, including a
re-treatment of \andxii, whose orbital properties were modelled in
\citet{chapman07}, but whose detailed properties were not assessed. We
also include a re-examination of the faint satellite, \andix, using the same
spectroscopic data as \citet{chapman05}, as there was an error in the
geometric modelling of the slit mask used to calculate the velocities,
which led to a misclassification of member stars; this has since been
corrected for, and we also present additional Subaru photometry for the
satellite. The outline of the paper is as follows: \S~2 presents the
observing strategy, the observations and how they were reduced ; \S~3
is dedicated to the analysis of \andxi, \andxii\, \andxiii\ and
\andix; we discuss our findings in \S~4 and provide conclusions in
\S~5.

\begin{figure*}
\begin{center}
\includegraphics[angle=0,width=0.33\hsize]{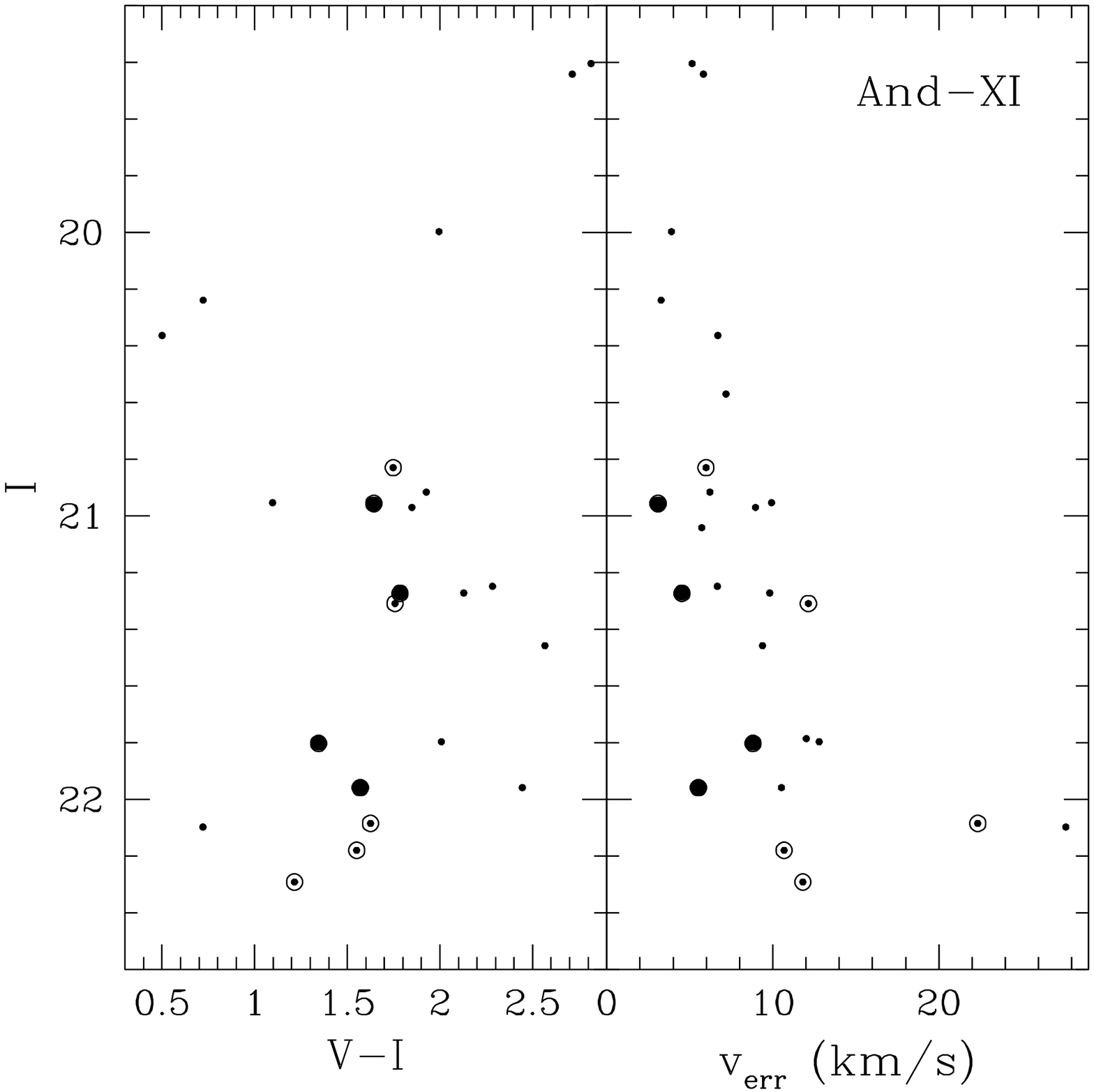}
\includegraphics[angle=0,width=0.33\hsize]{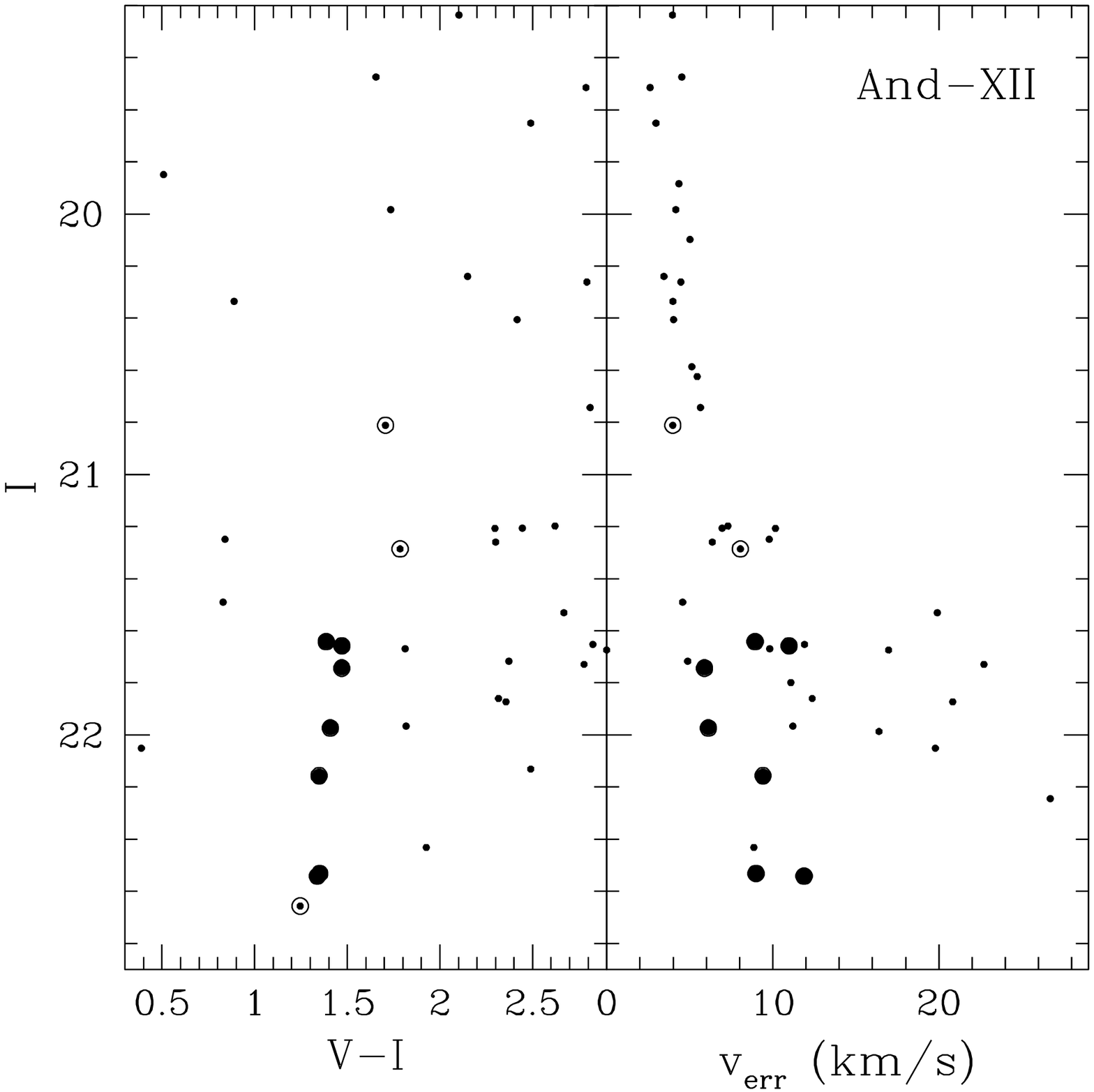}
\includegraphics[angle=0,width=0.33\hsize]{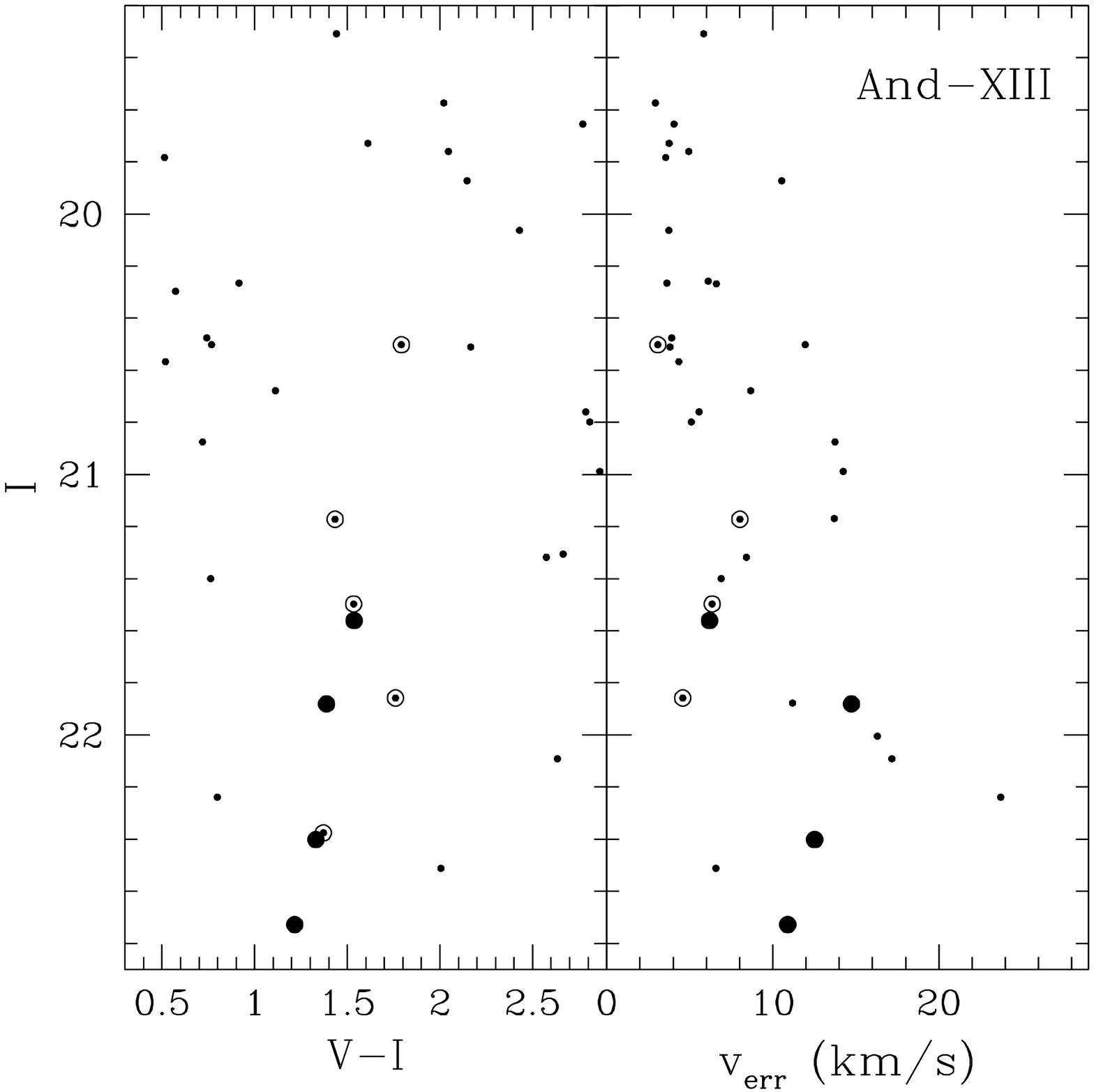}
\includegraphics[angle=0,width=0.33\hsize]{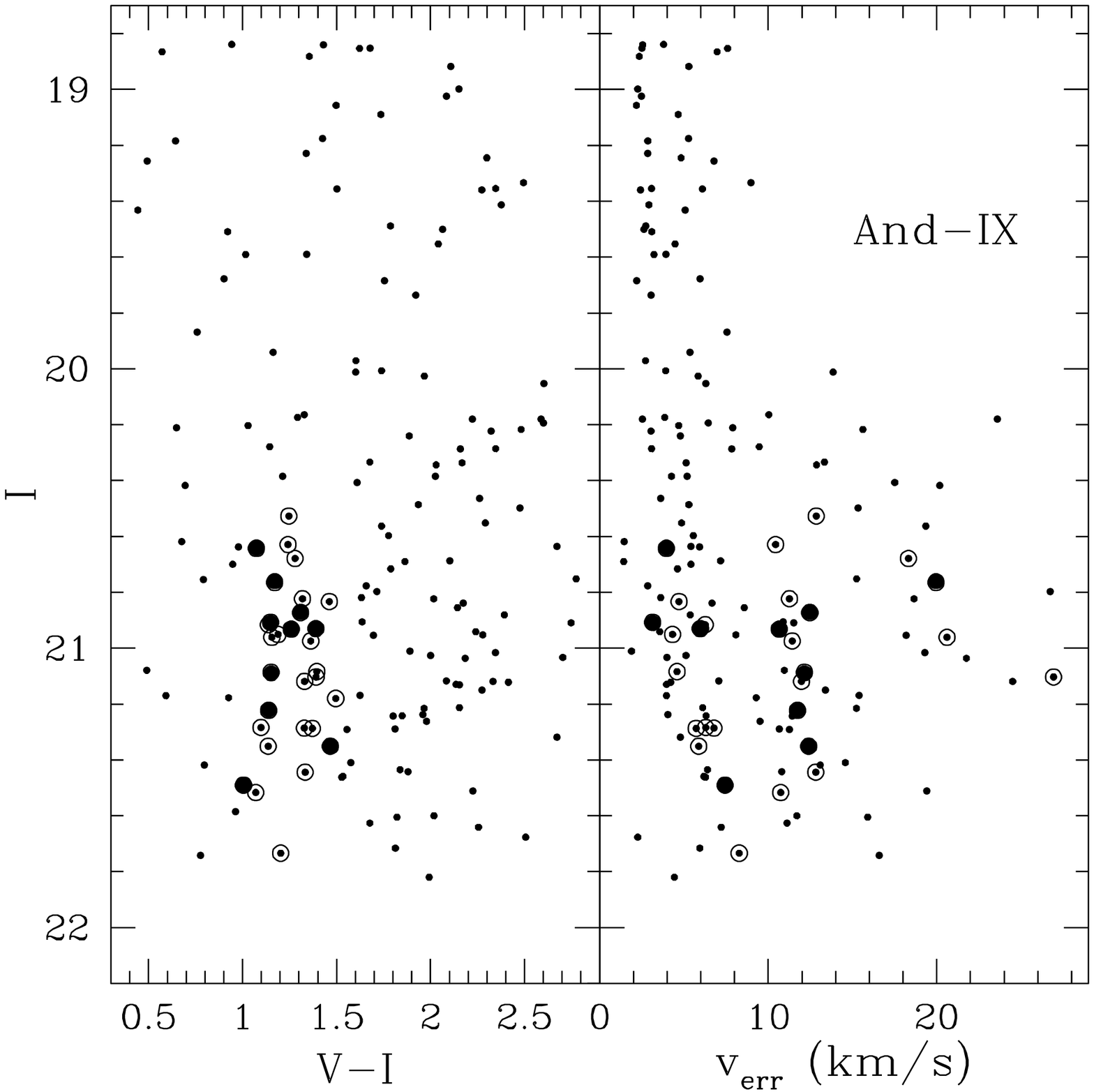}
\caption{CFHT-MegaCam color-magnitude diagrams (left panel of each
  Figure) , where the native MegaCam g and i bands have been converted
  into V and I magnitudes using transformations detailed in
  \citet{mcconnachie04b,ibata07}, and radial velocity uncertainties
  (right panel of each Figure) of the observed stars in the four
  objects surveyed in the paper.  The top left panel corresponds to
  \andxi, the top right to \andxii\ and the bottom left and right
  correspond to \andxiii\ and \andix\ respectively. The small dots
  represent CFHT objects in the region, filled circles represent
  target stars selected in the satellites CMD Red Giant Branch
  features. Circled big points represent stars that were initially
  selected as field stars but have the radial velocity and the CMD
  position of dwarf stars and thus were used to derive the dwarf
  parameters.}
\label{cmds}
\end{center}
\end{figure*}

\begin{table*}
\begin{center}
\caption{Details of the Keck/DEIMOS observations of \andxi, XII, XIII and IX}
\label{masks}
\begin{tabular}{llllccccc}
\hline
Date observed & Mask $\alpha_{J2000}$  & Mask $\delta_{J2000}$  & Object & Grating & Field P.A. & Exp. time (s) & No. targets & No. reduced  \\
\hline
 23/09/2006     & 00:46:21 & +33:48:21 &\andxi    & 600 line/mm  & 0$\deg$ & 3$\times$1200 & 35  & 32 \\
 21/09/2006     & 00:47:27 & +34:22:30 & \andxii  & 600 line/mm  & 0$\deg$ & 3$\times$1200 & 50 & 48 \\
 23/09/2006     & 00:47:27 & +34:22:30 & \andxii  & 600 line/mm  & 0$\deg$ & 3$\times$1200 & 50 & 48 \\
 23/09/2006     & 00:51:51 & +33:00:16 & \andxiii & 600 line/mm  & 0$\deg$ & 3$\times$1200 & 49 & 48 \\
 13/09/2004     & 00:52:51 & +43:11:48 &\andix    & 1200 line/mm & 0$\deg$ & 3$\times$1200 & 648 & 277 \\
\hline
\end{tabular}
\end{center}
\end{table*}

\section{Observations}

\subsection{Photometric Observations}

Each of the satellites were observed photometrically using the
Suprime-Cam instrument, mounted on the primary focus of the 8-metre
Subaru telescope, located at the summit of Mauna Kea. This
instrument is ideal for probing the spatial extent of these faint
dwarfs as it can reach the Horizontal Branch (HB) of populations at
distances greater than 500~kpc in a reasonable exposure time. Also, it
has a wide field of view (34$'$ x 27$'$) which is more than sufficient
to cover the entire region of the dwarfs at the distance of M31.  

Observations of \andxi, \andxii\ and \andxiii\ were made on the night
of September 22, 2006, in photometric conditions with an average
seeing of 0.6\scnp and 0.5\scnp in the g and i bands respectively. For
these satellites we exposed for a total of 2100s in g-band (split into
5 x 420s dithered subexposures) and 4800s in the i-band (split into 20
x 240s subexposures) in an attempt to reach below the HB to an equal
depth in both bands. The data were processed using the CASU general
purpose pipeline for processing wide-field optical CCD data
\citep{irwin01}. The images were debiased and trimmed before being
flat-fielded and gain-corrected to a common internal system using
master flats constructed from twilight sky observations.  Catalogues
were generated for every science image and used to refine the
astrometric alignment of the images.  The images were then grouped for
individual objects and passbands and stacked to form the final science
images based on the updated astrometry. A catalogue was then generated
for each final stacked science image, objects morphologically
classified as stellar, non-stellar or noise-like, and the g,i band
catalogue data merged.  As a final step, for And XI, XII, XIII the g,i
Subaru data was calibrated with respect to our MegaCam system g,i
using all stellar objects in common with signal-to-noise $>10$.
\andix\ was observed in service mode during the period 9--10 October
2004 with typically subarcsecond seeing, for a total of 600s in each
of g, r and i bands (split into 5 x 120s dithered subexposures for
each filter). The data were reduced in a similar manner to those for
\andxi, XII and XIII, although here, the photometry from the CASU
Subaru pipeline was matched to the SDSS M31 scan data.

The data for all satellites are extinction corrected and dereddened 
using the \citealt{schlegel98} dust maps. We find E(B-V)=0.08 for 
\andxi, XIII and IX, and E(B-V)=0.12 for \andxii.

\subsection{Spectroscopic Observation}

For each of the dSphs in question, candidate members were selected
using colour magnitude diagrams (CMDs) from the Canada France Hawaii
Telescope (CFHT) as shown in Fig.~\ref{cmds} where we present the
MegaCam colours (without correcting for extinction and reddening) for
the stars within these objects, transformed into Landolt V and I using
a two stage process detailed in \citet{ibata07} and
\citet{mcconnachie04b}. Colour magnitude boxes were used to select RGB
stars that lay within 3--5$'$ regions surrounding each of the
satellites, which is equivalent to several half-light radii of these
objects. DEIMOS is a multi-slit, double-beam spectrograph mounted on
the Nasmyth focus of the Keck II telescope on Mauna Kea, ideal for
taking optical spectra of faint-objects such as these. For \andxi,
\andxii, and \andxiii\ multi-object observations were taken from
21--24 September 2006 with DEIMOS in photometric conditions and with
excellent 0.6$''$ seeing.  Our chosen instrumental setting covered the
observed wavelength range from 5600--9800\AA\ with exposure time of
3x20 minute integrations. The Red Giant Branch (RGB) stars in these
three new dwarfs are faint ($i$ = 21.5--23.0), so to obtain a useful
S:N in a reasonable exposure time, all the spectra employ the low
resolution ($\sim3.5$\AA) 600 line/mm grating. \andix\ was observed on
the night of 13 September, 2004 under photometric conditions with
seeing from 0.5$''$ to 0.8$''$. We employed the 1200 line/mm grating,
covering the wavelength range 6400-9000\AA, also with exposure time of
3x20 minute integrations. Data reduction was performed using the
DEIMOS-DEEP2 pipeline \citep{deep2}, which included debiassing,
flat-fielding, extracting, wavelength-calibrating and sky subtracting
the spectra. In table \ref{masks}, we detail the positions targeted by
each mask, the number of target stars, and the total number of stars
for which accurate velocities were derived (discussed further below).

The kinematic data for \andix\ were originally presented in
\citet{chapman05}. However the geometric model used for the slit mask
was incorrect, causing a shift in the measured velocities of the
observed stars of order 5kms$^{-1}$, which has been corrected in this
work. This systematic error caused a misclassification of member stars
within the sample, which in turn affected the measurements of systemic
velocity, dispersion and metallicity for the satellite in Chapman et
al. (2005). The velocities for all observed stars has since been re-measured
allowing us to correctly derive the properties of \andix\ in this
work.

{\subsubsection{Velocity calculations and error analysis}
\begin{figure}
\begin{center}
\includegraphics[angle=0,width=0.8\hsize]{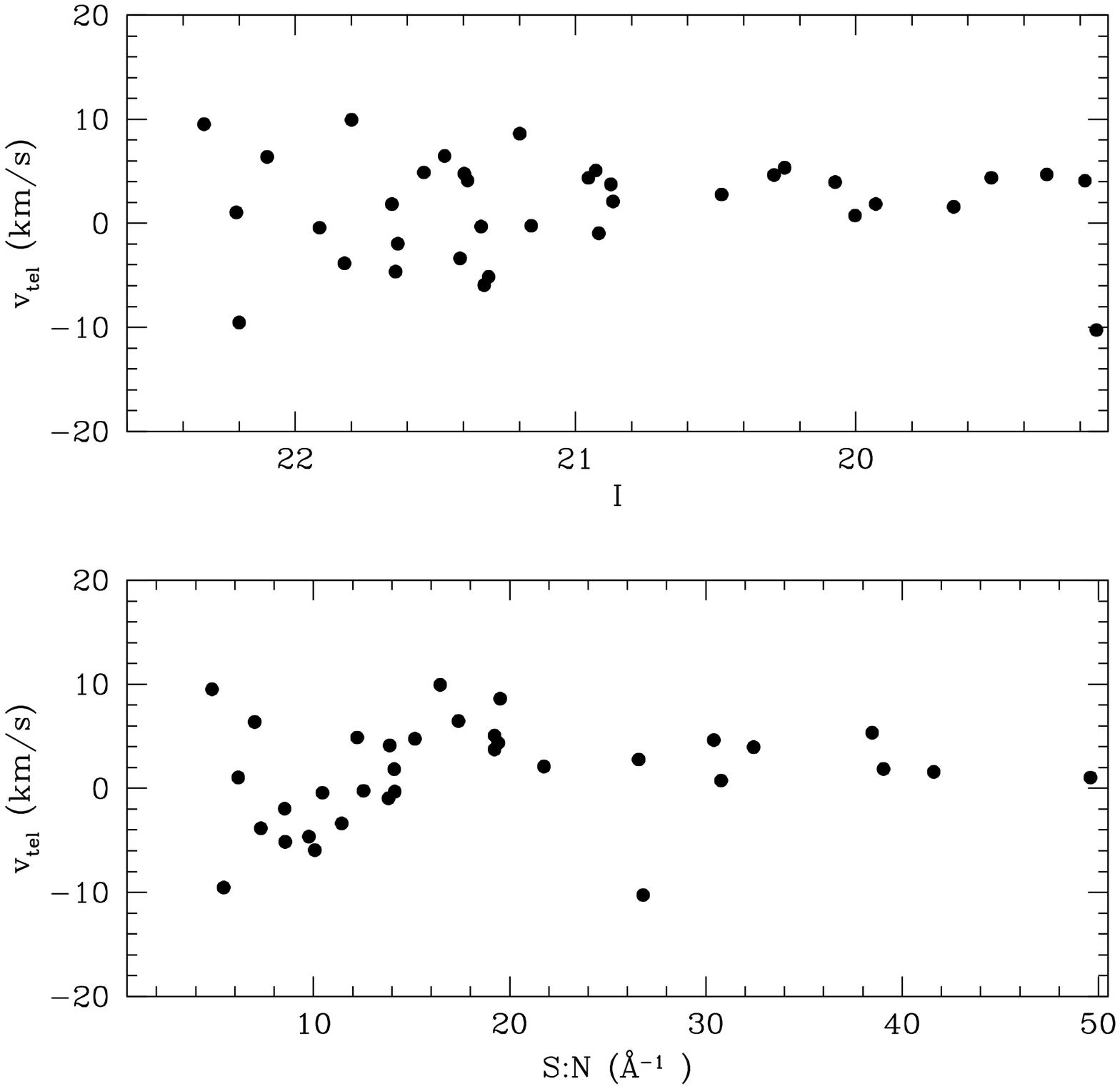}
\caption{Telluric velocity corrections for all stars in \andxii\ as a function of $I-$band magnitude (top) and S:N (bottom). The median telluric correction for this mask is 2.8~kms$^{-1}$, and for bright ($I<21$), high S:N (S:N$>12$\AA$^{-1}$) targets, this correction remains approximately constant. However for fainter stars with noisier spectra, this correction shows increased scatter about the median value.}
\label{telluric}
\end{center}
\end{figure}

\begin{figure}
\begin{center}
\includegraphics[angle=0,width=0.8\hsize]{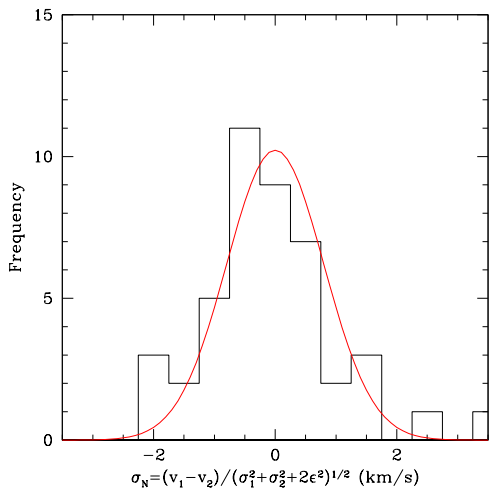}
\caption{A histogram showing the normalized error distribution for
  repeat measurements of the same stars in our \andxii\ data set. The
  normalized error, $\sigma_N$ incorporates the velocity differences
  between the repeat measurements (v$_1$ and v$_2$), and the
  Monte-Carlo uncertainties calculated for each observation
  ($\sigma_1$ and $\sigma_2$). In order to reproduce a unit Gaussian
  distribution for our uncertainties, we also include an additional
  error term, $\epsilon$, which accounts for any systematic
  uncertainties we have not included. For the \andxii\ data (observed
  with the 600l grating), we find $\epsilon$=6.2 kms$^{-1}$. For
  \andix, which was observed with the 1200l grating, we set
  $\epsilon$=2.2 kms$^{-1}$, which was the value derived by
  \citet{simon07} from repeat observations of stars in their
  Keck/DEIMOS survey, which used a similar observational setup to that
  of \andix.}
\label{error12}
\end{center}
\end{figure}

\begin{figure}
\begin{center}
\includegraphics[angle=0,width=0.8\hsize]{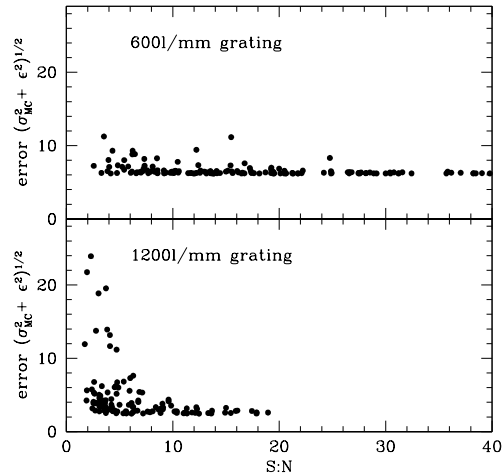}
\caption{The total error in velocity of all stars observed within
  \andxi, XII and XIII (top panel) and \andix\ (lower panel) as a
  function of S:N. The total error is calculated by combining the
  Monte-Carlo error ($\sigma_{MC}$) discussed in \S2.2.1, and the
  systematic error, $\epsilon$ (6.2 kms$^{-1}$ for \andxi, XII and
  XIII, 2.2 kms$^{-1}$ for \andix), in quadrature. It can be seen that
  we can measure the velocities of our sample to accuracies of
  typically less than 15 kms$^{-1}$ down to S:N$\ge$3\AA$^{-1}$.}
\label{errors}
\end{center}
\end{figure}

Our instrumental setup covered the wavelength range of the Calcium II
(Ca II) triplet, a strong absorption feature centered at
$\sim8550$\AA. To determine velocities for our targeted stars, we
cross-correlated each spectrum with a model template of the feature at
the rest-frame wavelength positions of the triplet lines (the
technique used is similar to that discussed by Wilkinson et
al. 2004). Briefly, we perform cross-correlations for each star with
our template and compute the corrected heliocentric velocity for each
star. We also compute the Tonry-Davis R-value (TDR) for each
cross-correlation, which is an indication of the reliability of the
resulting velocity. We consider any velocity cross-correlation with a
TDR value$\ge$3.5 to be reliable. In Tables 2 -- 5, we note the TDR
values for each of our likely member stars (discussed in greater
detail in \S3). We also estimated the velocities for each star using a
Doppler shift method, fitting Gaussians to each of the triplet lines
individually and computing the mean velocity shift from these. The
results of both calculations were generally found to be consistent
(within $\sim2-5$kms$^{-1}$) with one another. In a few instances, the
Gaussian technique revealed cases where the cross-correlation
technique had fixed onto a noise-feature, rather than the true Ca II
triplet, and these cases, we have taken the velocities derived from
the Gaussian technique. However, as it is generally accepted that
simple Doppler-shift velocities are less reliable in low S:N data
(e.g. \citealt{prieto07}), we refer to velocities computed by the
cross-correlation method throughout the text, unless otherwise stated.

For the Ca II triplet lines, the spectral features of interest overlap in
location with strong night sky OH recombination lines for some of the
dSph members. While we are able to detect continuum at S:N$\ge$3\AA$^{-1}$ for even 
the faintest ($i=23$) stars observed, the velocity errors are
subject to larger systematic errors when they lie close to these OH
lines. This effect is seen most prominently in the case of \andxii,
where the CaII$_{8498}$ line coincides with a strong sky feature. This
effect is seen in the night to night variations in the velocities of
the \andxii\ stars, for which we obtained repeat observations. A
systematic shift in the velocities between the two nights of
30~kms$^{-1}$ is observed for some of the stars, while other stars in
the mask at more favourable velocities have night to night errors
$<$5~kms$^{-1}$. For those stars where this effect is very prominent, we
mask out the affected line and repeat the cross-correlation on the
two unobscured lines.

There are a number of systematic errors that can be introduced into
velocity calculations from the instrument setup, such as the effect of
slight mask rotation, field mis-alignment and improperly cut masks,
all of which can cause the star to be mis-centred within the slit,
resulting in an incorrect velocity measurement that can range from a
few to tens of kms$^{-1}$. To correct for this error, we make use of
telluric absorption features present within our spectra, particularly
the strong Fraunhofer A-band telluric absorption feature located
between 7600 and 7630\AA. These features are the result of absorption
of stellar light by various gases (O$_2$, H$_2$O, CO$_2$, etc.) within
the Earth's atmosphere. These lines are therefore not Doppler shifted,
and should be present at the same wavelength in each
spectrum. Following the techniques of \citet{simon07} and
\citet{sohn07}, we construct a template telluric spectrum which can be
cross-correlated with each of our science spectra to determine the
effective velocity shift caused by these systematic errors. For the
fields observed in this work, we found that the average velocity shift
for each mask derived by this method ranged from +2.8 kms$^{-1}$ to
+8.1 kms$^{-1}$. To correct our observed velocities for this, we add
the velocity shift calculated for each star to our cross-correlation
velocities determined from the Ca II triplet. In Fig.\ref{telluric} we
show the telluric corrections derived for \andxii\ (average correction
= 2.8 kms$^{-1}$) as a function of $I-$band magnitude and S:N. For
brighter stars ($I<21$) and stars with higher S:N (S:N$>15$\AA$^{-1}$), the
velocity corrections are approximately constant, however for fainter
stars (stars with lower S:N) the magnitude of this telluric correction
shows significant scatter about the median value. It is therefore
unclear whether this method can accurately compute the telluric shift
in fainter objects.

We estimate the errors on our velocities by following the procedures
of \citet{simon07} and \citet{kalirai10}. First, we make an estimate
of our velocity uncertainties for each observed star using a Monte
Carlo method, whereby noise is randomly added to each pixel in the
spectrum, assuming a Poisson distribution for the noise. We then
recalculate the velocity and telluric offset of each star, using the
same cross-correlation method detailed above. This procedure is
repeated 1000 times, and then the error is calculated to be the square
root of the variance of the resulting mean velocity. Secondly, we make
use of our \andxii\ dataset, where we were able to obtain repeat
observations for our entire mask, giving us 47 stars with two velocity
measurements, $v_1$ and $v_2$. We then attempt to measure our velocity
errors directly by comparing these velocities and defining a
normalised error, $\sigma_N$ which incorporates our Monte Carlo error
estimates for each measurement pair, $\sigma_{1}$ and $\sigma_{2}$,
plus a random error, $\epsilon$, which represents any uncertainties we
have not accounted for. $\epsilon$ is defined as the additional random
error required in order to reproduce a unit Gaussian distribution with
our data (shown in Fig.~\ref{error12}).  For the \andxii\ stars, which
were observed with the 600 line/mm grating, we find $\epsilon=6.2$
kms$^{-1}$. While this is roughly double that found in other works
(e.g. 2.2 kms$^{-1}$ in \citealt{simon07} and $\sim3$kms$^{-1}$ in
\citealt{kalirai10}), we note that they use the 1200 line/mm grating,
which has a resolution of $\sim1.3$\AA\ compared to the 600 line/mm
gratings resolution of $\sim$3.5\AA. Therefore, for Andromeda's XI,
XII and XIII, we set $\epsilon=6.2$ kms$^{-1}$, and we use
\citet{simon07} value of $\epsilon=2.2$ kms$^{-1}$ for \andix. We then
combine this systematic error in quadrature with our calculated errors
from the Monte Carlo technique ($\sigma_{MC}$), giving us our final
error estimate. We display the errors for all observed stars in the
four dSphs as a function of S:N in Fig.~\ref{errors}. The dSphs
observed using the 600 line/mm grating (\andxi, XII and XIII) are
shown in the top panel, and \andix (observed with the 1200 line/mm
grating) is shown in the lower panel.  It can be seen in both cases
that we can measure velocities to accuracies of $\sim5-15$kms$^{-1}$
down to a S:N of 3 or better, while the higher signal-to-noise stars
have velocity errors that are consistent with the 2.2 and 6 kms$^{-1}$
systematic uncertainties for the 1200 and 600 line/mm gratings
respectively. We find a median error of $\sigma_{tot}=$ 3.6 kms$^{-1}$
for stars in the \andix\ mask, and $\sigma_{tot}=$ 6.6 kms$^{-1}$ for
the stars in the \andxi, XII and XIII masks.

\section{Satellite Properties}

\subsection{\andxi}

\andxi\ is the most luminous of the three dSphs
reported in \citet{martin06}. It was found as an over-density of stars
in the CFHT/MegaCam survey that follows a well-defined RGB. The
satellite lies at a projected distance of 106~kpc from M31 
(assuming a distance of 785 kpc for M31). Below we
present the structural and kinematic properties of this faint galaxy.

\begin{figure*}
\begin{center}
\includegraphics[angle=0,width=0.8\hsize]{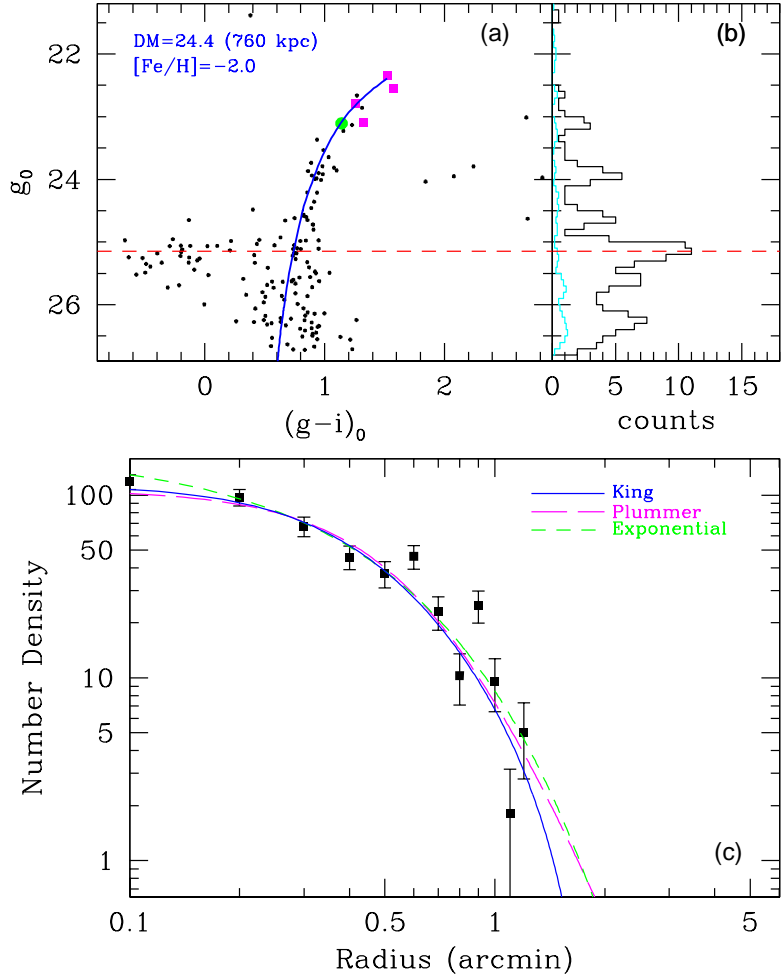}
\caption{{\bf Top left:} Here we display the de-reddened Subaru CMD
  for \andxi\, and highlight the stars for which DEIMOS spectroscopy
  is available with magenta squares. The outlier at -432.3$\kms$ is
  highlighted in green. A Dartmouth isochrone with [$\alpha$/Fe]=+0.4,[Fe/H= -2.0
  and age 13 Gyrs \citep{dart08} is overlaid at a distance modulus of
  24.5 (equivalent to 760 kpc) with [Fe/H]=-2.0.
  {\bf Top right:} The luminosity function for \andxi\ in the g-band
  is displayed as a histogram, and a clear peak at 25.1 for the HB is
  observed compared with the normalised luminosity function of a
  comparison region in the surrounding area of the dwarf (blue
  histogram). {\bf Bottom panel:} The radial density plot for
  \andxi\ with King (solid line), Plummer (long dashed line) and
  exponential (short dashed line) overlaid, with corresponding r$_{1/2}$ =
  0\mcnd64$\pm0.05$, 0\mcnd72$\pm0.06$ and 0\mcnd71$\pm0.03$.  }
\label{and11phot}
\end{center}
\end{figure*}

\begin{figure*}
\begin{center}
\includegraphics[angle=0,width=0.45\hsize]{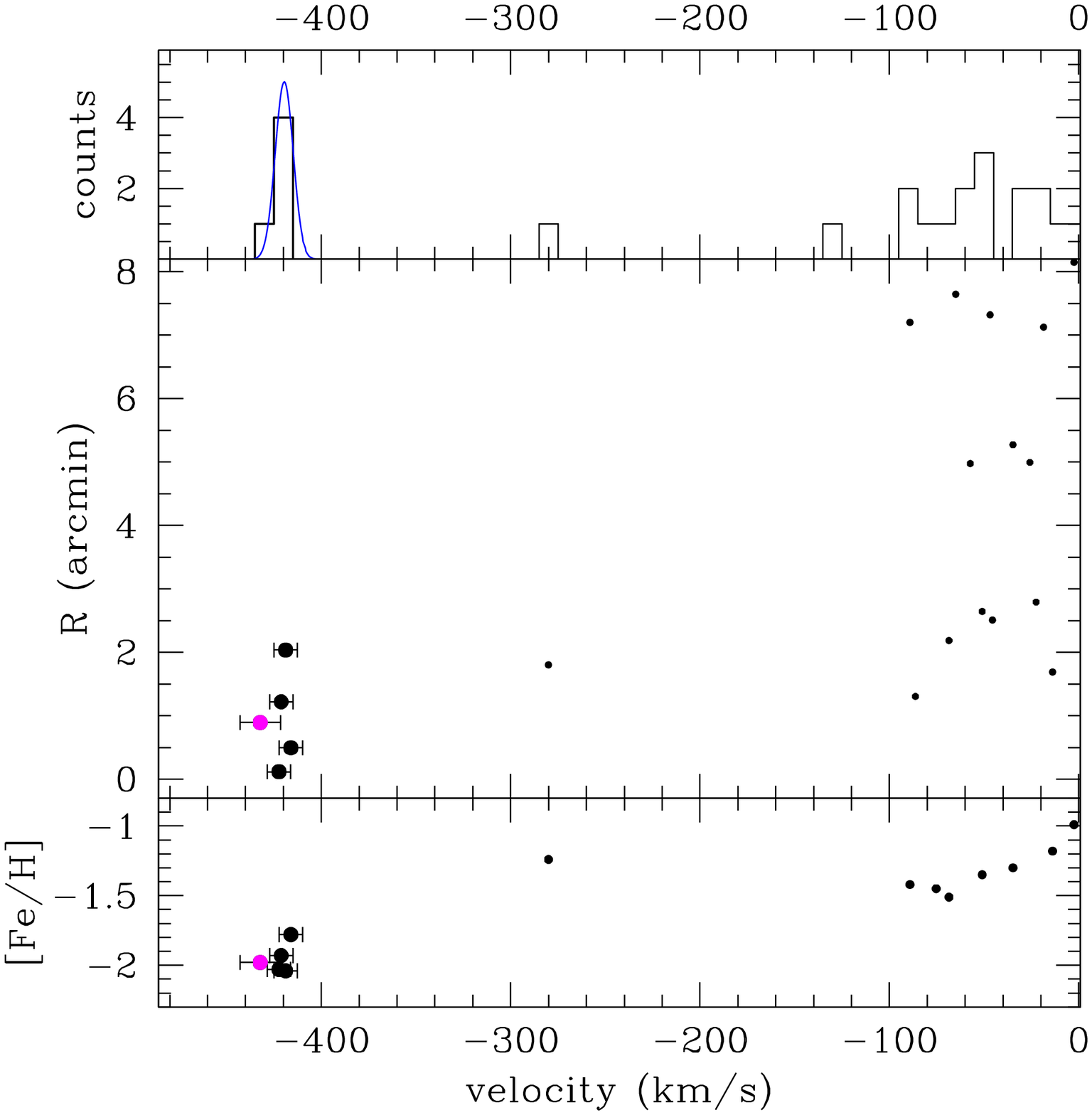}
\includegraphics[angle=0,width=0.45\hsize]{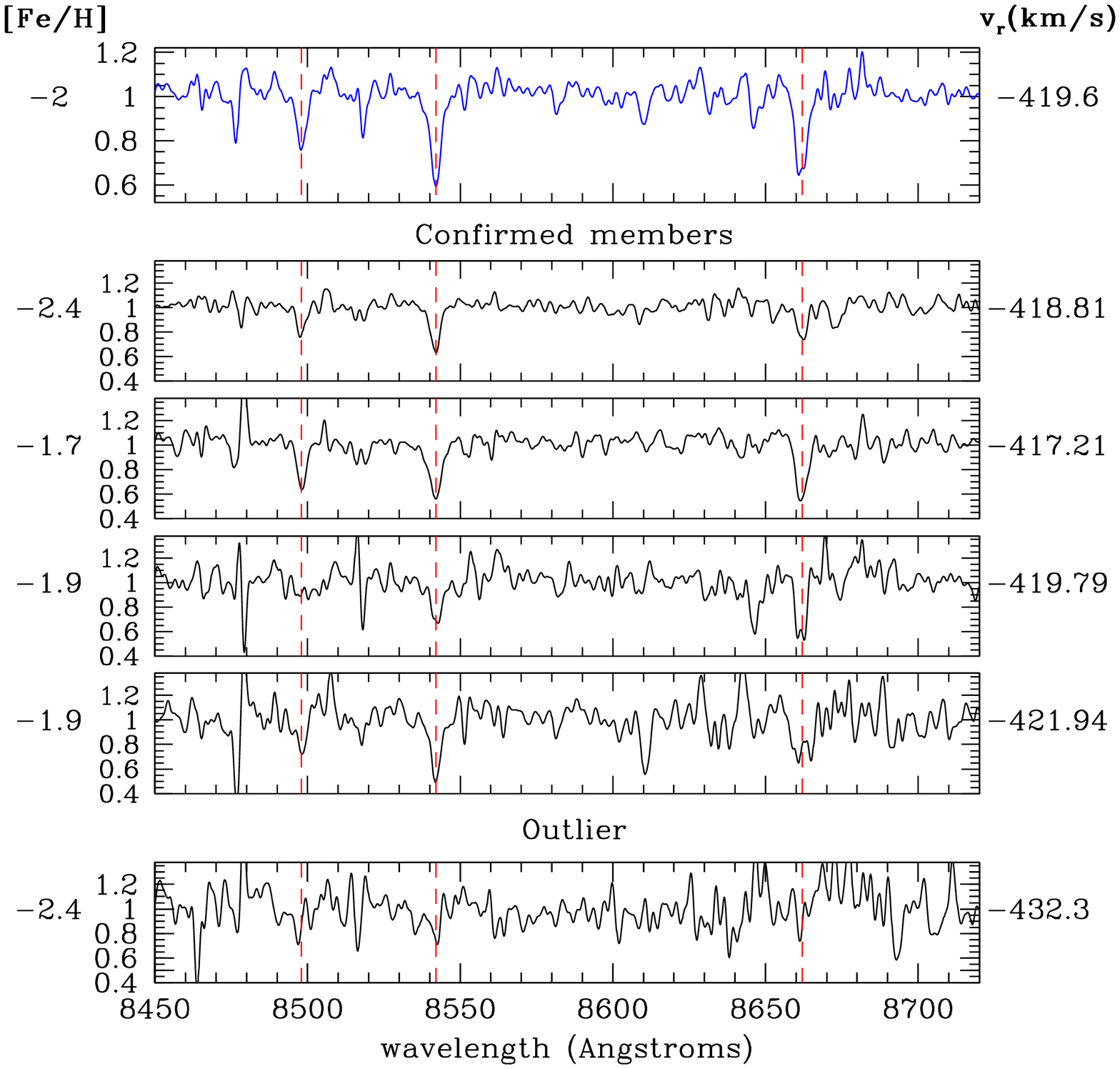}
\caption{{\bf Left panel:} The velocities of observed stars in the
  \andxi\ field are shown as a histogram, with the 5 candidate
  \andxi\ member stars (seen as a kinematic grouping at approximately
  -420 kms$^{-1}$) highlighted as a heavy histogram. To differentiate
  \andxi\ stars from the field, we additionally plot the velocities
  against their radius from the \andxi\ center, and in the lower
  panel, we display their photometric metallicities as a function of
  radial velocity, which were derived by interpolating between
  Dartmouth isochrones \citep{dart08} with [$\alpha$/Fe]=+0.4 and
  age=13 Gyrs.  {\bf Right panel:} The normalised spectra for each of
  the 4 plausible member stars of \andxi\ and the outlier (bottom
  spectrum) are shown, with the combined spectrum of the 4 plausible
  members plotted in the top panel. The spectroscopic [Fe/H] for each
  candidate member is shown also, and we derive [Fe/H]=-2.0$\pm0.2$
  from the composite spectrum.  }
\label{and11spec}
\end{center}
\end{figure*}

\begin{figure*}
\begin{center}
\includegraphics[angle=0,width=0.8\hsize]{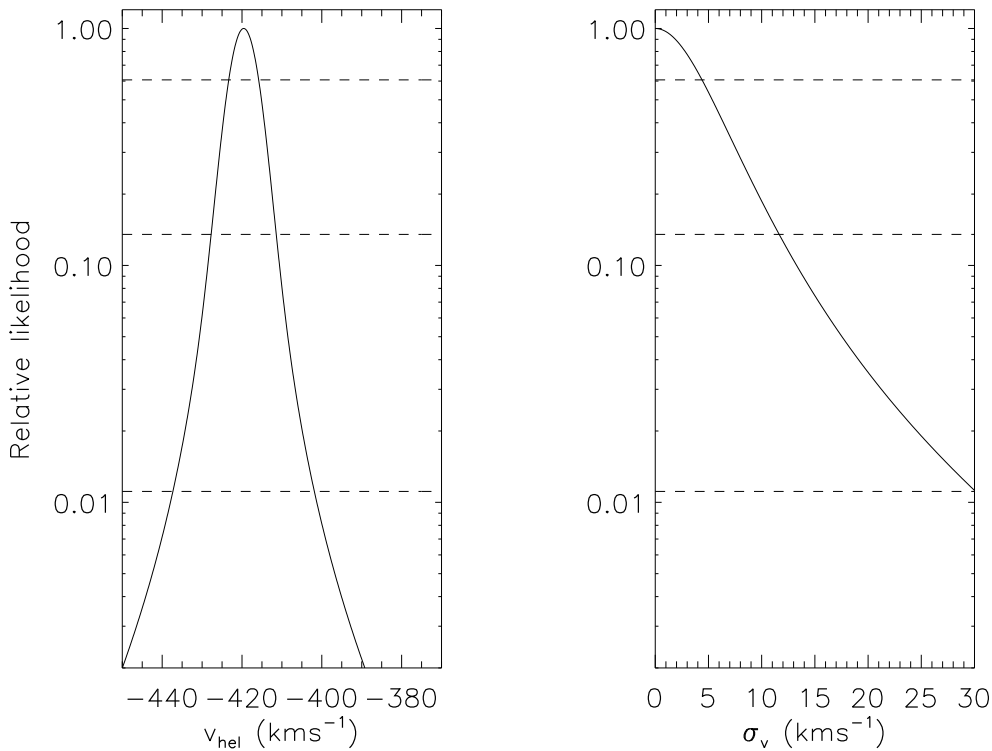}
\caption{ Likelihood distributions of member stars in \andxi, showing
  the most likely values of radial velocity (left panel) and velocity
  dispersion (right panel) for our 4 confirmed members. The dashed
  lines indicate the 1, 2, 3$\times\sigma$ uncertainties.  The
  most-likely velocity for \andxi\ is v$_r$ =-419.6~kms$^{-1}$ and the
  dispersion is not quite resolved. We use the value of 4.6~kms$^{-1}$
  from the 1$\sigma$ confidence level as an upper limit for the
  dispersion of \andxi.}
\label{and11contour}
\end{center}
\end{figure*}

\begin{table*}
\begin{center}
\caption{Properties of candidate member stars in \andxi, centered at $\alpha$=00:46:21, $\delta$=+33:48:22. Outlier at $v=-432.3$kms$^{-1}$ is also included as the last entry.}
\label{and11stars}
\begin{tabular}{lllcccccc}
\hline
star $\alpha$  & star $\delta$  & vel (kms$^{-1}$) & [Fe/H]$_{spec}$ &  [Fe/H]$_{phot}$ & S:N (\AA$^{-1}$) & TDR & $I$-mag & $V$-mag  \\
\hline
 00:46:22.65 & +33:46:21.6 & -418.8$\pm6.2$ & -2.42 & -2.0 & 16.7 &15.37 & 20.5 & 21.9 \\
 00:46:19.10 & +33:48:4.1 &  -417.2$\pm6.3$ & -1.56 & -1.5 & 13.5 & 17.2 &20.8 & 22.3 \\
 00:46:15.14 & +33:48:26.7 & -419.8$\pm6.2$ & -0.61 & -1.9 & 9.3  & 8.4 &21.3 & 22.5 \\
 00:46:21.56 & +33:48:22.4 & -421.9$\pm6.4$ & -1.69 & -1.4 & 7.8  & 7.3 & 21.5 & 22.9 \\
 \hline
 0:46:16.96 & +33:48:3.8  & -432.3$\pm10.8$ & - & -1.8 & 6.0 & 3.4  & 21.8 & 22.9 \\
\hline
\end{tabular}
\end{center}
\end{table*}

\subsubsection{Photometric properties}

First, we discuss the updated structural parameters derived for the
\andxi\ dwarf using the data from Subaru/Suprime-Cam. The extinction 
corrected and deredenned CMD for this data is shown in 
Fig.~\ref{and11phot}(a), containing only stars within a radius of 1.5' 
of the centre of \andxi, defined as the mean of the positions of the stars within 3$\mcnp$, 
giving ($\alpha$,$\delta$)=($00^{h} 46^{m} 21^{s}$, $+33\deg$ $48\mcnp$
$22\scnp$). The Subaru data probe much further down the 
luminosity function than the original CFHT data used by \citet{martin06} and 
shown in Fig. 1, and is complete down to the level of the HB at 
g$\sim$25. To assess the half-light and tidal radii of \andxi, we 
construct a background corrected radial surface density profile of
\andxi\ using circular annuli; this can be seen in
Figure~\ref{and11phot}(c). A background level of 8.1$\pm0.4$ stars per
square arc minute was determined from a series of comparison regions
in the area surrounding the dwarf. The half-light radius is then
calculated by using a chi-squared fitting routine to apply King (with
r$_t=2\mcnd23\pm0.2$), Plummer and exponential profiles to the radial
density data. For \andxi\ this yields half-light radii (r$_{1/2}$) of
$0\mcnd64\pm0.15$, $0\mcnd72\pm0.06$ and $0\mcnd71\pm0.03$
respectively, with the exponential profile providing the best fit
(with a reduced chi-squared of 2.6) to the data. Alternatively,
applying the algorithm of \citet{martin08} produces a similar result
of $0\mcnd68\pm0.04$ and converges on structural parameters without a
strong ellipticity, $\eta=0.08$ (N. Martin, private communication),
justifying the use of circular annuli in this case.

The CFHT data analysed by \citet{martin06} were not deep enough 
to measure the distance to \andxi, however our deeper, more complete Subaru 
observations allow us to place constraints on the distance to the satellite by 
combining the photometric data with our spectroscopic data (which
we discuss in more depth below). One can define the apparent
magnitude of the Tip of the Red Giant Branch (TRGB) for the dwarf by
assuming that our brightest spectroscopically confirmed member
represents the tip, allowing us to calculate the distance to the
satellite. Using the brightest DEIMOS star (which has an apparent
magnitude, I$_{0, Vega}$=20.393), and assuming the absolute magnitude
of the TRGB is M$_{I_{0},Vega}=-4.04$ \citep{bellazzini04}, we find
the maximum distance to be D$_{max}$= 770~kpc. To determine the
minimum distance, we must assess what the maximum possible offset
could be from this brightest RGB star to the true TRGB. To do this, we
follow the procedure of \citet{chapman07}, analysing the well
populated RGB of the dwarf spheroidal, And~II
\citep{mcconnachie06b}. With 1000 random samples of 4 of the brightest
stars from the top magnitude of the And~II RGB, we constrain a
probability distribution for the true TRGB, corresponding to a most
likely offset of 0.5 mag. Using this value, we find the range of
distances to \andxi\ is 610-770~kpc. We note that this calculated
range is based on assumption that the brightest DEIMOS star is found
within 0.5 mags of the TRGB. While this seems reasonable, the distance
range we calculate would vary if this assumption were
changed. Therefore, as a sanity check for these values, we can also
look for the peak in the luminosity function due to the HB of \andxi\ in the luminosity function,
shown in Fig.~\ref{and11phot}(b). An inspection of this reveals a
clear peak at g-magnitude of $\sim25.1$ (red dashed line). Given that the
absolute magnitude of the HB in the g-band is M$_{HB}$=0.6
\citep{irwin07} -- though we note that this assumed value is sensitive
to age and metallicity effects, see \citealt{chen09} for a
discussion -- we can estimate a distance modulus for the satellite of
$\sim24.5$ (~830 kpc), which sits outside of our calculated range for
the satellite. This result demonstrates the difficulty in determining
distances for faint satellites such as \andxi. A companion paper,
analysing HST data for \andxi, XII and XIII (Martin et al. in prep),
will discuss the distances to these galaxies in more detail. For now,
we use an intermediate distance of 760$^{+70}_{-150}$ kpc (where the upper and
lower error bounds correspond to the maximum and minimum distances
estimated within this work) for the remainder of our analysis.

In order to assess a mean photometric metallicity for the satellite,
we use the Dartmouth isochrones models of \citet{dart08}, at an alpha
element abundance of [$\alpha$/Fe]=+0.4 and an age of 13 Gyrs. We use
[$\alpha$/Fe]=+0.4 as it has been shown in numerous studies
(e.g. \citealt{marcolini08}), that the alpha abundance for metal-poor stars in
dSph populations such as those studied herein, tend to scatter between
[$\alpha$/Fe]=+0.2 and +0.5. We overlay isochrones at distances of 610
kpc, 830 kpc and 760 kpc (corresponding to the upper, lower and HB
distances calculated above), and note the best observed fit in each
case. This gives a metallicity range of [Fe/H]=-1.5 to -2.0 for the
dwarf. In Fig.~\ref{and11phot}, we overlay the isochrone corresponding
to the best observed fit at 760 kpc ([Fe/H]=-2.0). We note
that using a less alpha-enhanced isochrone model (with
[$\alpha$/Fe]=+0.2), gives a slightly more metal rich value, with a
best observed fit at the HB distance of [Fe/H]=-1.9.

We use the Subaru data to recalculate the luminosity of the satellite
by summing the flux of the likely member stars lying on the RGB and HB
in the CMD of the satellite. By summing the flux of stars populating
the CMD of the well studied globular cluster M92\footnote{We use M92
  as a comparison, rather than another M31 dSph as its CMD is well
  populated past the HB level, allowing us to better estimate the
  fraction of luminosity contained within this region. In addition, the
  satellites studied in this paper all appear to be relatively old,
  making the comparison with a globular cluster
  reasonable.}  within its half light radius (r$_{1/2}$=19.6$''$ \citealt{ferraro00}),
using archival HST data, we calculate the total flux contained in this 
radius from the TRGB down to the HB. As our own data start 
come 
incomplete at roughly the HB level, we calculate a range for the 
contained flux by choosing two different cuts for the HB of M92. 
We find that 58-68\% of the total luminosity of the
satellite is contained within this region, and we calculate a
value for the total luminosity of \andxi\ of
$4.9^{+0.4}_{-0.2}\times10^4\lsun$. Using the revised distance
estimate above and this luminosity, we update the derived parameters
of \andxi, with the half-light radius, r$_{1/2}$, becoming
$145^{+15}_{-29}$ pc using the half-light radius from the exponential
profile (previously estimated as 115~pc), and
M$_{v}=-6.9_{-0.5}^{+0.1}$, in agreement with the findings of
\citet{martin06}.

\begin{figure*}
\begin{center}
\includegraphics[angle=0,width=0.8\hsize]{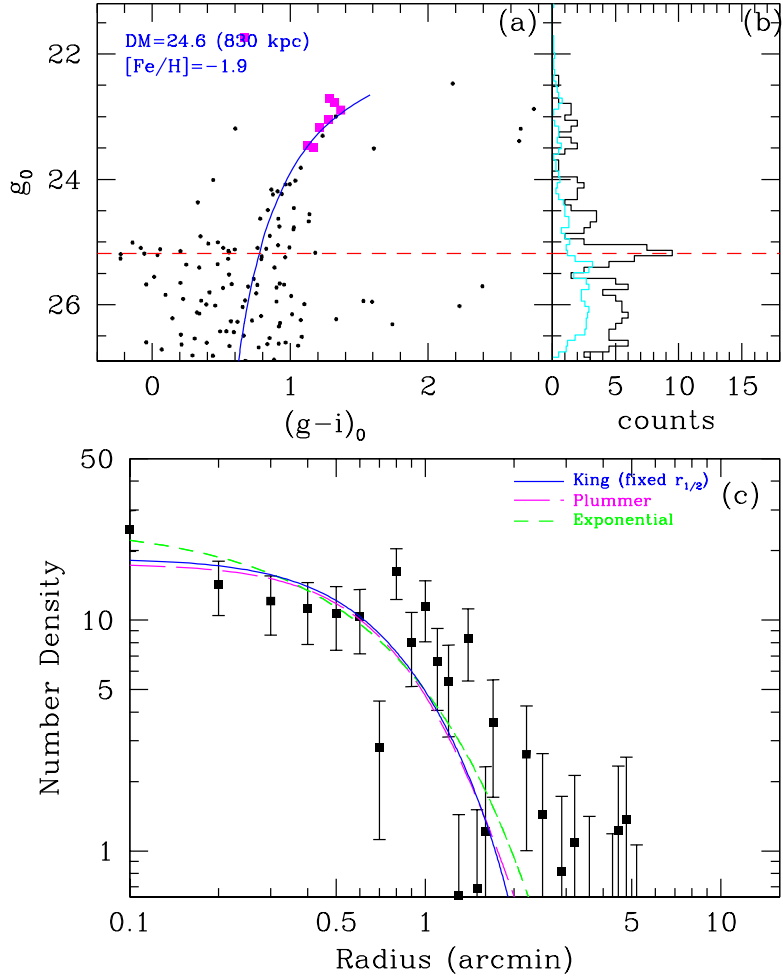}
\caption{As Fig.~\ref{and11phot} but for \andxii. A Dartmouth isochrone is overplotted with [Fe/H]=-1.9 and [$\alpha$/Fe]=+0.4 at
  distance modulus of 24.6 (830 kpc). In plot (c) Plummer and
  exponential profiles are overlaid, with respective half-light radii
  of $1\mcnd2\pm0.2$ and $1\mcnd1\pm0.2$, with the exponential profile
  providing the best fit to the data. . A King profile (with
  r$_{1/2}$ fixed at 1.2$'$ is overplotted also.}
\label{and12phot}
\end{center}
\end{figure*}

\begin{figure*}
\begin{center}
\includegraphics[angle=0,width=0.45\hsize]{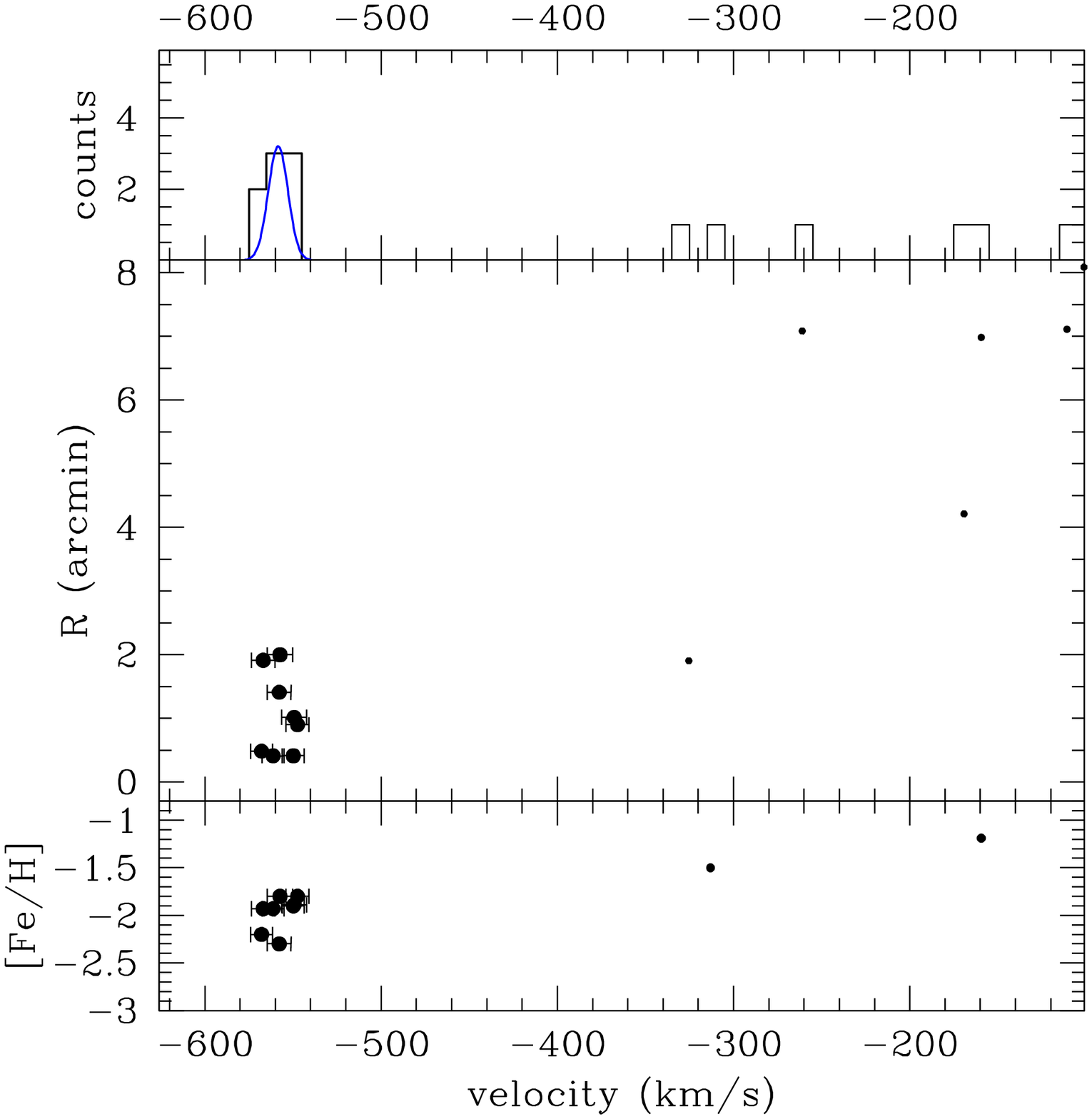}
\includegraphics[angle=0,width=0.45\hsize]{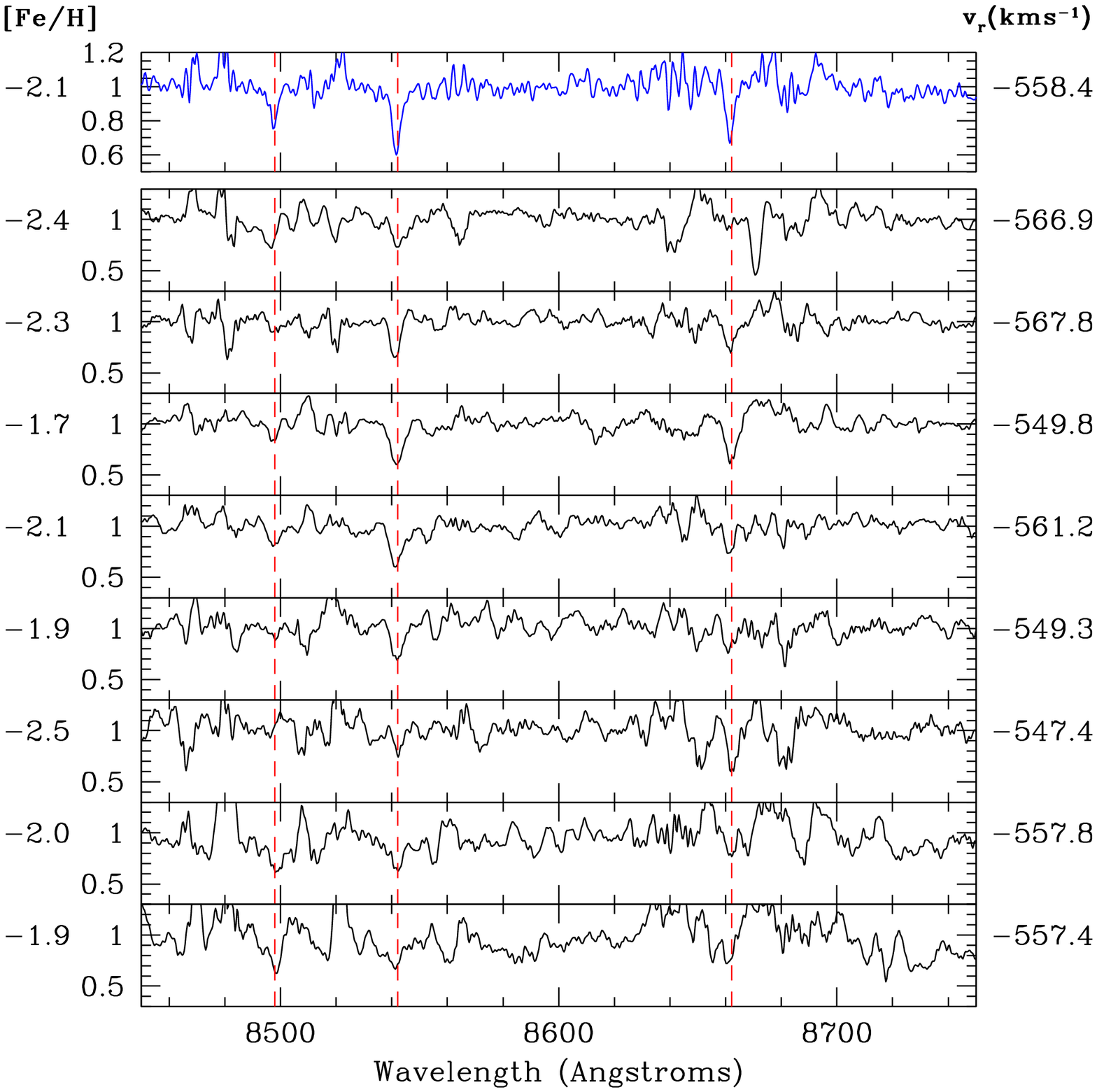}
\caption{As Fig.~\ref{and11spec}, but for \andxii. In the left panel,
  \andxii\ can be clearly identified as a kinematic grouping of 8
  stars at approximately -560~kms$^{-1}$, all located within 2' of the
  dwarf centre. The individual stars are all found to be metal poor,
  both photometrically (lower left panel) and spectroscopically (right
  panel). Analysis of the composite spectra of the confirmed members
  yields [Fe/H]=-2.1$\pm0.2$}
\label{and12spec}
\end{center}
\end{figure*}

\begin{figure*}
\begin{center}
\includegraphics[angle=0,width=0.8\hsize]{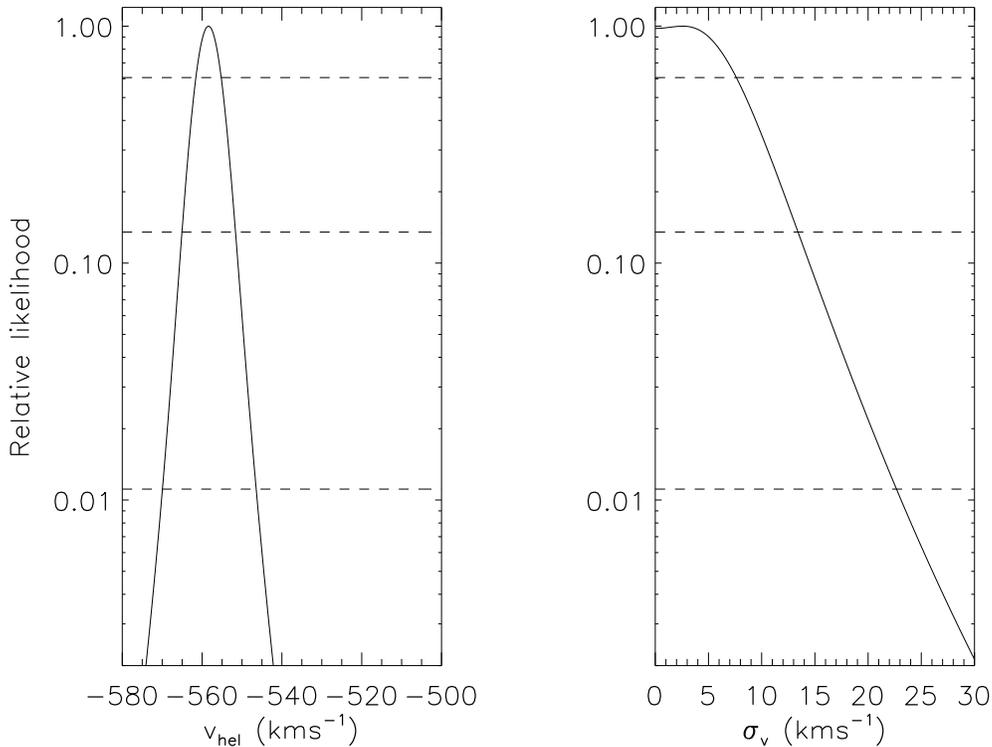}
\caption{Likelihood distributions for \andxii. Dashed lines represent
  the conventional 1,2 and 3$\sigma$ confidence levels. The mean
  velocity for \andxii\ is calculated to be -558.4~kms$^{-1}$, and we
  (marginally) resolve a velocity dispersion of
  2.6$^{+5.1}_{-2.6}$~kms$^{-1}$, though we note that this is
  consistent with zero within its 1$\sigma$ uncertainties.}
\label{and12contour}
\end{center}
\end{figure*}

\begin{table*}
\begin{center}
\caption{Properties of candidate member stars in \andxii, centered at $\alpha$=00:47:27, $\delta$=+34:22:29.}
\label{and12stars}
\begin{tabular}{lllcccccc}
\hline
star $\alpha$  & star $\delta$  & vel (kms$^{-1}$) & [Fe/H]$_{spec}$ &  [Fe/H]$_{phot}$ & S:N (\AA$^{-1}$)  & TDR & $I$-mag & $V$-mag  \\
\hline
00:47:31.04 & +34:24:12.1 & -566.9$\pm$6.6 & -2.4 & -2.0 &  8.9 & 5.0 & 21.1 & 22.4 \\
00:47:24.69 & +34:22:23.9 & -567.8$\pm$6.3 & -2.3 & -1.7 & 10.1 & 10.7 &21.2 & 22.5 \\
00:47:27.76 & +34:22:6.2  & -549.8$\pm$6.2 & -1.7 & -1.7 & 11.4 & 11.0 & 21.2 & 22.6 \\
00:47:28.63 & +34 22 43.1 & -561.2$\pm$6.3 & -2.1 & -1.8 & 9.7  & 8.9 &21.5 & 22.8 \\
00:47:31.34 & +34 22 57.6 & -549.3$\pm$7.0 & -1.9 & -1.6 & 7.3  & 6.75 &21.7 & 22.9 \\
00:47:26.65 & +34 23 22.8 & -547.4$\pm$6.5 & -2.5 & -1.7 & 6.2  & 5.97 & 22.0 & 23.3 \\
00:47:27.18 & +34 23 53.5 & -557.8$\pm$6.7 & -2.0 & -1.7 & 5.4  &5.3 &  22.0 & 23.2 \\
00:47:30.60 & +34 24 20.3 & -557.4$\pm$7.3 & -1.9 & -1.5 & 4.8  & 3.7 & 22.2 & 23.3 \\ 
\hline
\end{tabular}
\end{center}
\end{table*}

\begin{figure*}
\begin{center}
\includegraphics[angle=0,width=0.8\hsize]{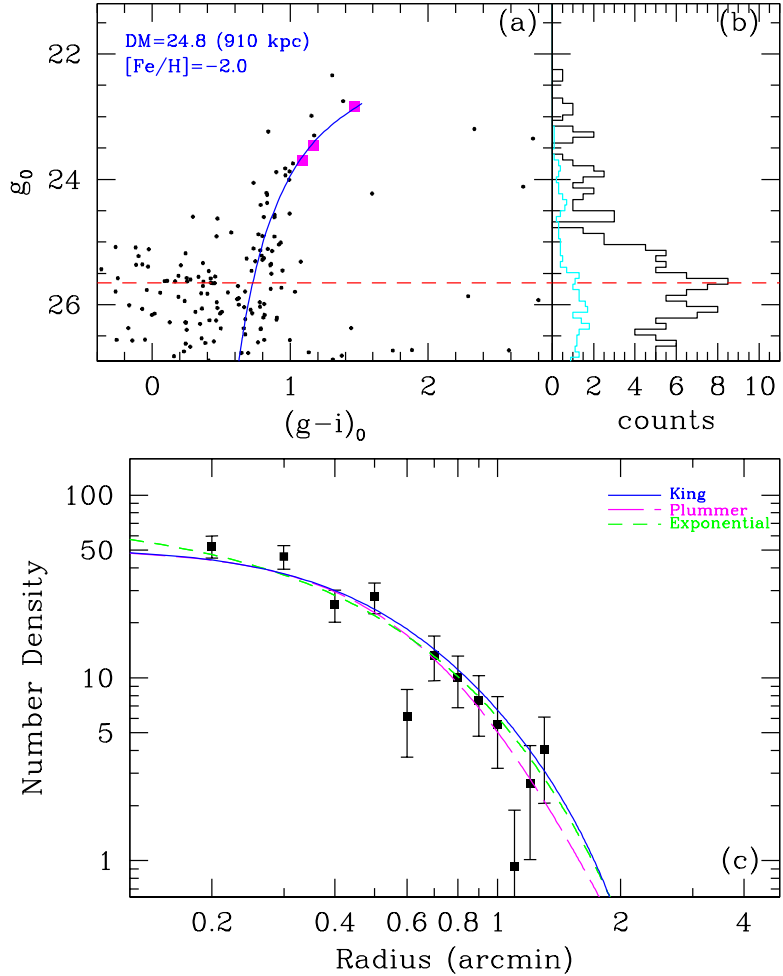}
\caption{As Fig.~\ref{and11phot} but for \andxiii. Dartmouth
  isochrones \citep{dart08} is overlaid with [Fe/H]=-2.0 and
  [$\alpha$/Fe]=+0.4 at the HB distance modulus of 24.8 (910 kpc as
  seen in the luminosity function of \andxiii, plot (b)) In plot (c),
  King (with r$_{t}=2.8$'), Plummer and exponential profiles are
  overlaid, with respective half-light radii of $0\mcnd78\pm0.08$,
  $0\mcnd81\pm0.09$ and $0\mcnd76\pm0.1$, with the King profile
  providing the best fit to the data. }
\label{and13phot}
\end{center}
\end{figure*}

\begin{figure*}
\begin{center}
\includegraphics[angle=0,width=0.45\hsize]{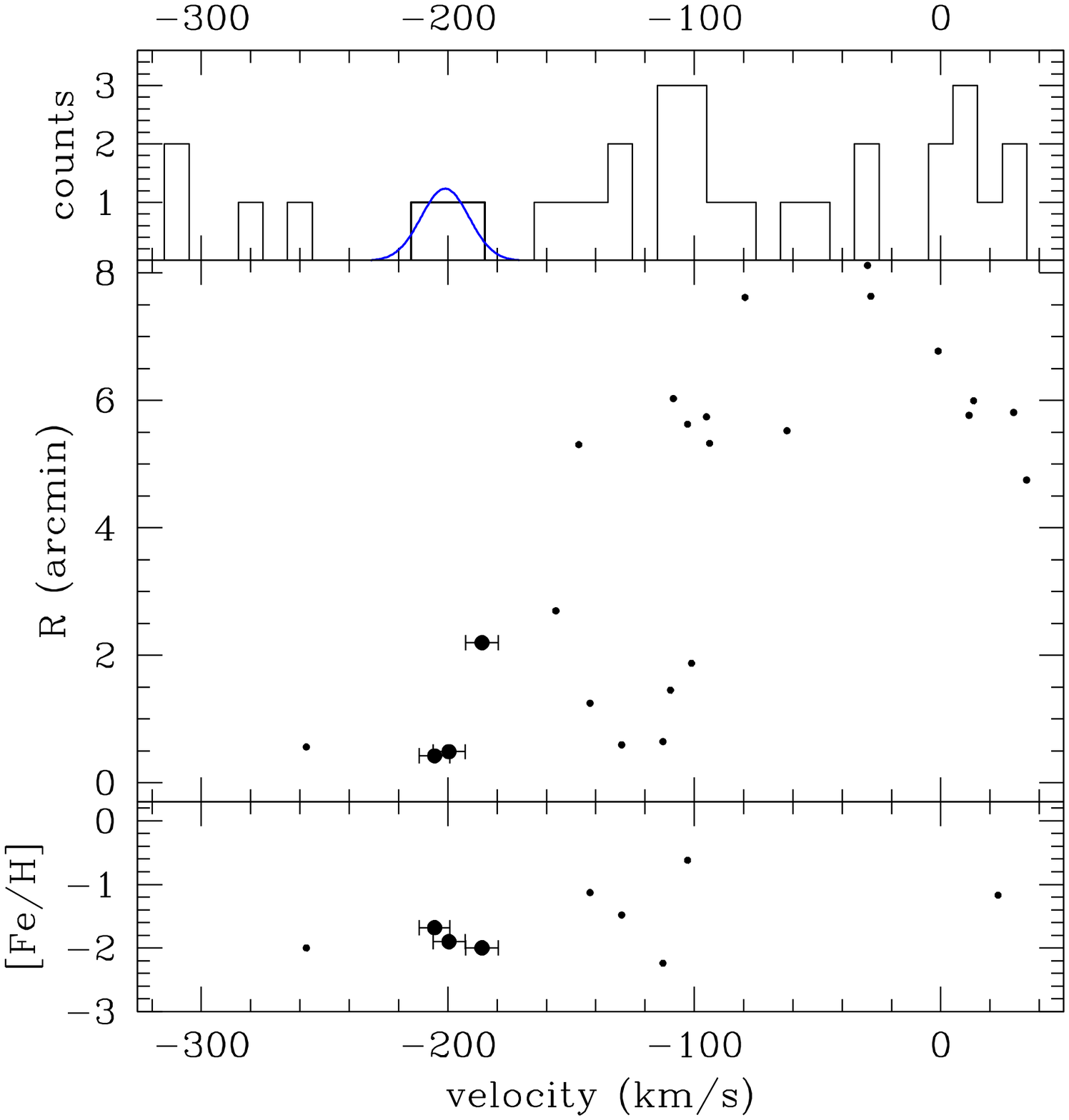}
\includegraphics[angle=0,width=0.45\hsize]{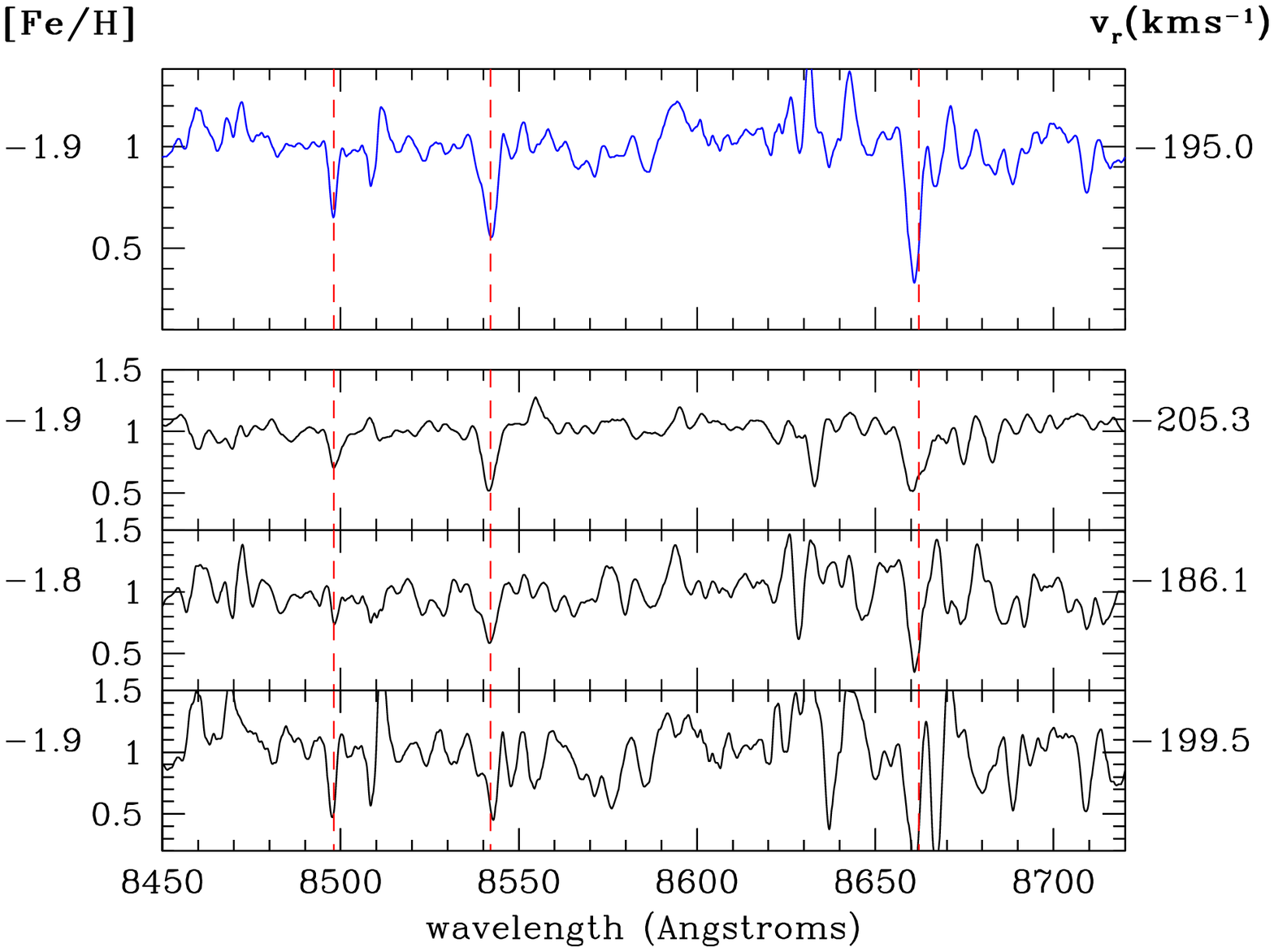}
\caption{As Fig.~\ref{and11spec}, but for \andxiii. In the left panel,
  \andxiii\ can be identified as a kinematic grouping of 3 stars at
  approximately -200~kms$^{-1}$, all of which are located within 2' of
  the dwarf centre. The individual stars are all found to
  be metal poor, both photometrically (lower left panel) and
  spectroscopically (right panel). Analysis of the composite spectra
  of the confirmed members yields [Fe/H]=-1.9$\pm0.2$}
\label{and13spec}
\end{center}
\end{figure*}

\begin{figure*}
\begin{center}
\includegraphics[angle=0,width=0.8\hsize]{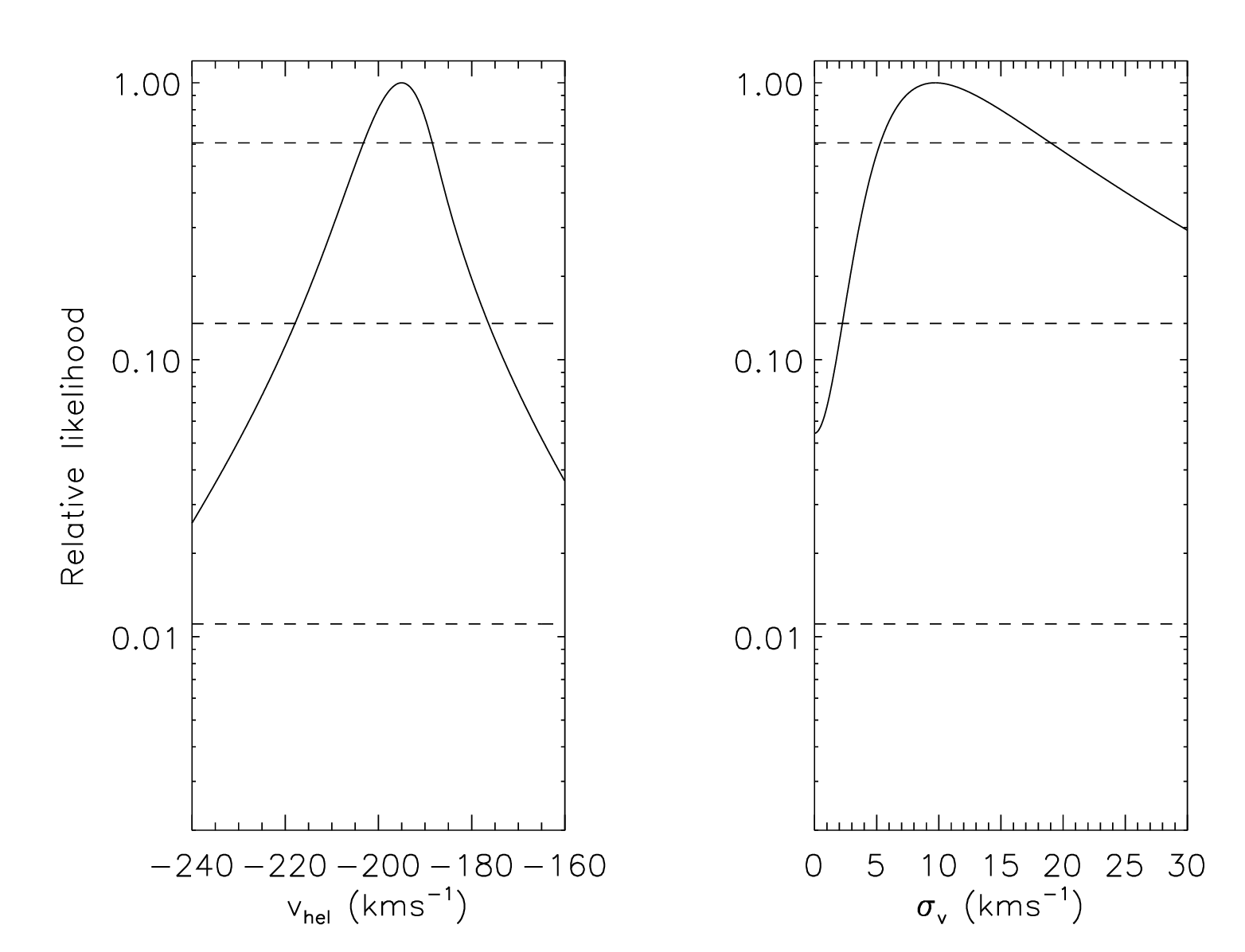}
\caption{Likelihood distributions for \andxiii. Dashed lines represent
  the conventional 1,2 and 3$\sigma$ confidence levels.  The mean
  velocity for \andxiii\ is determined to be -195.0~kms$^{-1}$, and
  the velocity dispersion is resolved as $\sigma_v=9.7^{+8.9}_{-4.5}~\kms$.}
\label{and13contour}
\end{center}
\end{figure*}

\subsubsection{Spectroscopic properties}

Here we present the spectroscopic results from \andxi; the data
for each spectroscopic member star can be found in
Table~\ref{and11stars}. Of the stars that were identified as potential
\andxi\ members from the CFHT CMD and targeted with DEIMOS, four
appear to be kinematically associated with the satellite. As shown in
the top left panel of Fig.~\ref{and11spec}, where we outline the
spectroscopic results for \andxi, one can clearly identify a kinematic
peak lying at a heliocentric velocity of $\sim-420$kms$^{-1}$. These
four stars have a mean velocity of v$_r=-419.4$kms$^{-1}$ and an
unusually low dispersion of $\sigma_v=2.4$kms$^{-1}$ (which we note is
uncorrected for instrumental errors). We also initially identify one
other star as a potential member of \andxi\, with a velocity of
-432.3~kms$^{-1}$. While it is clearly an outlier in velocity, it is
important to assess its potential membership to the satellite. The
inclusion of this star as a member decreases the uncorrected values of
v$_r$ to -422~kms$^{-1}$ and $\sigma_v$ to 4.0~kms$^{-1}$, however
the velocity error on this fifth star is greater than the other 4
candidate members (10.8~kms$^{-1}$ vs. average v$_{err}$ of 6.2
~kms$^{-1}$). This is partially due 
to the fact that the CaII$_{8662}$ line is
completely buried in the noise of the spectrum, as shown in the bottom
right panel of Fig.~\ref{and11spec}. As a result, when
cross-correlating the spectrum with the CaII template we achieve a peak
in the correlation function of less than 0.1  and a TDR value of only 3.4, 
implying that our velocity measurement for this star is unreliable. For 
this reason, we define this star as a tentative member of \andxi, 
but we exclude it from the following kinematic analysis. 

It is important for us to determine whether any of our remaining stars
are contaminants in the form of either foreground G dwarf stars or M31
halo stars.  Because \andxi\ has such a highly negative velocity
($\sim-420$kms$^{-1}$), contamination from MW G dwarfs is unlikely, as
these are predominantly found at velocities of greater than -160
kms$^{-1}$ \citep{robin04,ibata05}, so the chances of us finding these
stars at the velocity of \andxi\ are slim. Nevertheless, we can make
use of the Na I doublet, a pressure sensitive absorption feature
observed at $\sim8100$\AA. The strength of this feature is strongly
correlated with the surface gravity of stars, therefore it is
prominent in G dwarfs, and much weaker in RGB stars. As our spectra
cover the region of this doublet, we can measure the EWs of the
lines. Any star for which the EW of the Na I doublet is greater than
1.8\AA\ we classify as a foreground contaminant.
As there is some overlap in the EWs of these two populations, this
will inevitably lead to the occasional misclassification, but it
should ensure that the majority of foreground stars are removed. For
the four stars we classify as members of \andxi\ we measure the EW of
the Na I doublet to be less than 1.5\AA\ in all cases. Contaminating
M31 halo stars are more difficult to cleanly define. The M31 halo has
a systemic velocity of $\sim-300$kms$^{-1}$ and an average dispersion
of $\sim90$kms$^{-1}$, \citep{chapman06,kalirai06} so it is possible
that stray halo stars in the tails of the distribution could be found
at a velocity similar to that of \andxi. However, given the position
of \andxi, found in a low-density region of M31's halo, finding such a
star within a few half-light radii of the object is unlikely. To
limit any such contamination, we impose a radial cut on our members,
requiring them to sit within the tidal radius of the dSph, 2.13$'$. We plot the
distance of all observed stars from the centre of \andxi\ as a
function of their velocity in Fig.~\ref{and11spec}. It can be seen
that all our candidate members lie within this radius. As a final 
check for contaminants, we plot the photometric
metallicity of each star as a function of velocity, derived by
interpolating between the same Dartmouth isochrones \citep{dart08} as
used in \S~3.1.1. Halo stars are expected to have metallicities of
[Fe/H]$\sim-1.4$, and foreground dwarfs analysed with these isochrones
would likely appear to have metal-rich metallicities of
[Fe/H]$\sim-0.6$. Our 4 member stars and the outlier are all found to
be similarly metal poor, with a median [Fe/H]=-1.9. Combining these
factors, we find no reason to reject any of our four plausible members
as contaminants.

The spectra for each of the candidates (including the outlier) are
displayed in Fig.~\ref{and11spec}, along with a composite spectrum for
the four most plausible members (weighted by their signal-to-noise),
which is shown in the top right panel, highlighted in blue. As each of
the spectra has a high enough S:N, we can determine their metallicity
individually and also calculate an average value from their sum, from
the Calcium II triplet (as described in \citealt{ibata05}), giving a
resulting measure of [Fe/H] for the satellite which is independent of
photometric calculations. This relation between the EWs of the Ca
  II triplet and the resulting metallicity is dependent on the
  apparent V-band magnitude of the observed star. By summing together
  stars of different magnitudes we introduce an error into our
  calculations as we must assume an average value for the V-band
  magnitude of our sample. As the stars analysed within this work are
  all of a similar luminosity, this error is typically small (less
  than 0.1 dex), and we factor this into our error calculations. The
average value determined from the combined spectrum is
[Fe/H]$=-2.0\pm0.2$. This is considerably more metal poor than that
calculated by \citet{martin06} of [Fe/H]=-1.3 from the CFHT
photometry. This is likely due to the use of \citet{girardi04}
isochrones in their study, which produce more metal-rich values in the
MegaCam filter sets than other isochrone models, such as the Dotter
isochrones used in this paper. This effect has been noted in the
globular clusters NGC 2419 and M92 \citep{girardi04,girardi08}, where
the Girardi et al. isochrone models in the g-i colours produce more
metal-rich values for these objects than other published values.

Following the maximum likelihood algorithm outlined in
\citet{martin07}, we calculate the mean radial velocity, $v_r$, and
the intrinsic velocity dispersion, $\sigma$, for \andxi\ by sampling a
coarse grid in $(v_r,\sigma)$ space and determining the parameter
values that maximise the likelihood function (ML), defined as:

\begin{equation} 
ML(v_r,\sigma)=\prod_{i=1}^{N}\frac{1}{\sigma_{\mathrm{tot}}}
\exp\Big[-\frac{1}{2}\left(\frac{v_r-v_{r,i}}{\sigma_{\mathrm{tot}}}\right)^2\Big]
\end{equation}

\noindent with $N$ the number of stars in the sample, $v_{r,i}$ the
radial  velocity measured for the $i^\mathrm{th}$ star, $v_{err,i}$
the corresponding uncertainty and
$\sigma_{\mathrm{tot}}=\sqrt{\sigma^2+v_{err,i}^2}$. In this way, we
are able to separate the intrinsic dispersion of \andxi\ from the
dispersion introduced by our measurement uncertainties. We display the
one dimensional likelihood distributions for v$_r$ and $\sigma_v$ in
Fig.~\ref{and11contour} where the dashed lines represent the
conventional 1, 2 and 3$\sigma$ (68\%, 95\% and 99.7\%) uncertainties
on the values. We recover a systemic velocity for \andxi\ of
$v_r$=-419.6$^{+4.4}_{-3.8}$~kms$^{-1}$ (where the errors indicate the
1$\sigma$ uncertainties). Owing to our low sample size and S:N, the
dispersion is unresolved, however the 1$\sigma$ value suggests a
maximum value of 4.6~kms$^{-1}$ from the data, which we overlay as a
Gaussian curve in the top panel of Fig.~\ref{and11spec}a. We note that
these values are only marginally affected if we include the tentative
member at -432.3.~kms$^{-1}$, with $v_r$ increasing to
-422.5~kms$^{-1}$, and $\sigma_v$ is still unresolved at 0~kms$^{-1}$,
with a maximum set by the 1$\sigma$ uncertainty of $\sim5$kms$^{-1}$.

These results suggest that the velocity dispersion of \andxi\ is
unusually low, comparable to the curious MW objects Willman 1 and Leo
V, which have measured dispersions of $4.2_{-1.1}^{+2.1}$~kms$^{-1}$
and 2.4$_{-1.4}^{+2.4}$~kms$^{-1}$ respectively
\citep{martin07,walker09a}, and the M31 dSph, And X, which has a
dispersion of 3.9$\pm1.2\kms$ \citep{kalirai09}. We caution that
  the calculation of dispersions for cold systems such as \andxi\ rely
  heavily on accurately knowing the errors attributed to each velocity
  measurement. Whilst we have attempted to investigate our errors as
  thoroughly as possible (as detailed in \S2.2.1), if we have
  overestimated our errors, we will have underestimated our velocity
  dispersion (and vice-versa), and the true velocity dispersion could be
  higher than the dispersion we calculate here.

Using the spectral data obtained above, we can make an estimate for
the mass of \andxi\ by assuming it is a spherical system in virial
equilibrium. Throughout the literature, several different methods for
calculating the mass of dispersion-supported galaxies -- such as \andxi --
are employed which do not necessarily produce similar values when
applied to the same data sets. This makes comparisons of these
values for Local Group dSphs difficult, as one is not
necessarily comparing like with like. In this work, we shall use three
different methods to estimate the mass of our dSphs; the
\citet{richstone86} and \citet{illingworth76} methods, which are the
two most common approaches, and we also apply a new method, presented
in \citet{walker09b}, which we discuss in more detail below.

First, we apply the core-fitting method of \citet{richstone86}. This
method assumes that the system is in virial equilibrium with a flat
core, an isotropic velocity distribution and a constant mass-to-light
ratio throughout. One can then derive the mass-to-light ratio $(M/L)$
of \andxi\ from:

\begin{equation}
M/L=\eta\frac{9}{2\pi G}\frac{\sigma_0^2}{S_0 r_{hb}}
\end{equation}

\noindent where $S_0$ is the central surface brightness of the dwarf,
$r_{hb}$ is the value of the radius where the surface brightness drops
to half the peak surface brightness (approximately $2/3\times r_{1/2}$)
and $\sigma_0$ its central velocity dispersion. $\eta$ is a model
dependent, dimensionless constant that \citet{richstone86} showed to
be very nearly unity for a wide range of mass models, e.g. the
isothermal sphere, where they found $\eta$=1.013. For \andxi, we
derive $S_0=0.455\pm0.09\lsun\pc^{-2}$ when using $\mu_V=27.35
$~mag/arcsec$^2$, and taking $r_{1/2}=145^{+15}_{-29}\pc$ we find
r$_{hb}=97^{+10}_{-19}\pc$.  We further assume
$\sigma_0=\sigma_v=4.7\kms$ (taking the 1$\sigma$ limit from our
maximum likelihood calculation) and use this to derive an effective
upper limit on the mass-to-light ratio for \andxi. This leads to
$(M/L)<152\msun/\lsun$. Combining this with the updated luminosity of
$4.9^{+0.4}_{-0.2}\times10^4\lsun$ calculated above, we deduce an
upper limit on the mass of \andxi\ of $M<7.4\times10^6\msun$.

The prescription of \citet{illingworth76}, uses an r$^{1/4}$ law to
describe the surface brightness profile of the system, rather than the
core-fitting technique of \citet{richstone86}. It again assumes virial
equilibrium and an isotropic velocity distribution for the system. One
can then derive the mass using:

\begin{equation}
M=\eta r_{1/2}\sigma_v^2
\end{equation}

\noindent where $\eta$=850$M_{\odot}$pc$^{-1}$km$^{-2}$s$^2$, we calculate
M$<$2.6$\times10^6$ M$_{\odot}$ and M/L=53, which is less than half the
value derived above, highlighting the difficulty faced in comparing
the masses of dSphs within the Local Group.

Recently, papers by \citet{walker09b} and \citet{wolf09} have shown that
the mass within the half-light radius of dispersion-supported
galaxies, such as dSphs, can be accurately estimated with only mild
assumptions about the spatial variation of the stellar velocity
dispersion anisotropy ($\beta$), giving it an advantage over other
methods (such as those described above), which are more sensitive to
the adopted value of $\beta$. Following the
approach of \citet{walker09b}, we can derive the mass contained within
r$_{1/2}$ (M$_{half}$) from:

\begin{equation}
M_{half}=\mu r_{1/2} \sigma_v^2
\end{equation}

\noindent where $\mu=580$M$_{\odot}$pc$^{-1}$km$^{-2}$s$^2$. For
\andxi, this gives us an upper limit on M$_{half}$ for the dSph of
1.7$\times10^6\msun$. We do note that, as 2 of our 4 \andxi\ members
sit outside of the half-light radius for the dwarf, we are assuming
that there is no variation of $\sigma_v$ with radius, which may not be
the case. This could cause us to underestimate the mass of
\andxi\ within this region.

\subsection{\andxii}

\begin{table*}
\begin{center}
\caption{Properties of candidate member stars in \andxiii, centered at $\alpha$=00:51:51, $\delta$=+33:00:16. }
\label{and13stars}
\begin{tabular}{lllllcccccc}
\hline
star $\alpha$  & star $\delta$  & vel (kms$^{-1}$) & [Fe/H]$_{spec}$ &  [Fe/H]$_{phot}$ & S:N (\AA$^{-1}$) & TDR & $I$-mag & $V$-mag  \\
\hline
00:51:53.00 & +33:00:18.2 & -202.6$\pm$6.2  & -1.9  & -1.7  & 9.9 & 16.2 & 21.0 & 22.4 \\
00:51:49.81 & +33:00:41.1 & -195.7$\pm$6.5  & -1.9  & -1.6 & 3.3 & 7.2 & 22.1 & 23.3 \\
00:51:42.34 & +32:58:05.3  & -184.7$\pm$6.6  & -1.8  & -1.9 & 4.7 & 5.9 & 21.9 & 23.1 \\
\hline
\end{tabular}
\end{center}
\end{table*}

\andxii\ was the faintest of the trio of dwarfs presented in
\citet{martin06}, with an absolute magnitude of M$_{v}$= -6.4. Its
orbital properties were discussed in a previous paper by
\citet{chapman07}, where it was shown that due to its large radial
velocity, it was likely only experiencing the gravitational effects of
M31 for the first time. In this section, we give a complete overview
of the photometric and spectroscopic properties of \andxii\ in the
same format as for \andxi. We list the properties of our
spectroscopically confirmed members in Table~\ref{and12stars}.

\subsubsection{Photometric properties}

First, we discuss the structural properties derived for \andxii\ from
the Subaru data. We determine the centre of \andxii\ as
($\alpha$,$\delta$)=($00^{h} 47^{m} 27^{s}$, $+34\deg$ $22\mcnp$
$29\scnp$) using the same technique as for \andxi, and we display the
Subaru CMD for all stars within a 1.5$\mcnp$ radius of the centre of
\andxii\ in Fig.~\ref{and12phot}(a) (pink squares represent the DEIMOS
candidate members in this data set). The background-corrected radial
profile can be seen in Figure~\ref{and12phot}(c), with a background
level of 9.9$\pm0.4$ stars per square arc minute. Plummer and
exponential models were fitted and are shown over-plotted on the data. 
The Plummer and exponential fits yield half-light radii of
$1\mcnd2\pm0.2$ and $1\mcnd1\pm0.2$ respectively, with the plummer
profile showing the best fit to the data (reduced chi-squared of 3.7
vs. 3.9). The King profile would not converge on a physical fit for
these data, yielding results where the tidal radius is over an order
of magnitude smaller than the half-light radius. We therefore fit a
King profile, with the half light radius fixed to the half-light
radius given by the Plummer profile. The resulting best fit tidal
radius is r$_t$=3.1$'\pm0.3$, and we use this value as a cut for membership
 in \S3.2.2. As can be seen in Fig.~\ref{and12phot},
the radial profile for \andxii\ is not as clean as for \andxi, and
this may be due to the presence of variable extinction in the field of
And XII, which appears (based on visible cirrus) to vary on angular
scales of $\sim1'$, and is reflected by the larger errors in r$_{1/2}$
(of order $\sim10-20\%$ compared with $\sim5\%$ for \andxi) and the
inability to successfully fit a King profile to the data. This small
scale extinction is also present in a nearby globular cluster, MGC1
\citep{mackey09}, and is found to vary on scales much smaller than the
6' x 6' resolution of the Schlegel et al. (1998) dust maps of this
region. Conversely, these poor fits could be indicative of tidal
disruption in the outer regions of \andxii; however, given the large
systemic velocity of -558.4~kms$^{-1}$ (discussed in \S~3.2.2), this
is unlikely to have been as a result of an interaction between
\andxii\ and M31, as \andxii\ is likely experiencing the
potential of its host for the first time \citep{chapman07}.

Adopting the same TRGB Monte Carlo procedure as with \andxi\ we find
D$_{max}$=1~Mpc, and assuming a maximum offset of 0.5 mag from the
true TRGB, we compute a range of distances to \andxii\ of
800-1000~kpc, consistent with the most likely distance for \andxii\ of
830~$\pm50$~kpc found by \citet{chapman07}. An inspection of the
luminosity function of \andxii\ shown in the top right panel of
Figure~\ref{and12phot}(b) reveals a clear peak in the HB at
g-magnitude just below $\sim25.3$, implying a distance modulus for the
satellite of $\mu$$\sim24.7$ (~870 kpc), which sits within our
calculated range. We use the most-likely distance of 830$^{+170}_{-30}$ kpc
from \citet{chapman07} for the remainder of our analysis. Again, we
recalculate the luminosity of the satellite, finding a value of
$L=3.1^{+0.2}_{-0.1}\times10^4\lsun$, corresponding to an absolute
magnitude of M$_v$=-6.4$^{+0.1}_{-0.5}$ and a half-light radius (using
the value from the exponential model), r$_{1/2}$, becomes
289$^{+67}_{-49}$~pc (previously estimated as 125 pc).

We analyse the CMD of \andxii\ to ascertain a mean value of [Fe/H] for
the dSph. We overlay \citet{dart08} isochrones (with age and abundance
as detailed in \S~3.1.1) for the range of distances calculated for
\andxii\ above (800-1000 kpc), and conclude that the metallicity of
\andxii\ lies between [Fe/H]=-1.7 and [Fe/H]=-2.0. In
Fig~\ref{and12phot} we overlay the best observed fit for the
most-likely distance (830 kpc), with [Fe/H]=-1.9.

\subsubsection{Spectroscopic properties}

We now discuss the kinematic results for \andxii. As seen in
Fig~\ref{and12spec}, 8 stars are found to be kinematically associated
with \andxii\ at a velocity of $\sim-560$ kms$^{-1}$. Unlike \andxi\,
we have no ambiguity about which of these stars are members as they
are tightly distributed in velocity, all sit within
$2$'($<2$r$_{1/2}$) of the dwarf centre, and all show negligible absorption
at the position of the Na I doublet. Their photometric metallicities
(lower left panel of Fig.~\ref{and12spec}) are also in good agreement
with one another, ranging from [Fe/H]=-1.5 to -2.0, with median
[Fe/H]=-1.7. We display the individual spectra for each candidate
member in Fig.~\ref{and12spec}, with the combined spectrum for the
satellite shown in the top panel. Using the CaII approach as described
for \andxi, metallicity measurements are made for each of the
candidate stars, spanning [Fe/H]=-1.5 to -2.5, with a median
[Fe/H]=-2.0. To estimate an average metallicity for the dwarf, we
analyse the composite spectrum, and deduce a metallicity of
$\FeH=-2.1\pm0.2$. Eliminating the 3 stars with poor S:N (bottom 3
panels of Fig.~\ref{and12spec}) yields a value of
[Fe/H]=-2.0$\pm0.25$. Both of these values agree well with the
photometric metallicities discussed above.

Before correcting for measurement errors, the 8 stars imply
$\sigma_v=4.9$~kms$^{-1}$. The same maximum likelihood approach as
with \andxi\ yields a systemic velocity for the satellite of
-558.4$^{+3.2}_{-3.2}$ kms$^{-1}$, and we (marginally) resolve a velocity dispersion
for the satellite of $\sigma_v$= 2.6$^{+5.1}_{-2.6}$ kms$^{-1}$, where
the errors represent the 1$\sigma$ uncertainties. Again, this value is
uncommonly low for a dSph. 

As before, we apply eqns.~(2)-(4) to our \andxii\ data in an attempt
to constrain the mass and dark matter content of the galaxy.  Using
the \citet{richstone86} equation for mass-to-light (eqn.~(2)) and our
updated parameters for r$_{1/2}$ of 289$^{+67}_{-49}$~pc,
S$_0=0.21^{+0.07}_{-0.05}\lsun\pc^{-2}$ and the dispersion value from
the maximum likelihood analysis of 3.8$^{+3.9}_{-3.8}$ kms$^{-1}$, we
find M/L= 55$^{+107}_{-55}\msun/\lsun$, giving a mass estimate of
1.7$^{+3.3}_{-1.7}\times10^6\msun$. Eqn.~(3) \citep{illingworth76}
gives us M/L=54$^{+107}_{-55}$ and a mass of
1.7$^{+3.3}_{-1.7}\times10^6\msun$, in agreement with the
\citet{richstone86} approach. In both cases, this value falls below
the historical $\sim10^7\msun$ mass threshold typically observed for
brighter dwarfs, even when taking the large errors into account.
 
Finally, applying the same formula for the mass within the half-light
radius of \andxii\ as for \andxi\ (eqn.~(4) \citealt{walker09b}), we calculate
M$_{half}=1.1^{+2.1}_{-1.1}\times10^6\msun$. For \andxii, six of our 8 member stars
lie within the half-light radius of the dSph, making our measured
$\sigma_v$ a reasonable approximation for the dispersion within r$_{1/2}$.

\begin{figure*}
\begin{center}
\includegraphics[angle=0,width=0.9\hsize]{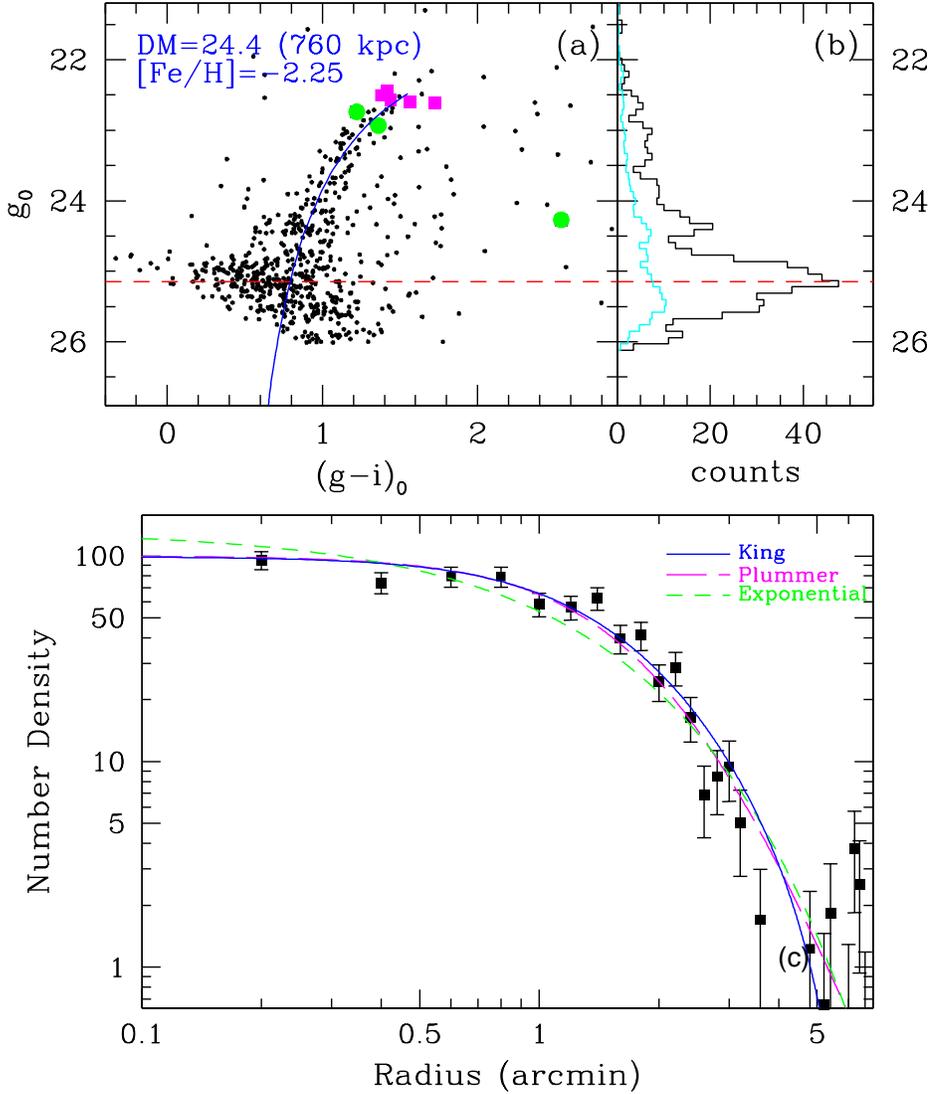}
\caption{As Fig.~\ref{and11phot} but for \andix. Isochrones are
  overlaid with [Fe/H]=-2.2 at the HB distance modulus of 24.5 (760
  kpc). In plot (c), King (with r$_{t}=6.3$), Plummer and exponential
  profiles are overlaid, with respective half-light radii of
  $2\mcnd7\pm0.2$, $2\mcnd6\pm0.1$ and $2\mcnd5\pm0.1$.  }
\label{and9phot}
\end{center}
\end{figure*}

\begin{figure*}
\begin{center}
\includegraphics[angle=0,width=0.45\hsize]{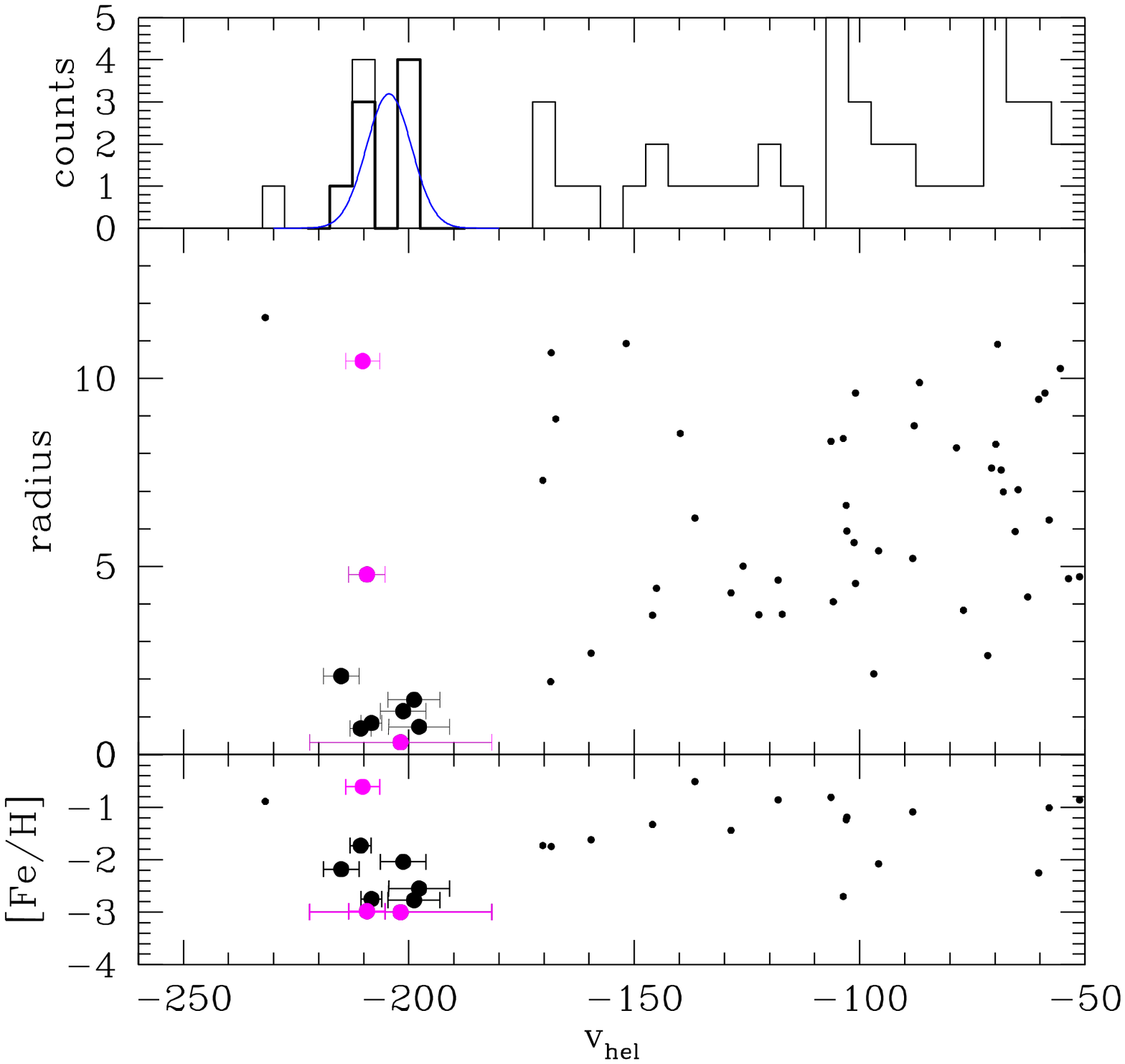}
\includegraphics[angle=0,width=0.45\hsize]{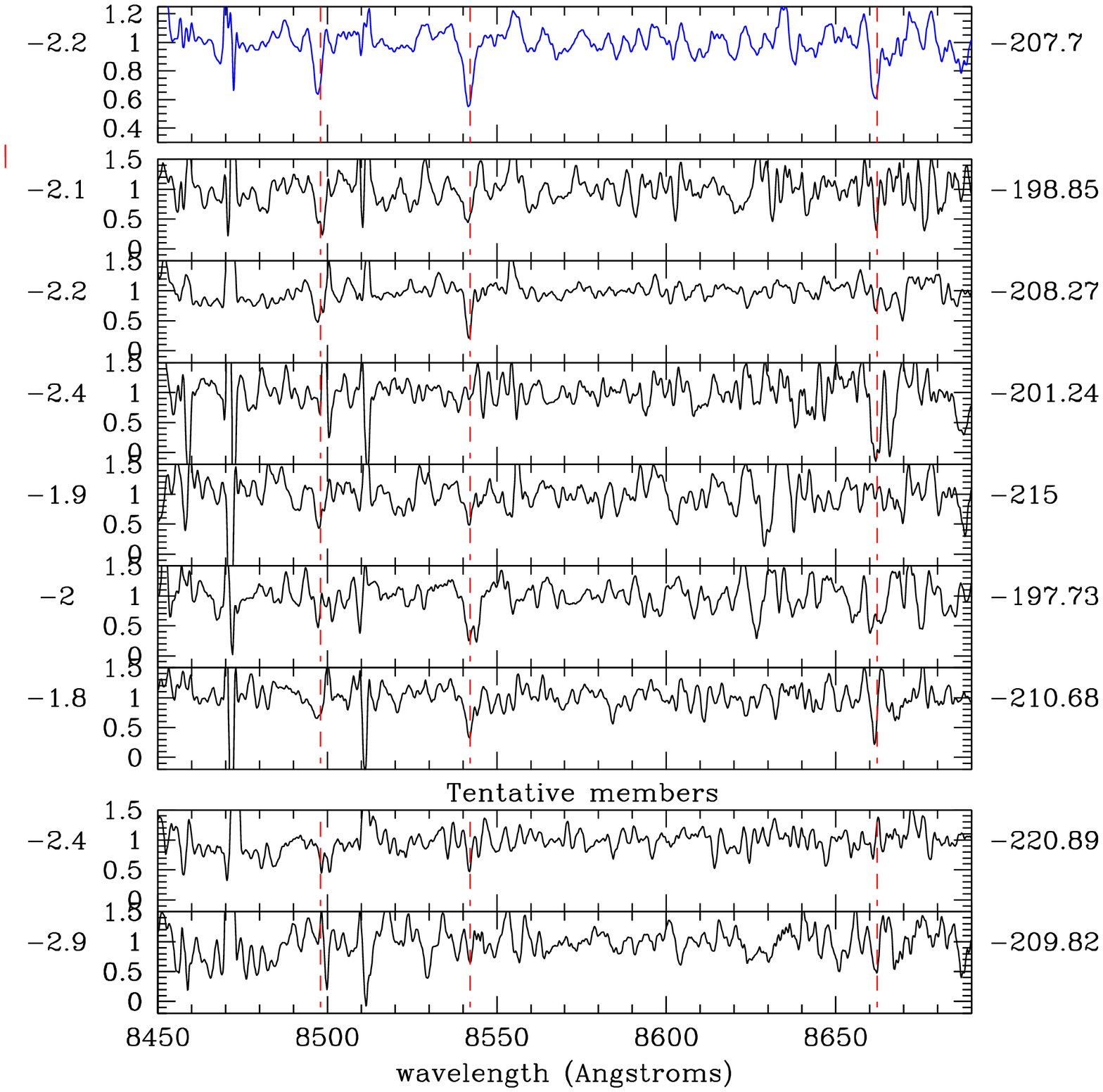}
\caption{As Fig.~\ref{and11spec}, but for \andix. \andix can clearly be identified as a kinematic grouping at $\sim$-200~kms$^{-1}$. Of the nine stars at this velocity, 8 are associated with \andix, with the 9th star lying well off the dwarf RGB (see Fig.~\ref{and9phot}, at a distance greater than 10 arcmins from the dwarf, and the two stars with low quality cross-correlations (TDR$<$3.5) highlighted in pink. The remaining 8 stars are all found to be metal poor both photometrically (lower left panel) and spectroscopically (right panel), however we note that 2 stars have unreliable velocity measurements. An analysis of the composite spectrum of the confirmed members yields [Fe/H]=-2.2$\pm0.2$.}
\label{and9spec}
\end{center}
\end{figure*}

\begin{figure*}
\begin{center}
\includegraphics[angle=0,width=0.8\hsize]{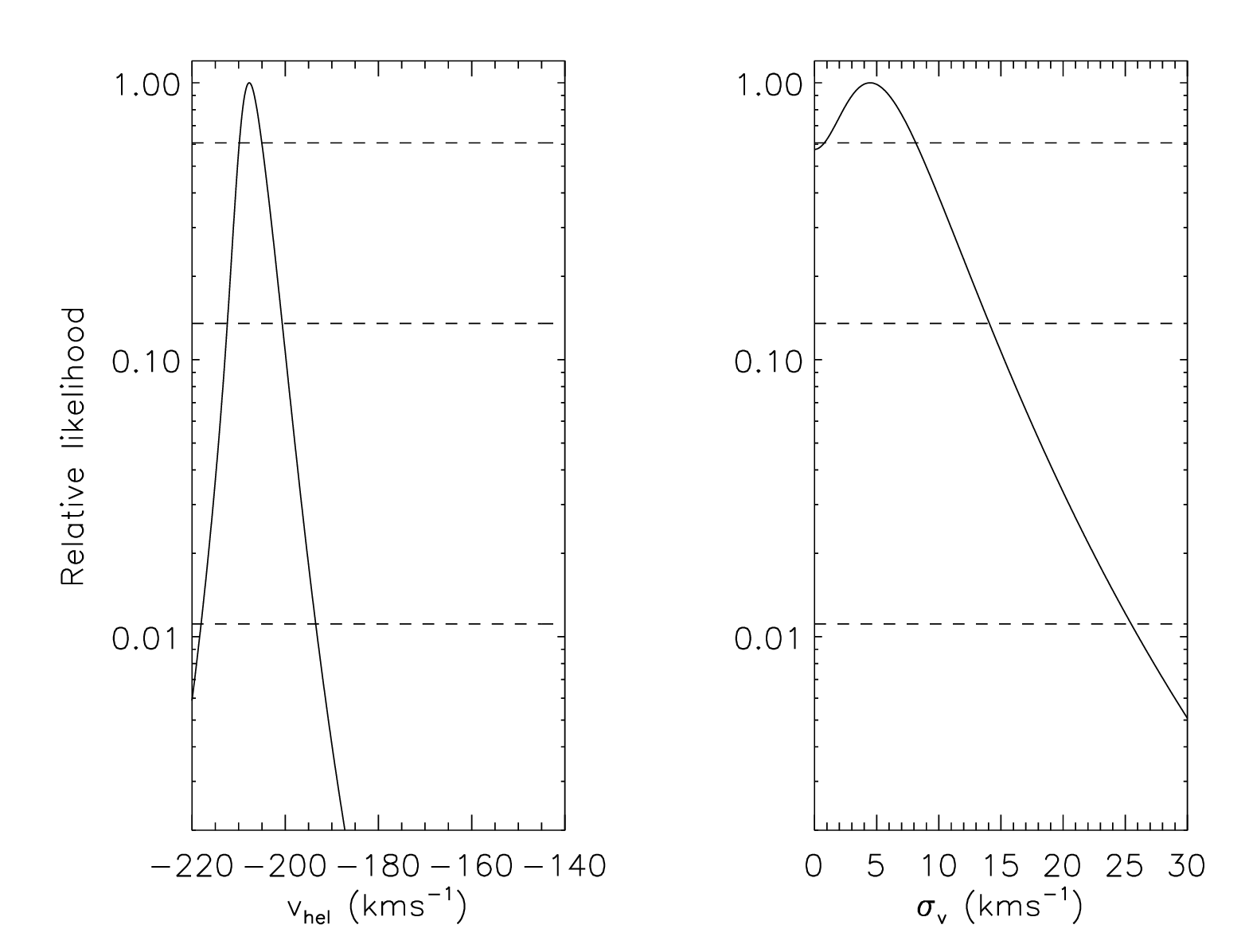}
\caption{Likelihood distributions for \andix. Dashed lines represent
  the conventional 1,2 and 3$\sigma$ confidence levels. The mean
  velocity for \andix\ is calculated to be -207.7~kms$^{-1}$, and the
  velocity dispersion is resolved by the technique as 4.5~$\kms$.}
\label{and9contour}
\end{center}
\end{figure*}

\subsection{\andxiii}

Discovered by \citet{martin06}, \andxiii\ suffered some chance
contamination with a background galaxy cluster, increasing the
apparent contrast of this dwarf in the imagery. Nonetheless, the
apparent brightness of \andxiii\ is similar to \andxi\ after
correction for this contamination. We present our observations of
\andxiii\ in the same format as the other M31 dwarfs, and list our
spectroscopic members in Table~\ref{and13stars}.

\subsubsection{Photometric properties}

Using the same Subaru photometric data set as for \andxi\ and \andxii,
we find the centre of \andxiii\ as ($\alpha$,$\delta$)=($00^{h}
51^{m} 50^{s}$, $+33\deg$ $00\mcnp$ $43\scnp$) and in
Fig.~\ref{and13phot}(a) we show the Subaru CMD for all stars within a
1.5$\mcnp$ radius of the centre of \andxiii\ (DEIMOS candidate members
are highlighted with pink squares, while tentative members are
highlighted by light blue squares). The background-corrected radial
profile is displayed here also (panel (c), with a background level of
11.2$\pm0.4$ stars per square arc minute). King (with
r$_{t}=2.8\pm0.2$ ), Plummer and exponential models are shown overplotted
on the data. These yield half-light radii of $0\mcnd78\pm0.08$,
$0\mcnd81\pm0.09$ and $0\mcnd66\pm0.1$ respectively, with the King
profile showing the best fit to the data. Again, this value of
r$_{1/2}$ is consistent with the r$_{1/2}$ resulting from the
application of the ML algorithm of \citet{martin08} (N. Martin,
private communication).

Adopting the same TRGB Monte Carlo procedure as with the previous 2
dwarfs, we find D$_{max}$=940~kpc for the satellite, and assuming a
maximum offset of 0.5 mag from the TRGB, a range of distances of
750-940~kpc. We also find a distance modulus from the HB of
\andxii\ of 24.8 (910~kpc), via inspection of the luminosity function,
which is in good agreement with the results from the TRGB
method. Using the HB distance, we calculate a total luminosity for the
satellite of $L=4.1^{+0.1}_{-0.2}\times10^4\lsun$. These revised distance and
luminosity values update the derived parameters of \andxiii\ with
M$_v$ becoming -6.7$^{+0.4}_{-0.1}$ (previously estimated as -6.9),
and the half-light radius, r$_{1/2}$, becomes 206$^{+27}_{-44}$ pc
(previously estimated as 115 pc).

As before, we overlay \citet{dart08} isochrones (with age and
abundance as detailed in \S~3.1) for the range of distances calculated
for \andxii\ above (750-940 kpc), giving a metallicity range of
[Fe/H]=-1.8 to -2.1 for \andxiii. In Fig~\ref{and13phot} we overlay the best
observed fit for the HB distance (910 kpc) which has [Fe/H]=-2.0.

\subsubsection{Spectroscopic properties}

Being the most distant of our 4 observed dSphs, \andxiii\ has
unsurprisingly provided us with the biggest challenge in identifying
probable member stars. When examining the individual spectra for the
stars in the \andxiii\ mask (a sample of which can be seen in
Fig.~\ref{and13spec}) we see that the CaII$_{8662}$ line has
experienced contamination from a skyline, which has significantly
affected its profile. As a result, the EW of the third line, which
should be $\sim3/4$ the EW of the second line, is larger than that of
the second in most cases. Such an effect will obviously change the
measured velocities and metallicities if we were to include it in our
cross-correlations and metallicity calculations. For this reason, we
derive velocities and metallicities for the satellite members based on
the CaII$_{8498}$ and CaII$_{8542}$ lines only. The resulting
kinematic data for \andxiii\ is shown in Fig.~\ref{and13spec}. We can
see that three RGB stars lying on the Subaru CMD
(Fig.~\ref{and13phot}) inhabit a similar region in velocity space,
with an average v$_r\sim$-190~kms$^{-1}$. This implies that
\andxiii\ is currently moving away from M31. These three stars have an
uncorrected dispersion of $\sigma_v$=4.3~kms$^{-1}$. The systemic
velocity of \andxiii\ places it much closer to the regime of Galactic
foreground contamination (v$_{hel}>-150$ kms$^{-1}$) than the previous
two dwarfs. With this being the case, it is particularly important for
us to assess whether any of our spectroscopic candidates could be
contaminants. Measuring the EW of the NaI doublet in each star, we
find that in each case the EW is significantly less than the
1.8\AA\ cut we defined in \S3.1.2, making them unlikely foreground
dwarf contaminants. To assess the likelihood of any of the three stars
being M31 halo stars that happen to be coincident with \andxiii\ we
invoke a radial cut equal to the tidal radius (2.8$'$). All our
potential member stars lie within this cut. Finally we assess the
quality of our derived velocities, and we note that all three stars
have TDR$>3.5$, meaning that they satisfy our quality
criteria. Therefore we use all three potential members in
our subsequent analysis.  Using the summed spectrum of these
members we calculate a metallicity for \andxiii\ of
[Fe/H]=-2.0$\pm0.3$, which is again more metal poor than that of
[Fe/H]=-1.4 reported in \citet{martin06}, but in good agreement with
our revised photometric metallicities detailed above.

The maximum likelihood distribution for the velocity and dispersion of
\andxiii\ is shown in Fig.~\ref{and13contour}. For our two probable
members, the technique suggests
$v_r$=-195.0$^{+7.4}_{-8.4}$~kms$^{-1}$ and a resolved dispersion of
$\sigma_v$=9.7$^{+8.9}_{-4.5}$~kms$^{-1}$. The corresponding errors are large ($\gta50\%$) owing to our small sample size (3 stars) and the low S:N (9.9, 4.7 and 3.3 \AA$^{-1}$) of our objects. If we compare this value
with dispersions for MW objects of a comparable luminosity such as
CVnI ($\sigma_v$=7.9 kms$^{-1}$, M$_V$=-7.1,
\citealt{martin08,simon07}) and Bootes I ($\sigma_v$=6.5
kms$^{-1}$,M$_V$=-6.3, \citealt{martin08,martin07}), we find that the
dispersion of \andxiii\ is consistent with its
counterparts within its $1\sigma$ errors.

We combine these results with the structural properties of
\andxiii\ derived from the Subaru data, to place a constraint
the dark matter content of \andxiii, using the same approach as for
the previous dwarfs. Using eqn.~(2) and a central surface brightness
of S$_0$=0.276$^{+0.05}_{-0.08}L_{\odot}$~\pc$^{-2}$ and the 
dispersion from our maximum likelihood analysis of $\sigma_v=$
9.7$\kms$, we calculate a mass to light ratio of
$M/L=821^{+1065}_{-538} M_{\odot}/L_{\odot}$ which implies a high dark matter
dominance. The upper limit for the mass is therefore
3.4$^{+4.4}_{-2.2}\times10^{7}$~M$_\odot$. The formula of \citet{illingworth76}
(eqn.~(3)), produces an upper limit on the mass of \andxiii\ of
1.6$^{+2.2}_{-1.0}\times10^7\msun$ (corresponding to M/L$=$390$^{+505}_{252}$), which is less than
half the value derived using the method of \citet{richstone86}, as was
the case for \andxi. Applying the formula for M$_{half}$ as before, we find
M$_{half}=$1.1$^{+1.4}_{-0.7}\times10^7\msun$.

\begin{table*}
\begin{center}
\caption{Properties of candidate member stars in \andix, centered at $\alpha$=00:52:51.1, $\delta$=+43:11:48.6. The two stars with unreliable velocities (TDR$<3.5$ and the contaminant star at $>$r$_{t}$ are also included as the last three entries.}
\label{and9stars}
\begin{tabular}{lllcccccc}
\hline
star $\alpha$  & star $\delta$  & vel (kms$^{-1}$) & [Fe/H]$_{spec}$ &  [Fe/H]$_{phot}$ & S:N (\AA$^{-1}$) & TDR  & $I$-mag & $V$-mag  \\
\hline
00:52:45.74 & +43:12:53.6 & -198.8$\pm5.8$ & -2.10 & -2.77 & 2.6 & 4.5 & 20.8 & 21.9 \\
00:52:53.18 & +43 12 48.2 & -208.3$\pm2.4$ & -2.20 & -2.75 & 4.7 & 7.5 & 20.9 & 22.1 \\
00:52:58.04 & +43:11:40.6 & -201.2$\pm5.1$ & -2.34 & -2.04 & 2.7 & 4.1 & 20.9 & 22.2 \\
00:52:58.96 & +43:13:39.1 & -215.0$\pm4.0$ & -1.90 & -2.18 & 2.5 & 4.3 & 21.0 & 22.2 \\
00:52:53.35 & +43:12:41.3 & -197.7$\pm6.0$ & -2.00 & -2.55 & 3.1 & 8.3 & 21.1 &  22.2 \\
00:52:52.30 & +43:12:41.1 & -210.7$\pm2.3$ & -1.75 & -1.73 & 3.7 & 12.1 & 20.9 & 22.3 \\
\hline
00:53:15.36 & +43:14:11.1 & -209.3$\pm12.1$ & -2.84 & -2.98 & 3.2 & 2.7 & 21.5 & 22.5 \\
00:52:50.32 & +43 11 54.4 & -201.7$\pm20.3$ & -2.90 & -3.00 & 2.5 & 1.6 & 21.5 & 22.6 \\
00:53:49.38 & +43 12 20.0 & -202.3$\pm4.5$  & -0.60 & -0.61 & 3.1 & 5.0 & 21.6 & 23.9 \\
\hline
\end{tabular}
\end{center}
\end{table*}

\subsection{\andix}

First reported by the SDSS team in \citet{zucker04}, \andix\ has
previously been studied both photometrically
\citep{zucker04,harbeck05,mcconnachie05a} and spectroscopically
\citep{chapman05}. In this work we intend to expand on previous
results by including new Subaru photometry (extinction corrected as
for the previous dwarfs, with E(B-V)=0.08 from the \citet{schlegel98}
maps), and update the kinematic results obtained with DEIMOS in
\citet{chapman05}, having corrected for the error in the initial
geometric models as discussed in \S~2. We present our
observations of \andix\ in the same format as the other M31 dwarfs,
and list all observed stars in Table~\ref{and9stars}.

\subsubsection{Photometric properties}

Using the Subaru data for the satellite, the centre of \andix\ is
derived as ($\alpha$,$\delta$)=($00^{h} 52^{m} 51.1^{s}$, $+43\deg$
$11\mcnp$ $48.6\scnp$) using the same technique as for the previous
dSphs. In Fig.~\ref{and9phot}(a) we show the Subaru CMD for all stars
within a 2.5' radius of the centre of \andix (pink squares represent
the DEIMOS candidate members in this data set). The
background-corrected radial profile was calculated using circular
annuli centred on these coordinates and can be seen in
Figure~\ref{and9phot} (with a background level of 23$\pm 4.8$ stars
per square arc minute). King (with r$_{t}=6\mcnd6\pm0.3$ ), Plummer
and exponential models have been plotted over the data. These
yield half-light radii of $2\mcnd7\pm0.2$, $2\mcnd6\pm0.1$ and
$2\mcnd5\pm0.1$ respectively, with the exponential profile showing the
best fit to the data (reduced chi-squared of 1.8 vs. 1.9 and 2.0 for
King and Plummer profiles). Again, this is consistent with the results
of the ML algorithm of \citet{martin08} (N. Martin, private
communication).

Adopting the same TRGB Monte Carlo procedure as with \andxi, we find
D$_{max}$=770~kpc for the satellite, and we calculate a range of
distances of 615-770~kpc for the dwarf. From the luminosity
function, we infer a distance modulus based on the HB of 24.4 (760~kpc),
also shown in Figure~\ref{and9phot}, which is consistent with the
findings of \citet{mcconnachie05a}. For this reason we use their
distance of D=765$^{+5}_{-150}$~kpc for the remainder of our
analysis. We calculate a total luminosity for the satellite at this
distance of $L=1.49^{+0.01}_{-0.07}\times10^5\lsun$. These revised
distance and luminosity values update the derived parameters of
\andix\ with M$_v$ becoming -8.1$^{+0.4}_{-0.1}$ (previously estimated
as -8.3), and using the result from the Plummer profile, r$_{1/2}$=
552$^{+22}_{-110}$ pc, previously estimated as 530~pc
\citep{chapman05} and 300~pc \citep{harbeck05}.

As for the previous dSphs, we estimate the mean photometric metallicity
of \andix\ by overlaying \citet{dart08} isochrones (with age and
abundance as detailed in \S~3.1) for the range of distances calculated
above (615-770 kpc), giving a metallicity range for \andix\ of
[Fe/H]=-1.9 to -2.3. We overlay the best observed fit for the HB
distance (765 kpc) in Fig~\ref{and9phot}, with [Fe/H]=-2.2.
 
\subsubsection{Spectroscopic properties}

We display our kinematic results for \andix\ in
Fig.~\ref{and9spec}. As with \andxiii, the third line of the Ca II
triplet in the stars observed in the \andix\ masks experiences
significant contamination from OH sky lines, so we derive velocities
using the first 2 lines only. We identify nine stars as potential
members of the satellite, at a velocity of $\sim$-200~kms$^{-1}$
(highlighted by the heavy histogram) and an uncorrected
$\sigma_v$=5.9~kms$^{-1}$. Like \andxiii, the systemic velocity of
\andix\ places it much closer to the regime of Galactic foreground
contamination. \andix\ also sits much closer (in projection) to M31
itself, so contamination from M31 halo stars is a larger problem than
for the other satellites, due to the increased density in this
region. Given both these factors, it is important for us to fully
check our sample for any contaminants. One star (highlighted in pink)
sits at a distance of roughly 10$'$ from the centre of the satellite,
which is beyond the tidal radius of \andix ($6.6'$). Further, when 
checking its position in the Subaru CMD
and its spectroscopic metallicity when compared with the other eight
candidates, we find that this anomalous 9th star lies well off the RGB
for the satellite (highlighted by the green circle in
Fig.~\ref{and9phot}) and is far more metal rich spectroscopically when
compared to the remaining eight stars ([Fe/H]=-0.6 versus
[Fe/H]$\sim$-2 to -3 for the other candidates). For these reasons, we
exclude this 9th star from the remainder of our analysis. Two other
stars with velocities consistent with the systemic velocity of
\andix\ at -220.9 kms$^{-1}$ and -209.8 kms$^{-1}$ have very low TDR
values of 2.7 and 1.6 respectively, falling well below our quality cut
of TDR$\ge$3.5, indicating that the velocities for these objects are
unreliable. Therefore, we do not include these objects in the rest of
our analysis, and label them as tentative members of \andix.

We display the spectra for each of the remaining 8 stars in
Fig.~\ref{and9spec} also, with the top panel displaying the summed
spectrum for the satellite. As was the case for \andxiii, it can be
seen that the CaII$_{8662}$ is obscured by OH skylines for the majority
of members. This is also evident in the composite spectrum, where the
equivalent width of the third line is measured to be significantly
smaller than that of the first line (whereas in \andxiii, the EW of
the 3rd line was enhanced). If we were to include this line in our
analysis, it would cause us to underestimate the value of [Fe/H] for
the dwarf. For this reason, we derive the metallicity for the
satellite based on the CaII$_{8498}$ and CaII$_{8542}$ lines only,
giving a metallicity for \andix\ of [Fe/H]=-2.2$\pm0.2$, which is a
more metal poor value than that found by \citet{chapman05} of
[Fe/H]=-1.5, and more consistent with the value of [Fe/H]=-2.0 found
by \citet{harbeck05} and the photometric metallicities we quote
above. We shall discuss this discrepancy further in \S~4. Such a low
metallicity would make this dwarf one of the most metal-poor M31 dSphs
observed to date, comparable to And V \citep{davidge02}.

A maximum likelihood distribution for \andix\ is shown in
Fig.~\ref{and9contour} showing the best value when correcting from
the instrumental measurements. From this technique we resolve
$v_r$=-207.7$^{+2.7}_{-2.2}$~kms$^{-1}$ and
$\sigma_v=4.5^{+3.6}_{-3.4}$~kms$^{-1}$ (overlaid as a Gaussian curve
in the top panel of Fig.~\ref{and9spec}). This dispersion is
significantly lower than the value quoted in \citet{chapman05} of
$\sigma_v$=6.8~kms$^{-1}$, and this is likely due to the inclusion of
non-members in their analysis as a result of the incorrect geometric
model.

As for \andxii, this measured velocity dispersion allows us to
constrain the dark matter content of \andix\ (rather than just assign
an upper limit, as for \andxi\ and \andxiii), assuming virial
equilibrium. As before, we first use eqn.~(2) to calculate the M/L of
\andix. We derive $\mu_0=$28.42 mags/arcsec$^2$ and
S$_0$=0.155$\pm0.01\lsun/pc^2$, and take
$\sigma_0=4.5^{+3.6}_{-3.4}$~kms$^{-1}$ from our ML analysis, giving a
most probable mass to light ratio $M/L=78^{+100}_{-78}
M_{\odot}/L_{\odot}$. The implied mass is therefore
1.2$^{+1.5}_{-1.2}\times10^{7}$~M$_\odot$, potentially low mass when
compared to other bright dwarfs (M$_V <-8$). Once again, applying
eqn.~(3) from \citet{illingworth76}, we calculate a mass of
9.9$^{+12.6}_{-9.9}\times10^{6}$~M$_\odot$ for \andxi, with a
corresponding M/L=66, which again highlights the inconsistencies
between the two methods.

Finally, applying equation (4) to our data, we estimate the mass
contained within the half-light radius of \andix\ to be
M$_{half}=6.5^{+8.3}_{-6.5}\times10^6\msun$. As these 6 \andix\ members 
sit within the half-light radius we determined in
\S~3.4.1, we can consider our velocity dispersion to be a good
approximation for the central dispersion of the dwarf.

\begin{table*}
\centering
\begin{minipage}{\textwidth}
\centering
\caption{Structural properties of \andxi, \andxii, \andxiii\ and
  \andix\ as derived in this work}
\begin{tabular}{lcccc}
\hline Property & \andxi\ & \andxii\ & \andxiii\ & \andix\   \\ \hline
\hline $\alpha$           & $0^h 46^m 21^s$       & $0^h 47^m 27^s$ &
$0^h 51^m 51^s$        &$0^h 52^m 51.1^s$ \\ $\delta$           &
$+33\deg\ 48' 22''$   & $+34\deg\ 22' 29''$    & $+33\deg\ 00' 16''$ &
$+43\deg\ 11' 48.6''$ \\ M$_{V,0}$          &-6.9$^{+0.5}_{-0.1}$
&-6.4$^{+0.1}_{-0.5}$      &-6.7$^{+0.4}_{-0.1}$ &-8.1$^{+0.4}_{-0.1}$
\\   r$_{1/2}$          &145$^{+24}_{-20}$ &289$^{+70}_{-47}$ pc
&203$^{+27}_{-44}$ pc &552$^{+22}_{-110}$ pc\\ Distance
&760$^{+10}_{-150}$ kpc &830$^{+170}_{-30}$ kpc           &910$^{+30}_{-160}$
kpc   &765$^{+5}_{-150}$ kpc \\   \hline
\end{tabular}
{\footnotesize
\begin{tabular}{l}
\end{tabular}}
\label{struct}
\end{minipage}
\end{table*}

\begin{table*}
\centering
\begin{minipage}{\textwidth}
\centering
\caption{Kinematic properties of \andxi, \andxii, \andxiii\ and
  \andix\ as derived in this work}
\begin{tabular}{lcccc}
\hline Property & \andxi\ & \andxii\ & \andxiii\ & \andix\   \\ \hline
\hline v$_r$ (kms$^{-1}$)  & -419.6$^{+4.4}_{-3.8}$
&-558.4$^{+3.2}_{-3.2}$ &-195.0$^{+7.4}_{-8.4}$ &-207.7$^{+2.7}_{2.2}$
\\ $\sigma_v$ (kms$^{-1}$  & 4.6$^*$            & 2.6$^{+5.1}_{-2.6}$
& 9.7$^{+8.9}_{-4.5}$              &4.5$^{+3.6}_{-3.4}$                  \\ 
 \FeH$_{spec}$ &-2.0$\pm
0.2$         &-2.1$\pm 0.2$            &-1.9$\pm 0.2$ &-2.2$\pm
0.2$\\ \FeH$_{phot}$      &-2.0                    &-1.9 &-2.0
&-2.2                \\ 
\hline
\end{tabular}
{\footnotesize
\begin{tabular}{l}
$^*$ Value quoted represents upper limit set by 1$\sigma$ uncertainty
  from maximum likelihood analysis\\
\end{tabular}}
\label{spec}
\end{minipage}
\end{table*}

\begin{table*}
\centering
\begin{minipage}{\textwidth}
\centering
\caption{Mass and M/L estimates for \andxi, \andxii, \andxiii\ and
  \andix\ as derived in this work}
\begin{tabular}{lcccc}
\hline Property & \andxi\ & \andxii\ & \andxiii\ & \andix\   \\ \hline
\hline
M/L$_{RT86}$ ($\msun$/$\lsun$)  & 152$^*$        &  55 $^{+152}_{-55}$    &
821$^{+1065}_{-538}$     & 117$^{+102}_{-100}$        \\ 
M$_{RT86}$ ($\msun$)& 7.4$\times10^{6*}$    &   1.7$^{+4.7}_{-1.7}\times10^{6}$ &  3.4$^{+4.4}_{-2.2}\times10^7$ &  1.2$^{+1.5}_{-1.2}\times10^{7}$ \\
M/L$_{I76}$ & 53$^*$     & 54 $^{+152}_{-54}$ & 401$^{+505}_{-252}$ & 66$^{+84}_{-66}$ \\
M$_{I76}$ ($\msun$) &  2.6$\times10^{6*}$   & 1.6$^{+4.6}_{-1.6}\times10^{6}$ & 1.6$^{2.1}_{1.0}\times10^{7}$ &   6.5$^{+8.3}_{-6.5}\times10^{6}$\\
M$_{half} (\msun)$  & 1.7$\times10^{6*}$   & 1.1$^{+3.1}_{-1.1}\times10^6$ &   1.1$^{+1.4}_{0.7}\times10^{7*}$ & 6.5$^{+8.3}_{-6.5}\times10^6$ \\
\hline
\end{tabular}
{\footnotesize
\begin{tabular}{l}
$^*$ Value quoted represents upper limit set by 1$\sigma$ uncertainty
  from maximum likelihood analysis\\
RT86 refers to \citet{richstone86}, eqn. (1)\\
I76 refers to \citet{illingworth76}, eqn. (2)\\
\end{tabular}}
\label{masses}
\end{minipage}
\end{table*}

\section{Discussion}

\subsection{Comparison with previous studies}

The 4 dSphs analysed in this work have all been previously examined
photometrically \citep{harbeck05,zucker04,martin06} and, in the case
of \andix\ and \andxii, spectroscopically
\citep{chapman05,chapman07}. Here we discuss how our results compare
with these other studies. For Andromeda's XI, XII and XIII, the deeper
photometric images from Subaru, combined with our spectroscopic
results from DEIMOS, have enabled us to significantly improve upon
previous values for structural parameters, such as half-light radius,
distance etc., as derived in \citet{martin06} from CFHT-MegaCam
data. We display our results for the structural properties of the
dwarfs in table~\ref{struct} and their kinematic properties are
displayed in table~\ref{spec}. Whilst the absolute magnitudes mirror
the initial values very closely, we find the other structural
parameters for the satellites all differ significantly. In all three
cases the half-light radii have increased from $\sim$115 pc to
$\sim$200 pc which is comparable to what is observed for MW dwarf
galaxies of a similar luminosity, e.g. LeoT with M$_{V}$=-7.1 and
r$_{1/2}$=170$\pm15$ \citep{irwin07,simon07}. This is in contrast to the
brighter M31 dSphs (M$_V<$-8), which are all observed to be larger
than their MW counterparts of a similar luminosity
\citep{mcconnachie05b}. We have also been able to derive distances to
each of the dwarfs individually, whereas in Martin et al. (2006),
photometry would only allow for an ensemble average. We find
\andxi\ to be the closest dwarf to us, lying at a distance of 760 kpc,
with both \andxii\ and \andxiii\ being more distant, located at $>$800
kpc. In addition, we have reanalysed \andxii\ spectroscopically, and
our findings are largely consistent with those of \citet{chapman07}.

For Andromeda IX, the structural properties we have derived from the
Subaru data closely mirror the findings of \citet{harbeck05},
\citet{zucker04} and \citet{chapman05}. Our range of distances
(615-770 kpc) and our HB distance estimate of 760 kpc agree well with
the \citet{mcconnachie05a} result of 765 kpc (using a TRGB
method). Our value for the half-light radius of 552$^{+22}_{-110}$ pc
is significantly larger than that quoted in \citet{harbeck05}, but
agrees with the findings of \citet{chapman06} and, like the remainder
of its bright M31 brethren, it is more extended than its Milky Way
counterparts of a similar luminosity (e.g. Draco, with M$_V$=-8.8 and
r$_{1/2}$=221 pc, \citealt{martin08}). However our re-reduction and
analysis of the spectroscopic data with the correct geometrical model
causes us to update a number of the parameters originally presented in
\citet{chapman05}. In our analysis, we resolve a velocity dispersion
of only $4.5^{+3.6}_{-3.4}~\kms$ , compared with their measured value
of $6.8\kms$. Reassessing the combined spectrum for \andix, we find
[Fe/H]=-2.2$\pm0.2$ for the satellite, significantly more metal poor
than [Fe/H]=-1.5 reported by \citet{chapman05}, largely due to
excluding the CaII$_{8662}$ line from our analysis, but also a result
of the misclassification of member stars in \citet{chapman05}. This
new value is in much better agreement with the work of
\citet{harbeck05} and \citet{zucker04} and is supported by our deep
Subaru photometry.

\begin{figure}
\begin{center}
\includegraphics[angle=0,width=0.8\hsize]{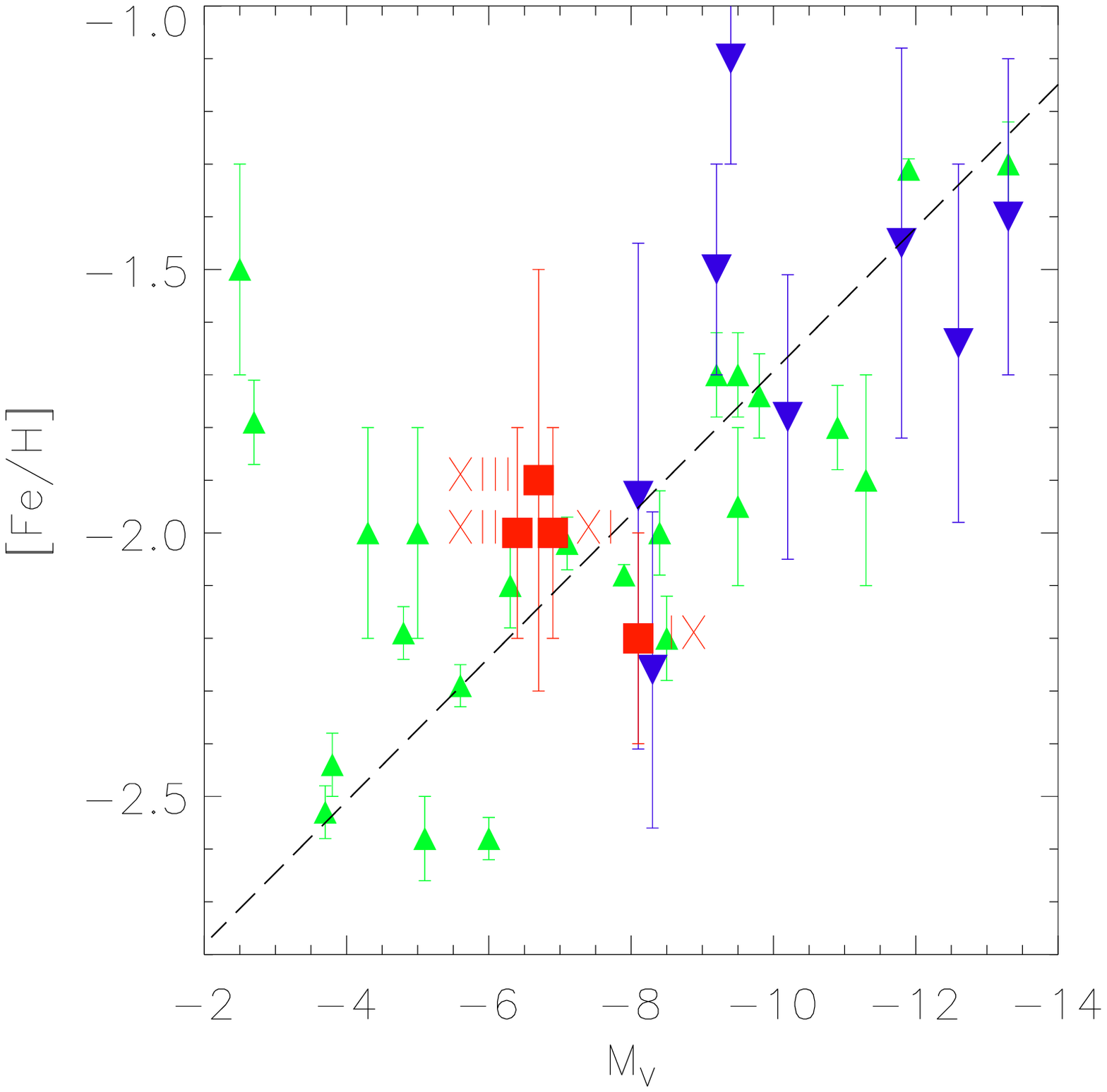}
\caption{Mean values for [Fe/H] for both M31 and MW satellites
  vs. absolute magnitude (M$_V$). A line of best fit to the MW dwarfs,
  taken from \citet{kirby08} is overplotted. Values for the M31 dwarfs
  were taken from
  \citet{dacosta02,dacosta00,davidge02,grebel99,mcconnachie05a,mcconnachie06b,mcconnachie08,irwin08,kalirai09,kalirai10},
  and for the MW, Koch et
  al. (2006,2007a,2007b),\citet{martin07,ibata06,kirby08,mateo98}. MW
  objects are plotted as green triangle, M31 objects as blue inverted
  triangles and our results are plotted as red squares.}
\label{feh}
\end{center}
\end{figure}

\subsection{Metallicities}

The metallicity for each dwarf has been reassessed, with all four
measured to be more metal-poor than in previous works. There is a good
agreement between Subaru photometry (using Dartmouth isochrones) and
spectroscopy for all the dSphs. In the past, much work has been
undertaken to determine the metallicities of faint dSphs in the MW  to
see if they follow the same pattern of decreasing metallicity with
decreasing luminosity as seen in their brighter counterparts
(e.g. \citealt{mateo98}). Until recently, this trend has appeared to
break down in the fainter regime of dSphs, with low luminosity
satellites measured to be significantly more metal-rich than
expected. However, work undertaken by \citet{kirby08}, measuring iron
abundances with high and moderate resolution spectra from the HIRES and
DEIMOS instruments on Keck, has shown that the faint satellites in the
MW do contain sizable populations of very metal-poor stars. This result places them in line with the
previously established trends, as shown in Figure \ref{feh},
where we plot the relation between absolute magnitude, M$_{v}$ and
[Fe/H] for dSphs. The green symbols represent the MW dwarfs and the
dashed line shows the best fit to these satellites as determined by
\citet{kirby08}. We have also plotted the available information for
the M31 dwarfs (blue inverted triangles) and it can be seen that they agree quite
well with the relation at higher luminosities. For all four dwarfs
analysed in this paper (plotted as red squares), it can be seen that
they appear to be consistent with the relation within their 1$\sigma$
error bars. This suggests that the relation between metallicity and
absolute magnitude does not break down at these fainter luminosities. 

Such a finding is perhaps unexpected, given work presenting evidence
for a common mass scale for the halos of the dSphs
population. \citet{strigari08} conclude that the dSphs of the Local
Group all have a common mass of $\sim10^7M_{\odot}$ within their
central 300~pc. Work by \citet{wolf09} measuring the mass
content of dSphs within their half-light radius, also presents
evidence for a common mass scale for the halos of the dSphs
population, concluding that even the faintest MW dSphs are consistent
with having formed within dark matter halos of
$\sim3\times10^9$M$_{\odot}$. Working within this framework of a
common mass scale for the halos of dSphs, one might expect to see
metallicity level off at fainter luminosities, rather than continuing
to decrease as, although the amount of luminous matter present in the
system is decreasing, the amount of dark matter remains roughly
constant. This would impose a limit to the amount of metals that could
escape the gravitational potential of the galaxy, preventing dSphs
from becoming overly metal poor. However, a similar study by
\citet{walker09b} (also measuring the mass content of dSphs within the
half-light radius) finds that, whilst the dSphs of the MW are broadly
consistent with having formed within a Universal dark matter halo,
there is significant scatter in this relationship at the faint end of
the luminosity spectrum which could indicate that these fainter dwarf
galaxies are residing in less massive halos. This would allow them to
expel more of their metals at early epochs, resulting in lower
metallicities, which could explain this continuation of the trend seen
in brighter dwarfs. Another explanation is provided by
\citet{salvadori09} who model the merger history of the Milky
Way from z=20 to the present day to investigate the nature of `Ultra
Faint' dSphs (such as those discussed in this work); they find that
the low metallicities of these objects are the result of lower gas
metallicity at the time of their formation, and suppression of star
formation during their evolution.

\begin{figure*}
\begin{center}
\includegraphics[angle=0,width=0.8\hsize]{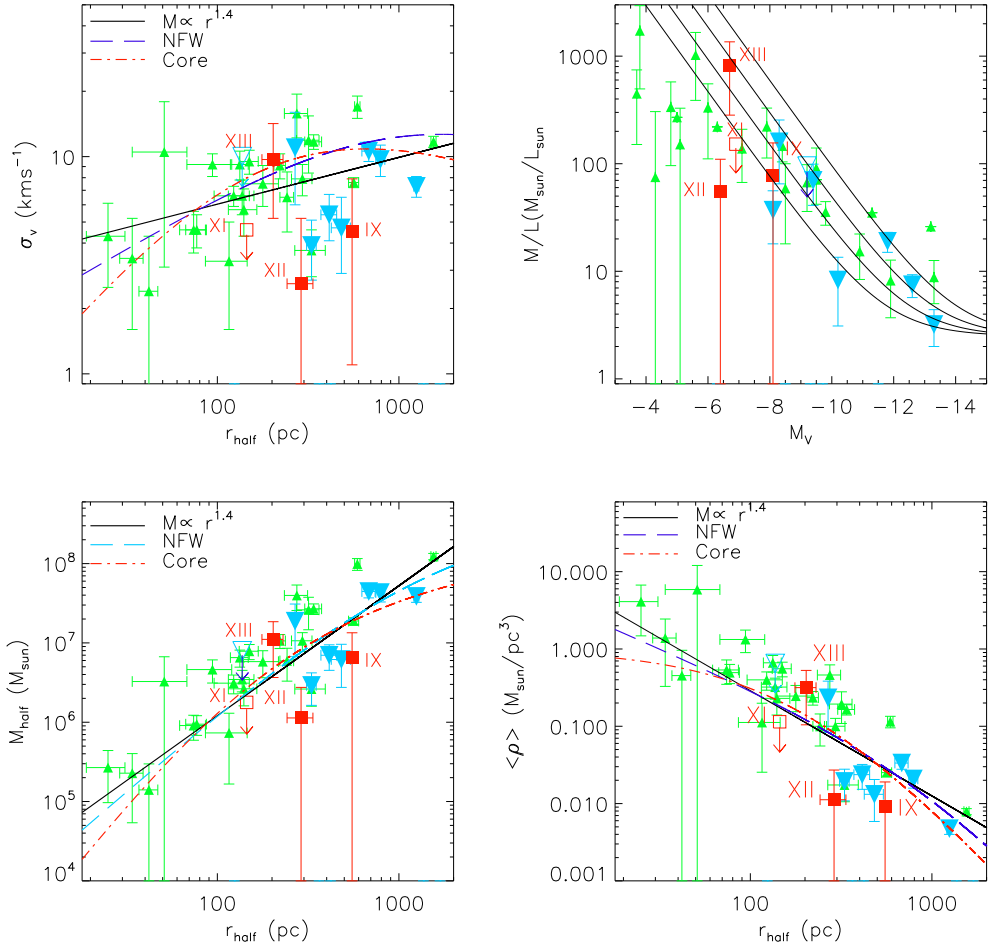}
\caption{ For all figures, the symbols used are as for
  Fig~\ref{feh}. In all cases, filled symbols represent resolved
  measurements, and open symbols represent upper limits set by
  1$\sigma$ uncertainties. References for MW and M31 data are as
  specified in Fig~\ref{feh}, plus \citet{walker09b} and references
  therein.{\bf Top left} Global velocity dispersion for MW and M31
  dSphs are plotted as a function of their half-light radius,
  The best-fitting mass profiles (based on power-law,
  NFW and cored halo profiles) for the MW dSph population, from
  \citet{walker09b} (transferred into the $\sigma_v$-r$_{1/2}$ plane)
  are overplotted. {\bf Top right:} Comparison of the mass-to-light ratios
  (M/L) and luminosity estimates (M$_v$) for the faintest known dwarf
  galaxies from both the MW and M31 with central velocity dispersion
  estimates. The Mateo (1998) relations for dwarf spheroidals are over
  plotted as black lines, representing curves of constant dark matter
  halo mass (1, 2, 4, 8$\times$ 10$^7$ M$_\odot$ from bottom to top),
  assuming a stellar mass-to-light ratio of 2.5
  M$_\odot$/L$_\odot$. \andxii\ appears to be an outlier to this
  relation. {\bf Bottom left:} Comparison of the mass contained within
  the half-light radius (M$_{half}$) as a function of r$_{1/2}$, with the
  \citet{walker09b} mass profiles overplotted. Both \andix\ and
  \andxii\ fall below these relations. {\bf Bottom right:} Mean
  density ($<\rho>$) within the half-light radius is plotted, with the
  same mass profiles as in the top right panel. Again, \andix\ and
  \andxii\ appear to be outliers to these relations.}
\label{mateo}
\end{center}
\end{figure*}

\subsection{The dynamics of M31's dSph population}

Recent work by \citet{walker09b} and \citet{wolf09}, investigating the
common mass scale for dSphs demonstrates a clear correlation between
the velocity dispersion and half-light radius for MW dSphs. In
Fig.~\ref{mateo} we plot $\sigma_v$ vs r$_{1/2}$ for all MW dwarfs
with published velocity dispersions (taken from \citealt{walker09b}),
the available M31 data points (\citealt{penarrubia08a,letarte09}, And
XVI is shown as an upper limit) and overplot the best fit power law,
NFW and cored profiles from \citet{walker09b} to these data. The upper
limit determined for \andxi\ is displayed as an open red
square, and the resolved measurements for \andxii, \andxiii\ and \andix\ are
shown as filled squares. Interestingly, \andxii\ and \andix\ both fall
below these relations within their 1$\sigma$ uncertainties. Whilst
these results could be an artifact of small sample size and low S:N,
given the scarcity of available kinematic data for the dSph population
of M31, it is useful to consider the implications of such findings. If
the dispersions of these dwarfs do fall significantly below
established trends, could this be an indication that the dSph
companions of M31 are dynamically colder than their MW counterparts?
The brighter dwarf galaxies of M31 are already observed to differ from
those orbiting the MW, with work by \citet{mcconnachie05b} showing
that M31 dSphs with M$_V<-8$ are spatially more extended than MW dSphs
of the same luminosity (though it is unclear what the origin of this
difference is). Curiously, if M31 and MW dSphs inhabit systematically
similar halos (i.e. the mass and extent of the stellar population is 
independent of the mass and extent of the dark matter halo), one would
expect to see the opposite trend, with the velocity dispersions of M31
satellites being roughly double those of MW dwarfs of the same
luminosity \citep{penarrubia08a}. It is also worth mentioning that a
recent study of And I, II, III, VII, X and XIV
\citep{kalirai09,kalirai10} finds that a number of these objects also
appear to be dynamically colder than their MW counterparts.

These colder dispersions translate to lower masses for \andix\ and
\andxii, as the three mass estimators we have used within this work
are all proportional to the square of the dispersion (as shown in
eqns.~(2)-(4)). Comparison of the masses of Local Group dSphs has
previously proven to be problematic. A variety of mass estimators (as
discussed in \S~3.1.2) used throughout the literature, and the
differing radial extent of the dSphs probed by kinematic surveys make
it difficult to ensure that, when making comparisons between dSphs
within the Local Group, we are truly comparing like with
like. Historically, the masses of dSphs have been discussed in terms
of the Mateo (1998) relation between M/L and the luminosity of dwarf
galaxies, which was established in relation to Local Group dSphs with
M$_V<$-8. This relation is displayed in the top right panel of
Fig.~\ref{mateo}, with M/L values for satellites from both the MW and
M31 taken from the literature plotted also. We display the upper
limits on M/L (using the \citealt{richstone86} mass estimator) for
\andxi\ (open red square), and plot the calculated values for \andix,
\andxii\ and \andxiii (filled red squares), where the error bars
represent their 1$\sigma$ uncertainties. We note that the lower limits
for three of the four dwarfs (\andxi, XII and IX) are consistent with
zero. \andix, XI and XIII are consistent with this relation within
their 1$\sigma$ error bars, but \andxii joins the ultra faint MW
dwarfs as an outlier.

The work by \citet{walker09b} and \citet{wolf09} has shown that the
mass within the half-light radius of dispersion supported galaxies,
such as dSphs, can be accurately estimated with only mild assumptions
about the spatial variation of the stellar velocity dispersion
anisotropy, unlike the two methods discussed above. Also, by
considering only the mass contained within r$_{1/2}$, one can negate
the effect of the differing radial extents probed by surveys, making
comparisons between the masses of the dSphs more meaningful. Their
conclusion that there is a tight correlation between velocity
dispersion (and hence, mass) and half-light radius for MW dSphs is
consistent with the idea that all the MW dSphs (including those with
M$_V>-8$) are formed with a ``Universal'' halo profile (as discussed
in \S~4.2). In the bottom left panel of Fig~\ref{mateo}, we plot
M$_{half}$ vs. r$_{1/2}$ for \andix, XI, XII and XIII along with the
other Local Group dSphs for which kinematic data are available, and
overplot the best fit power law, NFW and cored profiles from
\citet{walker09b} to these data. In this instance, we can see that
\andix\ and \andxii\ fall below these relations within their 1$\sigma$
errors, whereas \andxiii\ conforms to the MW trends.

In terms of its mass and dispersion, \andix\ in particular seems to
stand out. At M$_V=-8.1$ it is similar in terms of its absolute
brightness to the MW dwarfs, Draco and Ursa Minor (M$_V$= -8.8 and
-8.4 respectively, \citet{martin08}), and has a half-light radius that
is roughly twice that of these dwarfs (552 pc vs. $\sim200$pc),
however we measure a dispersion of only 4.5~kms$^{-1}$ for \andix,
compared with 9.1 and 9.6~kms$^{-1}$ for Draco and Ursa Minor
\citep{walker07,walker09b}. This suggests that \andix\ inhabits a
dark matter halo that is significantly different to those of the
brighter MW dSphs.

In the case of \andxii, whilst it does fall below the profiles of
\citet{walker09b}, it is quite similar in terms of both structure and
kinematics to several of its MW counterparts at a similar luminosity,
e.g. Hercules and Leo IV (with M$_V$=-6.0 and -5.1, r$_{1/2}$=330 and 116
pc, $\sigma_v$=3.7 and 3.3 kms$^{-1}$ respectively,
\citealt{walker09b}). Perhaps then, it is only these `Ultra Faint'
dSphs (M$_V>$-8) that reside in significantly different dark matter
halos, with \andix\ being a curious exception. It is worth noting that
the upper limit of $\sigma_v$=4.7 kms$^{-1}$ that we attach to
\andxi\ (M$_V$=-6.9, r$_{1/2}$=145~pc) suggests that this faint dwarf is
also an outlier to these relations.

If these satellites truly fall below these relations, the question
becomes, are they embedded in a less massive dark matter halos than
their MW counterparts, constituting a population of lower mass dwarfs
whose detection has thus far eluded us? These lower values of M/L and
M$_{half}$ are unlikely to be the result of tidal mass stripping, as
the process tends to increase rather than decrease values of M/L
\citep{penarrubia08b}. This is because the tightly bound central dark
matter ``cusp'' of a dwarf is more resilient to tidal disruption than
the stellar component of the galaxy, causing the object to become more
dark matter dominated as it undergoes tidal stripping. It should be
noted that the mass estimators of \citet{richstone86} and
\citet{illingworth76} are measures of the {\it central} population of
the galaxies, where the stars can be used as tracers, therefore they
are only true throughout the dwarf if the mass-to-light ratio is
constant throughout the system. It could be that these dwarfs reside
in more massive dark matter halos, such as is seen for brighter
galaxies, and these central low-mass estimates could imply lower
central densities for the two dwarfs, making them peculiar objects. To
examine this possibility further, we plot the mean total
(i.e. baryonic plus dark) mass density of the dSphs within the
half-light radius in the bottom right panel of Fig~\ref{mateo}, and it
can be seen that both \andix\ and \andxii\ have a lower density than
would be expected for their size.

\section{Conclusion}

We have studied the four faint M31 satellites, \andix, \andxi,
\andxii\ and \andxiii, using spectroscopic data from KeckII/DEIMOS and
Subaru/Suprime-Cam. Using our spectroscopic results, we have
identified probable members for each galaxy and constrained their
systemic velocities. Owing to our low S:N and small sample sizes, we
are unable to resolve a velocity dispersion for \andxi. Instead we use
the 1$\sigma$ confidence level from the maximum likelihood technique
as an upper limit for the satellite. In the case of \andix,
\andxii\ and \andxiii\ we resolve
$\sigma_v$=4.5$^{+3.6}_{-3.4}$kms$^{-1}$,
$\sigma_v$=3.8$^{+3.9}_{-3.8}$kms$^{-1}$ and
$\sigma_v$=9.7$^{+8.9}_{-4.5}\kms$ for each dwarf respectively, though
we note that \andxii\ is consistent with 0~kms$^{-1}$ within its
1$\sigma$ errors and the measurement of \andxiii's dispersion is based
on only three stars. Measuring the average spectroscopic metallicity
of each satellite from their composite spectra, we find that all four
satellites are metal-poor (\FeH$\sim$-2.0). Such metallicities are
comparable to what is observed in MW dSphs of a similar luminosity,
suggesting that the established relation between M$_V$ and
\FeH\ observed in both the brighter classical MW dwarfs
\citep{mateo98} and the fainter MW dSphs \citep{kirby08} continues
down to luminosities of M$_V>-6.4$ in the M31 dSphs population.

Finally, we place the velocity dispersions and masses for these
dSphs in the context of the universal dark matter halo profiles
established for the MW dwarf galaxies in \citet{walker09b} and find
that both \andix\ and \andxii\ have lower dispersions than would be
expected for dSphs of their size, similar to a number of dSphs studied 
by \citet{kalirai10}. If these results are verified, it
could imply that the dSph population of M31 is dynamically colder than
that of the MW, which is unexpected, given their larger spatial
extent. Such results also imply that the mass contained within the
half-light radii of these dwarfs is also lower than what is typically
observed in their MW counterparts of a similar size, and suggests that
these dSphs may reside in lower density dark matter halos. 

The results we have presented here indicate that, in terms of their 
dynamical properties, these satellites at
the faint end of the dwarf spheroidal spectrum do not behave as
expected from previous studies. Further spectroscopic observations of
faint objects similar to these (e.g., those found by
\citealt{mcconnachie08}) will undoubtedly help us to better understand
this low luminosity regime, and shed light onto the behaviour of these
faintest of galaxies.

\bibliography{mnemonic,michelle}{}

\begin{thebibliography}{}

\bibitem[\protect\citeauthoryear{{Allende Prieto}}{{Allende
  Prieto}}{2007}]{prieto07}
{Allende Prieto} C.,  2007, \aj, 134, 1843

\bibitem[\protect\citeauthoryear{{Bellazzini}, {Gennari}, {Ferraro} \&
  {Sollima}}{{Bellazzini} et~al.}{2004}]{bellazzini04}
{Bellazzini} M.,  {Gennari} N.,  {Ferraro} F.~R.,    {Sollima} A.,  2004,
  \mnras, 354, 708

\bibitem[\protect\citeauthoryear{{Belokurov} et~al.,}{{Belokurov}
  et~al.}{2007}]{belokurov07}
{Belokurov} V.,  et~al., 2007, \apj, 654, 897

\bibitem[\protect\citeauthoryear{{Belokurov}, {Walker}, {Evans}, {Faria},
  {Gilmore}, {Irwin}, {Koposov}, {Mateo}, {Olszewski} \& {Zucker}}{{Belokurov}
  et~al.}{2008}]{belokurov08}
{Belokurov} V.,  {Walker} M.~G.,  {Evans} N.~W.,  {Faria} D.~C.,  {Gilmore} G.,
   {Irwin} M.~J.,  {Koposov} S.,  {Mateo} M.,  {Olszewski} E.,    {Zucker}
  D.~B.,  2008, \apjl, 686, L83

\bibitem[\protect\citeauthoryear{{Chapman}, {Ibata}, {Lewis}, {Ferguson},
  {Irwin}, {McConnachie} \& {Tanvir}}{{Chapman} et~al.}{2005}]{chapman05}
{Chapman} S.~C.,  {Ibata} R.,  {Lewis} G.~F.,  {Ferguson} A.~M.~N.,  {Irwin}
  M.,  {McConnachie} A.,    {Tanvir} N.,  2005, \apjl, 632, L87

\bibitem[\protect\citeauthoryear{{Chapman}, {Ibata}, {Lewis}, {Ferguson},
  {Irwin}, {McConnachie} \& {Tanvir}}{{Chapman} et~al.}{2006}]{chapman06}
{Chapman} S.~C.,  {Ibata} R.,  {Lewis} G.~F.,  {Ferguson} A.~M.~N.,  {Irwin}
  M.,  {McConnachie} A.,    {Tanvir} N.,  2006, \apj, 653, 255

\bibitem[\protect\citeauthoryear{{Chapman}, {Pe{\~n}arrubia}, {Ibata},
  {McConnachie}, {Martin}, {Irwin}, {Blain}, {Lewis}, {Letarte}, {Lo}, {Ludlow}
  \& {O'neil}}{{Chapman} et~al.}{2007}]{chapman07}
{Chapman} S.~C.,  {Pe{\~n}arrubia} J.,  {Ibata} R.,  {McConnachie} A.,
  {Martin} N.,  {Irwin} M.,  {Blain} A.,  {Lewis} G.~F.,  {Letarte} B.,  {Lo}
  K.,  {Ludlow} A.,    {O'neil} K.,  2007, \apjl, 662, L79

\bibitem[\protect\citeauthoryear{{Chen}, {Zhao} \& {Zhao}}{{Chen}
  et~al.}{2009}]{chen09}
{Chen} Y.~Q.,  {Zhao} G.,    {Zhao} J.~K.,  2009, \apj, 702, 1336

\bibitem[\protect\citeauthoryear{{C{\^o}t{\'e}}, {Mateo}, {Olszewski} \&
  {Cook}}{{C{\^o}t{\'e}} et~al.}{1999}]{cote99}
{C{\^o}t{\'e}} P.,  {Mateo} M.,  {Olszewski} E.~W.,    {Cook} K.~H.,  1999,
  \apj, 526, 147

\bibitem[\protect\citeauthoryear{{Da Costa}, {Armandroff} \& {Caldwell}}{{Da
  Costa} et~al.}{2002}]{dacosta02}
{Da Costa} G.~S.,  {Armandroff} T.~E.,    {Caldwell} N.,  2002, \aj, 124, 332

\bibitem[\protect\citeauthoryear{{Da Costa}, {Armandroff}, {Caldwell} \&
  {Seitzer}}{{Da Costa} et~al.}{2000}]{dacosta00}
{Da Costa} G.~S.,  {Armandroff} T.~E.,  {Caldwell} N.,    {Seitzer} P.,  2000,
  \aj, 119, 705

\bibitem[\protect\citeauthoryear{{Davidge}, {Da Costa}, {J{\o}rgensen} \&
  {Allington-Smith}}{{Davidge} et~al.}{2002}]{davidge02}
{Davidge} T.~J.,  {Da Costa} G.~S.,  {J{\o}rgensen} I.,    {Allington-Smith}
  J.~R.,  2002, \aj, 124, 886

\bibitem[\protect\citeauthoryear{{Dotter}, {Chaboyer}, {Jevremovic}, {Kostov},
  {Baron} \& {Ferguson}}{{Dotter} et~al.}{2008}]{dart08}
{Dotter} A.,  {Chaboyer} B.,  {Jevremovic} D.,  {Kostov} V.,  {Baron} E.,
  {Ferguson} J.~W.,  2008, ArXiv e-prints, 0804.4473

\bibitem[\protect\citeauthoryear{{Faber} et~al.,}{{Faber}
  et~al.}{2003}]{deep2}
{Faber} S.~M.,  et~al., 2003, in {Iye} M.,  {Moorwood} A.~F.~M.,  eds, Society
  of Photo-Optical Instrumentation Engineers (SPIE) Conference Series Vol.~4841
  of Presented at the Society of Photo-Optical Instrumentation Engineers (SPIE)
  Conference, {The DEIMOS spectrograph for the Keck II Telescope: integration
  and testing}.
pp 1657--1669

\bibitem[\protect\citeauthoryear{{Ferraro}, {Paltrinieri}, {Rood}, {Fusi Pecci}
  \& {Buonanno}}{{Ferraro} et~al.}{2000}]{ferraro00}
{Ferraro} F.~R.,  {Paltrinieri} B.,  {Rood} R.~T.,  {Fusi Pecci} F.,
  {Buonanno} R.,  2000, \apj, 537, 312

\bibitem[\protect\citeauthoryear{{Geha}, {Willman}, {Simon}, {Strigari},
  {Kirby}, {Law} \& {Strader}}{{Geha} et~al.}{2009}]{geha09}
{Geha} M.,  {Willman} B.,  {Simon} J.~D.,  {Strigari} L.~E.,  {Kirby} E.~N.,
  {Law} D.~R.,    {Strader} J.,  2009, \apj, 692, 1464

\bibitem[\protect\citeauthoryear{{Girardi}, {Dalcanton}, {Williams}, {de Jong},
  {Gallart}, {Monelli}, {Groenewegen}, {Holtzman}, {Olsen}, {Seth}, {Weisz} \&
  {the ANGST/ANGRRR Collaboration}}{{Girardi} et~al.}{2008}]{girardi08}
{Girardi} L.,  {Dalcanton} J.,  {Williams} B.,  {de Jong} R.,  {Gallart} C.,
  {Monelli} M.,  {Groenewegen} M.~A.~T.,  {Holtzman} J.~A.,  {Olsen} K.~A.~G.,
  {Seth} A.~C.,  {Weisz} D.~R.,    {the ANGST/ANGRRR Collaboration} 2008,
  \pasp, 120, 583

\bibitem[\protect\citeauthoryear{{Girardi}, {Grebel}, {Odenkirchen} \&
  {Chiosi}}{{Girardi} et~al.}{2004}]{girardi04}
{Girardi} L.,  {Grebel} E.~K.,  {Odenkirchen} M.,    {Chiosi} C.,  2004, \aap,
  422, 205

\bibitem[\protect\citeauthoryear{{Grebel} \& {Guhathakurta}}{{Grebel} \&
  {Guhathakurta}}{1999}]{grebel99}
{Grebel} E.~K.,  {Guhathakurta} P.,  1999, \apjl, 511, L101

\bibitem[\protect\citeauthoryear{{Harbeck}, {Gallagher}, {Grebel}, {Koch} \&
  {Zucker}}{{Harbeck} et~al.}{2005}]{harbeck05}
{Harbeck} D.,  {Gallagher} J.~S.,  {Grebel} E.~K.,  {Koch} A.,    {Zucker}
  D.~B.,  2005, \apj, 623, 159

\bibitem[\protect\citeauthoryear{{Ibata}, {Chapman}, {Ferguson}, {Lewis},
  {Irwin} \& {Tanvir}}{{Ibata} et~al.}{2005}]{ibata05}
{Ibata} R.,  {Chapman} S.,  {Ferguson} A.~M.~N.,  {Lewis} G.,  {Irwin} M.,
  {Tanvir} N.,  2005, \apj, 634, 287

\bibitem[\protect\citeauthoryear{{Ibata}, {Chapman}, {Irwin}, {Lewis} \&
  {Martin}}{{Ibata} et~al.}{2006}]{ibata06}
{Ibata} R.,  {Chapman} S.,  {Irwin} M.,  {Lewis} G.,    {Martin} N.,  2006,
  \mnras, 373, L70

\bibitem[\protect\citeauthoryear{{Ibata}, {Martin}, {Irwin}, {Chapman},
  {Ferguson}, {Lewis} \& {McConnachie}}{{Ibata} et~al.}{2007}]{ibata07}
{Ibata} R.,  {Martin} N.~F.,  {Irwin} M.,  {Chapman} S.,  {Ferguson} A.~M.~N.,
  {Lewis} G.~F.,    {McConnachie} A.~W.,  2007, \apj, 671, 1591

\bibitem[\protect\citeauthoryear{{Illingworth}}{{Illingworth}}{1976}]{illingwo%
rth76}
{Illingworth} G.,  1976, \apj, 204, 73

\bibitem[\protect\citeauthoryear{{Irwin} \& {Lewis}}{{Irwin} \&
  {Lewis}}{2001}]{irwin01}
{Irwin} M.,  {Lewis} J.,  2001, New Astronomy Review, 45, 105

\bibitem[\protect\citeauthoryear{{Irwin} et~al.,}{{Irwin}
  et~al.}{2007}]{irwin07}
{Irwin} M.~J.,  et~al., 2007, \apjl, 656, L13

\bibitem[\protect\citeauthoryear{{Irwin}, {Ferguson}, {Huxor}, {Tanvir},
  {Ibata} \& {Lewis}}{{Irwin} et~al.}{2008}]{irwin08}
{Irwin} M.~J.,  {Ferguson} A.~M.~N.,  {Huxor} A.~P.,  {Tanvir} N.~R.,  {Ibata}
  R.~A.,    {Lewis} G.~F.,  2008, \apjl, 676, L17

\bibitem[\protect\citeauthoryear{{Kalirai}, {Beaton}, {Geha}, {Gilbert},
  {Guhathakurta}, {Kirby}, {Majewski}, {Ostheimer}, {Patterson} \&
  {Wolf}}{{Kalirai} et~al.}{2010}]{kalirai10}
{Kalirai} J.~S.,  {Beaton} R.~L.,  {Geha} M.~C.,  {Gilbert} K.~M.,
  {Guhathakurta} P.,  {Kirby} E.~N.,  {Majewski} S.~R.,  {Ostheimer} J.~C.,
  {Patterson} R.~J.,    {Wolf} J.,  2010, \apj, 711, 671

\bibitem[\protect\citeauthoryear{{Kalirai}, {Gilbert}, {Guhathakurta},
  {Majewski}, {Ostheimer}, {Rich}, {Cooper}, {Reitzel} \&
  {Patterson}}{{Kalirai} et~al.}{2006}]{kalirai06}
{Kalirai} J.~S.,  {Gilbert} K.~M.,  {Guhathakurta} P.,  {Majewski} S.~R.,
  {Ostheimer} J.~C.,  {Rich} R.~M.,  {Cooper} M.~C.,  {Reitzel} D.~B.,
  {Patterson} R.~J.,  2006, \apj, 648, 389

\bibitem[\protect\citeauthoryear{{Kalirai}, {Zucker}, {Guhathakurta}, {Geha},
  {Kniazev}, {Martinez-Delgado}, {Bell}, {Grebel} \& {Gilbert}}{{Kalirai}
  et~al.}{2009}]{kalirai09}
{Kalirai} J.~S.,  {Zucker} D.~B.,  {Guhathakurta} P.,  {Geha} M.,  {Kniazev}
  A.~Y.,  {Martinez-Delgado} D.,  {Bell} E.~F.,  {Grebel} E.~K.,    {Gilbert}
  K.~M.,  2009, ArXiv e-prints

\bibitem[\protect\citeauthoryear{{Kirby}, {Simon}, {Geha}, {Guhathakurta} \&
  {Frebel}}{{Kirby} et~al.}{2008}]{kirby08}
{Kirby} E.~N.,  {Simon} J.~D.,  {Geha} M.,  {Guhathakurta} P.,    {Frebel} A.,
  2008, \apjl, 685, L43

\bibitem[\protect\citeauthoryear{{Klypin}, {Kravtsov}, {Valenzuela} \&
  {Prada}}{{Klypin} et~al.}{1999}]{klypin99}
{Klypin} A.,  {Kravtsov} A.~V.,  {Valenzuela} O.,    {Prada} F.,  1999, \apj,
  522, 82

\bibitem[\protect\citeauthoryear{{Koch} \& {Grebel}}{{Koch} \&
  {Grebel}}{2006}]{koch06}
{Koch} A.,  {Grebel} E.~K.,  2006, \aj, 131, 1405

\bibitem[\protect\citeauthoryear{{Koposov}, {Belokurov}, {Evans}, {Hewett},
  {Irwin}, {Gilmore}, {Zucker}, {Rix}, {Fellhauer}, {Bell} \&
  {Glushkova}}{{Koposov} et~al.}{2008}]{koposov08}
{Koposov} S.,  {Belokurov} V.,  {Evans} N.~W.,  {Hewett} P.~C.,  {Irwin} M.~J.,
   {Gilmore} G.,  {Zucker} D.~B.,  {Rix} H.,  {Fellhauer} M.,  {Bell} E.~F.,
  {Glushkova} E.~V.,  2008, \apj, 686, 279

\bibitem[\protect\citeauthoryear{{Letarte}, {Chapman}, {Collins}, {Ibata},
  {Irwin}, {Ferguson}, {Lewis}, {Martin}, {McConnachie} \& {Tanvir}}{{Letarte}
  et~al.}{2009}]{letarte09}
{Letarte} B.,  {Chapman} S.~C.,  {Collins} M.,  {Ibata} R.~A.,  {Irwin} M.~J.,
  {Ferguson} A.~M.~N.,  {Lewis} G.~F.,  {Martin} N.,  {McConnachie} A.,
  {Tanvir} N.,  2009, ArXiv e-prints

\bibitem[\protect\citeauthoryear{{Mackey}, {Ferguson}, {Irwin}, {Martin},
  {Huxor}, {Tanvir}, {Chapman}, {Ibata}, {Lewis} \& {McConnachie}}{{Mackey}
  et~al.}{2009}]{mackey09}
{Mackey} A.~D.,  {Ferguson} A.~M.~N.,  {Irwin} M.~J.,  {Martin} N.~F.,  {Huxor}
  A.~P.,  {Tanvir} N.~R.,  {Chapman} S.~C.,  {Ibata} R.~A.,  {Lewis} G.~F.,
  {McConnachie} A.~W.,  2009, ArXiv e-prints

\bibitem[\protect\citeauthoryear{{Majewski}, {Beaton}, {Patterson}, {Kalirai},
  {Geha}, {Mu{\~n}oz}, {Seigar}, {Guhathakurta}, {Gilbert}, {Rich}, {Bullock}
  \& {Reitzel}}{{Majewski} et~al.}{2007}]{majewski07}
{Majewski} S.~R.,  {Beaton} R.~L.,  {Patterson} R.~J.,  {Kalirai} J.~S.,
  {Geha} M.~C.,  {Mu{\~n}oz} R.~R.,  {Seigar} M.~S.,  {Guhathakurta} P.,
  {Gilbert} K.~M.,  {Rich} R.~M.,  {Bullock} J.~S.,    {Reitzel} D.~B.,  2007,
  \apjl, 670, L9

\bibitem[\protect\citeauthoryear{{Marcolini}, {D'Ercole}, {Battaglia} \&
  {Gibson}}{{Marcolini} et~al.}{2008}]{marcolini08}
{Marcolini} A.,  {D'Ercole} A.,  {Battaglia} G.,    {Gibson} B.~K.,  2008,
  \mnras, 386, 2173

\bibitem[\protect\citeauthoryear{{Martin}, {de Jong} \& {Rix}}{{Martin}
  et~al.}{2008}]{martin08}
{Martin} N.~F.,  {de Jong} J.~T.~A.,    {Rix} H.-W.,  2008, \apj, 684, 1075

\bibitem[\protect\citeauthoryear{{Martin}, {Ibata}, {Chapman}, {Irwin} \&
  {Lewis}}{{Martin} et~al.}{2007}]{martin07}
{Martin} N.~F.,  {Ibata} R.~A.,  {Chapman} S.~C.,  {Irwin} M.,    {Lewis}
  G.~F.,  2007, \mnras, 380, 281

\bibitem[\protect\citeauthoryear{{Martin}, {Ibata}, {Irwin}, {Chapman},
  {Lewis}, {Ferguson}, {Tanvir} \& {McConnachie}}{{Martin}
  et~al.}{2006}]{martin06}
{Martin} N.~F.,  {Ibata} R.~A.,  {Irwin} M.~J.,  {Chapman} S.,  {Lewis} G.~F.,
  {Ferguson} A.~M.~N.,  {Tanvir} N.,    {McConnachie} A.~W.,  2006, \mnras,
  371, 1983

\bibitem[\protect\citeauthoryear{{Martin}, {McConnachie}, {Irwin}, {Widrow},
  {Ferguson}, {Ibata}, {Dubinski}, {Babul}, {Chapman}, {Fardal}, {Lewis},
  {Navarro} \& {Rich}}{{Martin} et~al.}{2009}]{martin09}
{Martin} N.~F.,  {McConnachie} A.~W.,  {Irwin} M.,  {Widrow} L.~M.,  {Ferguson}
  A.~M.~N.,  {Ibata} R.~A.,  {Dubinski} J.,  {Babul} A.,  {Chapman} S.,
  {Fardal} M.,  {Lewis} G.~F.,  {Navarro} J.,    {Rich} R.~M.,  2009, ArXiv
  e-prints

\bibitem[\protect\citeauthoryear{{Mateo}}{{Mateo}}{1998}]{mateo98}
{Mateo} M.~L.,  1998, \araa, 36, 435

\bibitem[\protect\citeauthoryear{{McConnachie}, {Huxor}, {Martin}, {Irwin},
  {Chapman}, {Fahlman}, {Ferguson}, {Ibata}, {Lewis}, {Richer} \&
  {Tanvir}}{{McConnachie} et~al.}{2008}]{mcconnachie08}
{McConnachie} A.,  {Huxor} A.,  {Martin} N.,  {Irwin} M.,  {Chapman} S.,
  {Fahlman} G.,  {Ferguson} A.,  {Ibata} R.,  {Lewis} G.,  {Richer} H.,
  {Tanvir} N.,  2008, ArXiv e-prints, 806

\bibitem[\protect\citeauthoryear{{McConnachie}, {Irwin}, {Chapman}, {Ibata},
  {Ferguson}, {Lewis} \& {Tanvir}}{{McConnachie} et~al.}{2005}]{mcconnachie05b}
{McConnachie} A.,  {Irwin} M.,  {Chapman} S.,  {Ibata} R.,  {Ferguson} A.,
  {Lewis} G.,    {Tanvir} N.,  2005, in {Jerjen} H.,  {Binggeli} B.,  eds, IAU
  Colloq. 198: Near-fields cosmology with dwarf elliptical galaxies {Andromeda
  and the seven dwarfs}.
pp 84--91

\bibitem[\protect\citeauthoryear{{McConnachie} et~al.,}{{McConnachie}
  et~al.}{2009}]{mcconnachie09}
{McConnachie} A.~W.,  et~al., 2009, \nat, 461, 66

\bibitem[\protect\citeauthoryear{{McConnachie} \& {Irwin}}{{McConnachie} \&
  {Irwin}}{2006a}]{mcconnachie06b}
{McConnachie} A.~W.,  {Irwin} M.~J.,  2006a, \mnras, 365, 1263

\bibitem[\protect\citeauthoryear{{McConnachie} \& {Irwin}}{{McConnachie} \&
  {Irwin}}{2006b}]{mcconnachie06a}
{McConnachie} A.~W.,  {Irwin} M.~J.,  2006b, \mnras, 365, 902

\bibitem[\protect\citeauthoryear{{McConnachie}, {Irwin}, {Ferguson}, {Ibata},
  {Lewis} \& {Tanvir}}{{McConnachie} et~al.}{2004}]{mcconnachie04b}
{McConnachie} A.~W.,  {Irwin} M.~J.,  {Ferguson} A.~M.~N.,  {Ibata} R.~A.,
  {Lewis} G.~F.,    {Tanvir} N.,  2004, \mnras, 350, 243

\bibitem[\protect\citeauthoryear{{McConnachie}, {Irwin}, {Ferguson}, {Ibata},
  {Lewis} \& {Tanvir}}{{McConnachie} et~al.}{2005}]{mcconnachie05a}
{McConnachie} A.~W.,  {Irwin} M.~J.,  {Ferguson} A.~M.~N.,  {Ibata} R.~A.,
  {Lewis} G.~F.,    {Tanvir} N.,  2005, \mnras, 356, 979

\bibitem[\protect\citeauthoryear{{Moore}, {Ghigna}, {Governato}, {Lake},
  {Quinn}, {Stadel} \& {Tozzi}}{{Moore} et~al.}{1999}]{moore99}
{Moore} B.,  {Ghigna} S.,  {Governato} F.,  {Lake} G.,  {Quinn} T.,  {Stadel}
  J.,    {Tozzi} P.,  1999, \apjl, 524, L19

\bibitem[\protect\citeauthoryear{{Mu{\~n}oz}, {Majewski} \&
  {Johnston}}{{Mu{\~n}oz} et~al.}{2008}]{munoz08}
{Mu{\~n}oz} R.~R.,  {Majewski} S.~R.,    {Johnston} K.~V.,  2008, \apj, 679,
  346

\bibitem[\protect\citeauthoryear{{Niederste-Ostholt}, {Belokurov}, {Evans},
  {Gilmore}, {Wyse} \& {Norris}}{{Niederste-Ostholt}
  et~al.}{2009}]{niederste09}
{Niederste-Ostholt} M.,  {Belokurov} V.,  {Evans} N.~W.,  {Gilmore} G.,  {Wyse}
  R.~F.~G.,    {Norris} J.~E.,  2009, \mnras, 398, 1771

\bibitem[\protect\citeauthoryear{{Oh}, {Lin} \& {Aarseth}}{{Oh}
  et~al.}{1995}]{oh95}
{Oh} K.~S.,  {Lin} D.~N.~C.,    {Aarseth} S.~J.,  1995, \apj, 442, 142

\bibitem[\protect\citeauthoryear{{Pe{\~n}arrubia}, {McConnachie} \&
  {Navarro}}{{Pe{\~n}arrubia} et~al.}{2008}]{penarrubia08a}
{Pe{\~n}arrubia} J.,  {McConnachie} A.~W.,    {Navarro} J.~F.,  2008, \apj,
  672, 904

\bibitem[\protect\citeauthoryear{{Pe{\~n}arrubia}, {Navarro} \&
  {McConnachie}}{{Pe{\~n}arrubia} et~al.}{2008}]{penarrubia08b}
{Pe{\~n}arrubia} J.,  {Navarro} J.~F.,    {McConnachie} A.~W.,  2008, \apj,
  673, 226

\bibitem[\protect\citeauthoryear{{Piatek} \& {Pryor}}{{Piatek} \&
  {Pryor}}{1995}]{piatek95}
{Piatek} S.,  {Pryor} C.,  1995, \aj, 109, 1071

\bibitem[\protect\citeauthoryear{{Richstone} \& {Tremaine}}{{Richstone} \&
  {Tremaine}}{1986}]{richstone86}
{Richstone} D.~O.,  {Tremaine} S.,  1986, \aj, 92, 72

\bibitem[\protect\citeauthoryear{{Robin}, {Reyl{\'e}}, {Derri{\`e}re} \&
  {Picaud}}{{Robin} et~al.}{2004}]{robin04}
{Robin} A.~C.,  {Reyl{\'e}} C.,  {Derri{\`e}re} S.,    {Picaud} S.,  2004,
  \aap, 416, 157

\bibitem[\protect\citeauthoryear{{Salvadori} \& {Ferrara}}{{Salvadori} \&
  {Ferrara}}{2009}]{salvadori09}
{Salvadori} S.,  {Ferrara} A.,  2009, \mnras, 395, L6

\bibitem[\protect\citeauthoryear{{Schlegel}, {Finkbeiner} \&
  {Davis}}{{Schlegel} et~al.}{1998}]{schlegel98}
{Schlegel} D.~J.,  {Finkbeiner} D.~P.,    {Davis} M.,  1998, \apj, 500, 525

\bibitem[\protect\citeauthoryear{{Simon} \& {Geha}}{{Simon} \&
  {Geha}}{2007}]{simon07}
{Simon} J.~D.,  {Geha} M.,  2007, \apj, 670, 313

\bibitem[\protect\citeauthoryear{{Sohn}, {Majewski}, {Mu{\~n}oz}, {Kunkel},
  {Johnston}, {Ostheimer}, {Guhathakurta}, {Patterson}, {Siegel} \&
  {Cooper}}{{Sohn} et~al.}{2007}]{sohn07}
{Sohn} S.~T.,  {Majewski} S.~R.,  {Mu{\~n}oz} R.~R.,  {Kunkel} W.~E.,
  {Johnston} K.~V.,  {Ostheimer} J.~C.,  {Guhathakurta} P.,  {Patterson} R.~J.,
   {Siegel} M.~H.,    {Cooper} M.~C.,  2007, \apj, 663, 960

\bibitem[\protect\citeauthoryear{{Somerville}}{{Somerville}}{2002}]{somerville%
02}
{Somerville} R.~S.,  2002, \apjl, 572, L23

\bibitem[\protect\citeauthoryear{{Strigari}, {Bullock}, {Kaplinghat}, {Simon},
  {Geha}, {Willman} \& {Walker}}{{Strigari} et~al.}{2008}]{strigari08}
{Strigari} L.~E.,  {Bullock} J.~S.,  {Kaplinghat} M.,  {Simon} J.~D.,  {Geha}
  M.,  {Willman} B.,    {Walker} M.~G.,  2008, \nat, 454, 1096

\bibitem[\protect\citeauthoryear{{Tollerud}, {Bullock}, {Strigari} \&
  {Willman}}{{Tollerud} et~al.}{2008}]{tollerud08}
{Tollerud} E.~J.,  {Bullock} J.~S.,  {Strigari} L.~E.,    {Willman} B.,  2008,
  \apj, 688, 277

\bibitem[\protect\citeauthoryear{{Walker}, {Belokurov}, {Evans}, {Irwin},
  {Mateo}, {Olszewski} \& {Gilmore}}{{Walker} et~al.}{2009}]{walker09a}
{Walker} M.~G.,  {Belokurov} V.,  {Evans} N.~W.,  {Irwin} M.~J.,  {Mateo} M.,
  {Olszewski} E.~W.,    {Gilmore} G.,  2009, \apjl, 694, L144

\bibitem[\protect\citeauthoryear{{Walker}, {Mateo}, {Olszewski}, {Gnedin},
  {Wang}, {Sen} \& {Woodroofe}}{{Walker} et~al.}{2007}]{walker07}
{Walker} M.~G.,  {Mateo} M.,  {Olszewski} E.~W.,  {Gnedin} O.~Y.,  {Wang} X.,
  {Sen} B.,    {Woodroofe} M.,  2007, \apjl, 667, L53

\bibitem[\protect\citeauthoryear{{Walker}, {Mateo}, {Olszewski},
  {Pe{\~n}arrubia}, {Wyn Evans} \& {Gilmore}}{{Walker}
  et~al.}{2009}]{walker09b}
{Walker} M.~G.,  {Mateo} M.,  {Olszewski} E.~W.,  {Pe{\~n}arrubia} J.,  {Wyn
  Evans} N.,    {Gilmore} G.,  2009, ArXiv e-prints

\bibitem[\protect\citeauthoryear{{Willman}, {Blanton}, {West}, {Dalcanton},
  {Hogg}, {Schneider}, {Wherry}, {Yanny} \& {Brinkmann}}{{Willman}
  et~al.}{2005}]{willman05a}
{Willman} B.,  {Blanton} M.~R.,  {West} A.~A.,  {Dalcanton} J.~J.,  {Hogg}
  D.~W.,  {Schneider} D.~P.,  {Wherry} N.,  {Yanny} B.,    {Brinkmann} J.,
  2005, \aj, 129, 2692

\bibitem[\protect\citeauthoryear{{Willman}, {Dalcanton}, {Martinez-Delgado},
  {West}, {Blanton}, {Hogg}, {Barentine}, {Brewington}, {Harvanek}, {Kleinman},
  {Krzesinski}, {Long}, {Neilsen} Jr., {Nitta} \& {Snedden}}{{Willman}
  et~al.}{2005}]{willman05b}
{Willman} B.,  {Dalcanton} J.~J.,  {Martinez-Delgado} D.,  {West} A.~A.,
  {Blanton} M.~R.,  {Hogg} D.~W.,  {Barentine} J.~C.,  {Brewington} H.~J.,
  {Harvanek} M.,  {Kleinman} S.~J.,  {Krzesinski} J.,  {Long} D.,  {Neilsen}
  Jr. E.~H.,  {Nitta} A.,    {Snedden} S.~A.,  2005, \apjl, 626, L85

\bibitem[\protect\citeauthoryear{{Willman}, {Governato}, {Dalcanton}, {Reed} \&
  {Quinn}}{{Willman} et~al.}{2004}]{willman04}
{Willman} B.,  {Governato} F.,  {Dalcanton} J.~J.,  {Reed} D.,    {Quinn} T.,
  2004, \mnras, 353, 639

\bibitem[\protect\citeauthoryear{{Wolf}, {Martinez}, {Bullock}, {Kaplinghat},
  {Geha}, {Munoz}, {Simon} \& {Avedo}}{{Wolf} et~al.}{2009}]{wolf09}
{Wolf} J.,  {Martinez} G.~D.,  {Bullock} J.~S.,  {Kaplinghat} M.,  {Geha} M.,
  {Munoz} R.~R.,  {Simon} J.~D.,    {Avedo} F.~F.,  2009, ArXiv e-prints

\bibitem[\protect\citeauthoryear{{Zucker} et~al.,}{{Zucker}
  et~al.}{2004}]{zucker04}
{Zucker} D.~B.,  et~al., 2004, \apjl, 612, L121

\bibitem[\protect\citeauthoryear{{Zucker} et~al.,}{{Zucker}
  et~al.}{2006a}]{zucker06b}
{Zucker} D.~B.,  et~al., 2006a, \apjl, 650, L41

\bibitem[\protect\citeauthoryear{{Zucker} et~al.,}{{Zucker}
  et~al.}{2006b}]{zucker06a}
{Zucker} D.~B.,  et~al., 2006b, \apjl, 643, L103

\bibitem[\protect\citeauthoryear{{Zucker} et~al.,}{{Zucker}
  et~al.}{2007}]{zucker07}
{Zucker} D.~B.,  et~al., 2007, \apjl, 659, L21

\end{thebibliography}
\bibliographystyle{mn2e}

\end{document}